\documentclass[12pt]{article}

\usepackage{axodraw}
\usepackage{epsfig}

\makeatletter

\def\paragraph{\@startsection{paragraph}{4}{\z@}{+2.00ex plus
 +1ex minus +.2ex}{0.5ex plus .2ex}{\it\normalsize}}

\def\section{\@startsection {section}{1}{\z@}{+3.0ex plus +1ex minus
  +.2ex}{2.3ex plus .2ex}{\normalsize\bf\boldmath}}
\def\subsection{\@startsection{subsection}{2}{\z@}{+2.5ex plus +1ex
minus +.2ex}{1.5ex plus .2ex}{\normalsize\bf\boldmath}}
\def\subsubsection{\@startsection{subsubsection}{3}{\z@}{+3.25ex plus
 +1ex minus +.2ex}{1.5ex plus .2ex}{\normalsize\it}}

\expandafter\ifx\csname mathrm\endcsname\relax\def\mathrm#1{{\rm #1}}\fi

\newcounter{saveeqn}

\@addtoreset{equation}{section}

\newcount\@tempcntc
\def\@citex[#1]#2{\if@filesw\immediate\write\@auxout{\string\citation{#2}}\fi
  \@tempcnta\z@\@tempcntb\m@ne\def\@citea{}\@cite{\@for\@citeb:=#2\do
    {\@ifundefined
       {b@\@citeb}{\@citeo\@tempcntb\m@ne\@citea
        \def\@citea{,\penalty\@m\ }{\bf ?}\@warning
       {Citation `\@citeb' on page \thepage \space undefined}}%
    {\setbox\z@\hbox{\global\@tempcntc0\csname
b@\@citeb\endcsname\relax}%
     \ifnum\@tempcntc=\z@ \@citeo\@tempcntb\m@ne
       \@citea\def\@citea{,\penalty\@m}
       \hbox{\csname b@\@citeb\endcsname}%
     \else
      \advance\@tempcntb\@ne
      \ifnum\@tempcntb=\@tempcntc
      \else\advance\@tempcntb\m@ne\@citeo
      \@tempcnta\@tempcntc\@tempcntb\@tempcntc\fi\fi}}\@citeo}{#1}}

\def\@citeo{\ifnum\@tempcnta>\@tempcntb\else\@citea
  \def\@citea{,\penalty\@m}%
  \ifnum\@tempcnta=\@tempcntb\the\@tempcnta\else
   {\advance\@tempcnta\@ne\ifnum\@tempcnta=\@tempcntb \else
\def\@citea{--}\fi
    \advance\@tempcnta\m@ne\the\@tempcnta\@citea\the\@tempcntb}\fi\fi}

\def\nl{\nonumber\\}

\newcommand{\lsim}
{\mathrel{\raisebox{-.3em}{$\stackrel{\displaystyle <}{\sim}$}}}
\newcommand{\gsim}
{\mathrel{\raisebox{-.3em}{$\stackrel{\displaystyle >}{\sim}$}}}
\def\asymp#1%
{\mathrel{\raisebox{-.4em}{$\widetilde{\scriptstyle #1}$}}}

\def\Nequal#1%
{\mathrel{\raisebox{-.5em}{$\stackrel{=}{\scriptstyle\rm#1}$}}}
\newcommand{\dsl}[1]{\not \hspace{-0.7mm}#1}
\def\dsl{\mathpalette\make@slash}
\def\make@slash#1#2{\setbox\z@\hbox{$#1#2$}%
  \hbox to 0pt{\hss$#1/$\hss\kern-\wd0}\box0}

\def\beq{\begin{equation}}
\def\eeq{\end{equation}}
\def\beqar{\begin{eqnarray}}
\def\eeqar{\end{eqnarray}}
\def\barr#1{\begin{array}{#1}}
\def\earr{\end{array}}
\def\bfi{\begin{figure}}
\def\efi{\end{figure}}
\def\btab{\begin{table}}
\def\etab{\end{table}}
\def\bce{\begin{center}}
\def\ece{\end{center}}
\def\nn{\nonumber}
\def\disp{\displaystyle}
\def\text{\textstyle}

\def\al{\alpha}
\def\be{\beta}
\def\Ga{\Gamma}
\def\ga{\gamma}
\def\de{\delta}

\def\eps{\epsilon}

\def\si{\sigma}

\def\refeq#1{\mbox{(\ref{#1})}}

\def\reffi#1{\mbox{Figure~\ref{#1}}}
\def\reffis#1{\mbox{Figures~\ref{#1}}}
\def\refta#1{\mbox{Table~\ref{#1}}}

\def\refse#1{\mbox{Section~\ref{#1}}}

\def\citere#1{\mbox{Ref.~\cite{#1}}}
\def\citeres#1{\mbox{Refs.~\cite{#1}}}

\newcommand{\TeV}{\unskip\,\mathrm{TeV}}
\newcommand{\GeV}{\unskip\,\mathrm{GeV}}
\newcommand{\MeV}{\unskip\,\mathrm{MeV}}

\newcommand{\fba}{\unskip\,\mathrm{fb}}

\newcommand{\ri}{{\mathrm{i}}}
\newcommand{\rd}{{\mathrm{d}}}

\newcommand{\M}{{\cal{M}}}

\def\mathswitchr#1{\relax\ifmmode{\mathrm{#1}}\else$\mathrm{#1}$\fi}

\newcommand{\PF}{F}
\newcommand{\PV}{V}
\newcommand{\PW}{\mathswitchr W}
\newcommand{\Pw}{\mathswitchr w}
\newcommand{\PZ}{\mathswitchr Z}

\newcommand{\Pg}{\mathswitchr g}
\newcommand{\PH}{\mathswitchr H}

\newcommand{\Pe}{\mathswitchr e}

\newcommand{\Pd}{\mathswitchr d}

\newcommand{\Pu}{\mathswitchr u}

\newcommand{\Ps}{\mathswitchr s}

\newcommand{\Pc}{\mathswitchr c}

\newcommand{\Pb}{\mathswitchr b}

\newcommand{\Pt}{\mathswitchr t}

\newcommand{\Pep}{\mathswitchr {e^+}}
\newcommand{\Pem}{\mathswitchr {e^-}}

\def\mathswitch#1{\relax\ifmmode#1\else$#1$\fi}

\newcommand{\MF}{\mathswitch {m_\PF}}
\newcommand{\MV}{\mathswitch {M_\PV}}
\newcommand{\MW}{\mathswitch {M_\PW}}

\newcommand{\MZ}{\mathswitch {M_\PZ}}
\newcommand{\MH}{\mathswitch {M_\PH}}
\newcommand{\Me}{\mathswitch {m_\Pe}}

\newcommand{\Mt}{\mathswitch {m_\Pt}}
\newcommand{\GW}{\Gamma_{\PW}}
\newcommand{\GZ}{\Gamma_{\PZ}}
\newcommand{\GV}{\Gamma_{\PV}}
\newcommand{\Gt}{\Gamma_{\Pt}}

\newcommand{\sw}{\mathswitch {s_\Pw}}
\newcommand{\cw}{\mathswitch {c_\Pw}}

\newcommand{\GF}{\mathswitch {G_\mu}}

\newcommand{\born}{{\mathrm{Born}}}

\renewcommand{\min}{{\mathrm{min}}}
\renewcommand{\max}{{\mathrm{max}}}

\newcommand{\ISR}{{\mathrm{ISR}}}

\newcommand{\spac}{\langle p_a p_c \rangle}
\newcommand{\spad}{\langle p_a p_d \rangle}
\newcommand{\spae}{\langle p_a p_e \rangle}
\newcommand{\spaf}{\langle p_a p_f \rangle}
\newcommand{\spag}{\langle p_a p_g \rangle}

\newcommand{\spba}{\langle p_b p_a \rangle}%
\newcommand{\spbc}{\langle p_b p_c \rangle}
\newcommand{\spbd}{\langle p_b p_d \rangle}
\newcommand{\spbe}{\langle p_b p_e \rangle}
\newcommand{\spbf}{\langle p_b p_f \rangle}

\newcommand{\spbh}{\langle p_b p_h \rangle}
\newcommand{\spca}{\langle p_c p_a \rangle}%
\newcommand{\spcd}{\langle p_c p_d \rangle}
\newcommand{\spce}{\langle p_c p_e \rangle}

\newcommand{\spcg}{\langle p_c p_g \rangle}

\newcommand{\spdf}{\langle p_d p_f \rangle}

\newcommand{\spdh}{\langle p_d p_h \rangle}
\newcommand{\spef}{\langle p_e p_f \rangle}
\newcommand{\speg}{\langle p_e p_g \rangle}

\newcommand{\spfh}{\langle p_f p_h \rangle}

\newcommand{\cspad}{\spad^*}

\newcommand{\cspaf}{\spaf^*}

\newcommand{\cspba}{\spba^*}%

\newcommand{\cspbd}{\spbd^*}

\newcommand{\cspbf}{\spbf^*}

\newcommand{\cspbh}{\spbh^*}
\newcommand{\cspca}{\spca^*}%
\newcommand{\cspcd}{\spcd^*}

\newcommand{\cspdf}{\spdf^*}

\newcommand{\cspdh}{\spdh^*}
\newcommand{\cspef}{\spef^*}

\newcommand{\cspfh}{\spfh^*}

\def\Re{\mathop{\mathrm{Re}}\nolimits}

\newcommand{\eeffff}{\Pep\Pem\to 4f}
\newcommand{\eeffffg}{\Pep\Pem\to 4f\ga}

\newcommand{\bk}{{\bf k}}
\newcommand{\bfr}{{\bf r}}

\def\draftdate{\relax}
\def\mda{\relax}
\def\mua{\relax}
\def\mla{\relax}
\def\draft{
\def\thtystars{******************************}
\def\sixtystars{\thtystars\thtystars}
\typeout{}
\typeout{\sixtystars**}
\typeout{* Draft mode!
         For final version remove \protect\draft\space in source file *}
\typeout{\sixtystars**}
\typeout{}
\def\draftdate{\today}
\def\mua{\marginpar[\boldmath\hfil$\uparrow$]%
                   {\boldmath$\uparrow$\hfil}%
                    \typeout{marginpar: $\uparrow$}\ignorespaces}
\def\mda{\marginpar[\boldmath\hfil$\downarrow$]%
                   {\boldmath$\downarrow$\hfil}%
                    \typeout{marginpar: $\downarrow$}\ignorespaces}
\def\mla{\marginpar[\boldmath\hfil$\rightarrow$]%
                   {\boldmath$\leftarrow $\hfil}%
                    \typeout{marginpar: $\leftrightarrow$}\ignorespaces}
\def\Mua{\marginpar[\boldmath\hfil$\Uparrow$]%
                   {\boldmath$\Uparrow$\hfil}%
                    \typeout{marginpar: $\uparrow$}\ignorespaces}
\def\Mda{\marginpar[\boldmath\hfil$\Downarrow$]%
                   {\boldmath$\Downarrow$\hfil}%
                    \typeout{marginpar: $\downarrow$}\ignorespaces}
\def\Mla{\marginpar[\boldmath\hfil$\Rightarrow$]%
                   {\boldmath$\Leftarrow $\hfil}%
                    \typeout{marginpar: $\leftrightarrow$}\ignorespaces}
\overfullrule 5pt
\oddsidemargin -15mm
\marginparwidth 29mm
}

\def\stars{\strut\leaders\hbox{*}\hfill\strut}
\def\starline{\hfil\strut\hfil\hbox to \textwidth {\stars}\hfil}

\begin{document}
\thispagestyle{empty}
\def\thefootnote{\fnsymbol{footnote}}
\setcounter{footnote}{1}
\null
\draftdate\hfill DESY 02-081 \\
\strut\hfill KA-TP-10-2002 \\
\strut\hfill hep-ph/0206070 
\vfill
\begin{center}
{\large \bf\boldmath
{\sc Lusifer:} a LUcid approach to SIx-FERmion production
\par} \vskip 2em
\vspace{1cm}
{\large
{\sc Stefan Dittmaier$^1$%
\footnote{Heisenberg fellow of the Deutsche Forschungsgemeinschaft}
and Markus Roth$^{2}$ } } 
\\[.5cm]
$^1$ {\it Deutsches Elektronen-Synchrotron DESY \\
D-22603 Hamburg, Germany}
\\[0.3cm]
$^2$ {\it Institut f\"ur Theoretische Physik, Universit\"at Karlsruhe\\
D-76131 Karlsruhe, Germany}
\par 
\end{center}\par
\vskip 2.0cm {\bf Abstract:} \par 
{\sc Lusifer} is a Monte Carlo event generator for all
processes $\Pep\Pem\to 6\,$fermions,
which is based on the multi-channel Monte Carlo integration technique
and employs the full set of tree-level diagrams.
External fermions are taken to be massless, but can be arbitrarily
polarized.
The calculation of the helicity amplitudes and of the squared matrix
elements is presented in a compact way.
Initial-state radiation is included at the leading logarithmic level 
using the structure-function approach.
The discussion of numerical results contains a 
comprehensive list of cross sections relevant for a $500\GeV$ collider,
including a tuned comparison to results obtained with the combination
of the {\sc Whizard} and {\sc Madgraph} packages as far as possible.
Moreover, for off-shell top-quark pair production
and the production of a Higgs boson in the intermediate mass range
we additionally discuss some
phenomenologically interesting distributions. Finally, we numerically
analyze the effects of gauge-invariance violation by comparing various
ways of introducing decay widths of intermediate top quarks, gauge
and Higgs bosons.
\par
\vskip 1cm
\vfill
\noindent
June 2002   
\null
\setcounter{page}{0}
\clearpage
\def\thefootnote{\arabic{footnote}}
\setcounter{footnote}{0}

\section{Introduction}
\label{se:intro}

At future $\Pep\Pem$ colliders, such as TESLA \cite{Accomando:1998wt},
some of the most interesting elementary particle reactions 
lead to final states involving six fermions.
Typically these multi-particle final states represent the final decay
stage of unstable particles that were produced as 
resonances in subprocesses. 
Because of the high precision of future colliders, predictions that
are entirely based on the narrow-width approximation (with possible 
improvements by spin correlations between production and decays
or by resonance expansions) are not sufficient in the most cases.
In practice, this means that the full set of Feynman diagrams 
(not only the resonant ones) has to be considered in perturbative calculations,
at least in lowest order.
The situation is very similar to four-fermion production, $\Pep\Pem\to 4f$, 
and the related radiative processes $\Pep\Pem\to 4f+\gamma$, at LEP2
\cite{Grunewald:2000ju}.
Although the complexity of the corresponding calculations 
increases when turning from four-fermion to six-fermion production, 
$\Pep\Pem\to 6f$, most of the existing results are obtained by 
applying and further developing the methods that had been worked out for 
four-fermion production. 
We briefly summarize these results according to the subprocesses of interest:

\paragraph{Top-quark pair production}
 
Since top quarks decay via the cascade $\Pt\to\Pb\PW^+\to\Pb f\bar f'$
into three fermions, the production of $\Pt\bar\Pt$ pairs corresponds
to a particular class of $\Pep\Pem\to 6f$ processes:
$\Pep\Pem\to\Pb\bar\Pb f_1 \bar f'_1 f_2 \bar f'_2$.
Here $f_i \bar f'_i$ denote two weak isospin doublets.
Some processes are already discussed in the literature.
The specific process $\Pep\Pem\to\Pb\bar\Pb \Pu\bar \Pd \mu^-\bar\nu_\mu$
was studied in \citere{Yuasa:1997fa} with the GRACE package
\cite{Ishikawa:1993qr}.
In \citeres{Accomando:1997gu,Accomando:1997gj} and \citere{Gangemi:1999qb} 
the cases of one and two hadronically decaying W~bosons were discussed
in more detail, respectively. The former results are based on a
generalization of the program {\sc PHACT} \cite{Ballestrero:1994jn},
the latter on the {\sc ALPHA} algorithm \cite{Caravaglios:1995cd}
for matrix elements.
Recently various resonance approximations for the
top quarks were compared with a calculation based on full sets of
$\Pep\Pem\to 6f$ diagrams in \citere{Kolodziej:2001xe};
these results underline the importance of calculations based on full
sets of Feynman diagrams.
However, to our knowledge, results have not yet been presented for all
final states, for instance, not yet for 
$\Pep\Pem\to\Pb\bar\Pb \Pem\bar\nu_\Pe\nu_\Pe\Pep$.

\paragraph{Vector-boson scattering and quartic gauge-boson couplings}

One of the most promising windows to electroweak symmetry breaking
is provided by investigating the scattering of massive vector bosons,
$V_1 V_2\to V_3 V_4$. This subprocess is initiated by the emission
of the vector bosons $V_{1,2}$ from the incoming $\Pep\Pem$ system
and leads to six fermions in the final state: two remnants from the
initial state and four fermions from the decays of the vector bosons
$V_{3,4}$. Thus, the corresponding reactions are of the form
$\Pep\Pem\to\Pep\Pem/\Pep\nu_\Pe/\bar\nu_\Pe\Pem/\bar\nu_\Pe\nu_\Pe+4f$.
Many studies of vector-boson scattering were presented in the literature
(see e.g.\ \citere{Accomando:1998wt} and references therein), but with
very few exceptions they were based on approximations with respect to
the kinematics of the incoming and/or the outgoing vector bosons.
In \citeres{Gangemi:2000sk,Chierici:2001ar} the sensitivity of the
processes $\Pep\Pem\to\nu_\Pe\bar\nu_\Pe+4\,$quarks to possible
anomalous quartic gauge-boson couplings was investigated
making use of full $\Pep\Pem\to 6f$
matrix elements. The matrix elements used in \citere{Gangemi:2000sk} 
were obtained with {\sc ALPHA}, while 
in \citere{Chierici:2001ar} the
package {\sc O'Mega} \cite{Moretti:2001zz} delivered the amplitudes
and the phase-space generation was performed with 
{\sc Whizard}~\cite{Kilian:2001qz}.

\paragraph{Higgs production for intermediate Higgs-boson masses}

If the Higgs boson of the Standard Model has an intermediate mass of
$\MH\gsim 150\GeV$, it predominantly decays via $\PH\to\PW\PW\to 4f$. 
Since the Higgs boson is either produced by Higgs-strahlung off Z~bosons, 
$\Pep\Pem\to\PZ\PH$, or vector-boson fusion, 
$\Pep\Pem\to\Pep\Pem\PH/\bar\nu_\Pe\nu_\Pe\PH$,
the search for the Higgs boson in the intermediate mass range also
proceeds via $\Pep\Pem\to 6f$ processes.
In \citeres{Accomando:1997gj,Montagna:1997dc,Accomando:1998ju,Gangemi:1998vc}
the Higgs-strahlung signal was studied for various
$6f$ final states by employing full matrix elements from {\sc PHACT}
and {\sc ALPHA}. \citere{Gangemi:1998vc} also contains 
some results for the vector-boson fusion channel.

\paragraph{Three-gauge-boson production}

Last but not least, all $6f$ final states in $\Pep\Pem\to 6f$
contribute to the signal of resonant three-gauge-boson production,
such as $\Pep\Pem\to\PW\PW\PZ/\PZ\PZ\PZ$, from which valuable
information on the quartic gauge-boson couplings can be deduced.
In \citeres{Accomando:1997gu,Accomando:1997gj} WWZ production was
investigated for some interesting final states using the full
$6f$ matrix elements from {\sc PHACT}. However, more detailed studies
including more final states are certainly wanted.%
\vspace{.8em}

It should be mentioned 
that the combination of the {\sc PHEGAS} and {\sc  HELAC}
packages \cite{Papadopoulos:2000tt} is also able to deal with
six-fermion production processes. However, no detailed results of 
these programs for $\Pep\Pem\to 6f$ have been presented yet in the 
literature.

In this paper we present the Monte Carlo event generator
{\sc Lusifer}, which is, in its first version, designed for all Standard
Model processes $\Pep\Pem\to 6\,$fermions in lowest order;
gluon-exchange diagrams can be optionally included for final states with
two leptons and four quarks (not yet for six-quark final states).%
\footnote{A full treatment of the production of four or more jets in
$\Pep \Pem\to 6\,$jet reactions additionally requires the inclusion
of gluon jets, as for instance done in \citere{Moretti:1997sn}.}
Technically the approach closely follows the structure of
{\sc Excalibur} \cite{Berends:1994pv} and 
the lowest-order part of {\sc RacoonWW} \cite{Denner:1999gp,Denner:2000kn}
for the processes $\Pep\Pem\to 4f(+\gamma)$.
This means the matrix elements are evaluated within the 
Weyl--van~der~Waerden (WvdW) spinor technique as described in
\citere{Dittmaier:1999nn} (see also references therein);
the external fermions are taken to be massless, but can be arbitrarily
polarized. The phase-space integration is performed with the
multi-channel Monte Carlo integration technique \cite{Berends:gf},
improved by adaptive weight optimization \cite{Kleiss:qy}.
The lowest-order predictions obtained this way are dressed by
initial-state radiation (ISR) in the leading logarithmic approximation
following the structure-function approach \cite{sf}, as summarized in 
the appendix of \citere{Beenakker:1996kt}.

Although the above list of topics shows that several studies of six-fermion
production processes have already 
been presented in the literature, no detailed
tuned comparison between results from different approaches is available
yet. We make a first step to fill this gap by giving a full list of
cross sections for $6f$ states with up to three neutrinos and up to
four quarks for a centre-of-mass (CM) energy of $500\GeV$. Moreover, 
we compare these cross sections with results obtained by
{\sc Whizard} \cite{Kilian:2001qz} and 
{\sc Madgraph} \cite{Stelzer:1994ta} as far as the combination of these
packages is applicable. In general, we find good numerical
agreement, but for several channels the limitation of these
multi-purpose programs becomes visible. We continue the tuned
comparison by comparing some distributions for topics of phenomenological
interest, such as invariant-mass and angular distributions for top-quark
and Higgs-boson production. We conclude the discussion of numerical
results by comparing various schemes for 
introducing the finite decay widths of unstable particles in the amplitudes. 
Already for CM energies in the TeV
range the gauge-invariance-breaking effects in some cases are clearly
visible, underlining the importance of this issue. Within {\sc Lusifer} 
several width schemes are implemented, comprising 
also the {\it complex-mass
scheme}, which was introduced in \citere{Denner:1999gp}
for tree-level predictions and maintains gauge invariance.
Hence, gauge-violating artefacts can be controlled by comparing a given 
width scheme with the complex-mass scheme.

The paper is organized as follows:
In \refse{se:amps} the calculation of the helicity amplitudes and of 
the squared matrix elements is described in detail, followed
by a description of the multi-channel phase-space integration
in \refse{se:psint}. 
The treatment of ISR is described in \refse{se:isr}.
Section~\ref{se:ee6fprocs} contains a classification of the processes
$\Pep\Pem\to 6f$ according to the underlying resonance subprocesses.
In \refse{se:numres} we present numerical results, 
including a comprehensive list of cross sections, their tuned
comparison with {\sc Whizard} and {\sc Madgraph} results, 
some distributions relevant for top-quark and Higgs-boson
production, and the discussion of gauge-invariance-breaking effects.
A summary is given in \refse{se:sum}.

\section{Matrix-element calculation}
\label{se:amps}

\subsection{Generic amplitudes for eight external fermions}

The Feynman diagrams contributing to a process with 8 external
fermions can be classified into different categories according
to the number of external fermion--antifermion pairs that
directly fuse to a boson. At tree level there are at least two such
pairs, called ``fermion currents'' in the following, so that
we have three categories: diagrams with 4, 3, or 2 fermion currents.
Assuming massless external fermions, the generic diagrams 
of these classes are shown in \reffis{fig:4Vdiags}--\ref{fig:2Vdiags}.
\bfi
\centerline{
\setlength{\unitlength}{1pt}
\begin{picture}(130,150)(0,0)
\ArrowLine( 25, 65)( 10, 80)
\ArrowLine( 10, 50)( 25, 65)
\ArrowLine(110, 80)( 95, 65)
\ArrowLine( 95, 65)(110, 50)
\ArrowLine( 60,100)( 75,115)
\ArrowLine( 45,115)( 60,100)
\ArrowLine( 75, 15)( 60, 30)
\ArrowLine( 60, 30)( 45, 15)
\Photon(25,65)(60, 65){2}{5}
\Photon(95,65)(60, 65){2}{5}
\Photon(60,65)(60, 30){2}{5}
\Photon(60,65)(60,100){2}{5}
\Vertex(25, 65){2.0}
\Vertex(95, 65){2.0}
\Vertex(60, 65){2.0}
\Vertex(60,100){2.0}
\Vertex(60, 30){2.0}
\put(33,50){$\PW$}
\put(65,38){$\PW$}
\put(75,75){$V_1$}
\put(43,83){$V_2$}  
\put(  0, 80){$a$}
\put(  0, 45){$b$}
\put( 38,  4){$c$}
\put( 75,  4){$d$}
\put(115, 80){$f$}
\put(115, 45){$e$}
\put( 38,122){$h$}
\put( 75,122){$g$}
\put(  0,135){\bf (4a)}
\end{picture}
\hspace{1em}
\setlength{\unitlength}{1pt}
\begin{picture}(130,150)(0,0)
\ArrowLine( 25, 30)( 10, 45)
\ArrowLine( 10, 15)( 25, 30)
\ArrowLine( 25,100)( 10,115)
\ArrowLine( 10, 85)( 25,100)
\ArrowLine( 95, 30)(110, 15)
\ArrowLine(110, 45)( 95, 30)
\ArrowLine( 95,100)(110, 85)
\ArrowLine(110,115)( 95,100)
\Photon(25, 30)(40,65){2}{5}
\Photon(25,100)(40,65){2}{5}
\Photon(95, 30)(80,65){2}{5}
\Photon(95,100)(80,65){2}{5}
\Photon(40, 65)(80,65){2}{5}
\Vertex(25, 30){2.0}
\Vertex(40, 65){2.0}
\Vertex(25,100){2.0}
\Vertex(95, 30){2.0}
\Vertex(95,100){2.0}
\Vertex(80, 65){2.0}
\put(36,83){$\PW$}
\put(36,39){$V_1$}
\put(71,39){$\PW$}
\put(71,83){$V_2$}
\put(54,50){$V_3$}
\put(  0,115){$a$}
\put(  0, 80){$b$}
\put(  0, 45){$c$}
\put(  0, 10){$d$}
\put(115, 45){$f$}
\put(115, 10){$e$}
\put(115,115){$h$}
\put(115, 80){$g$}
\put(  0,135){\bf (4b)}
\end{picture}
\hspace{1em}
\setlength{\unitlength}{1pt}
\begin{picture}(130,150)(0,0)
\ArrowLine( 25, 30)( 10, 45)
\ArrowLine( 10, 15)( 25, 30)
\ArrowLine( 25,100)( 10,115)
\ArrowLine( 10, 85)( 25,100)
\ArrowLine( 95, 30)(110, 15)
\ArrowLine(110, 45)( 95, 30)
\ArrowLine( 95,100)(110, 85)
\ArrowLine(110,115)( 95,100)
\Photon(25, 30)(40,65){2}{5}
\Photon(25,100)(40,65){2}{5}
\Photon(95, 30)(80,65){2}{5}
\Photon(95,100)(80,65){2}{5}
\DashLine(40, 65)(80,65){5}
\Vertex(25, 30){2.0}
\Vertex(40, 65){2.0}
\Vertex(25,100){2.0}
\Vertex(95, 30){2.0}
\Vertex(95,100){2.0}
\Vertex(80, 65){2.0}
\put(36,83){$V_1$}
\put(36,39){$V_2$}
\put(71,39){$V_3$}
\put(71,83){$V_4$}
\put(54,52){$S$}
\put(  0,115){$a$}
\put(  0, 80){$b$}
\put(  0, 45){$c$}
\put(  0, 10){$d$}
\put(115, 45){$f$}
\put(115, 10){$e$}
\put(115,115){$h$}
\put(115, 80){$g$}
\put(  0,135){\bf (4c)}
\end{picture}
} 
\vspace*{-.5em}
\caption{Generic diagrams with 4 fermion currents}
\label{fig:4Vdiags}
\vspace{3em}
\centerline{
\setlength{\unitlength}{1pt}
\begin{picture}(130,150)(0,0)
\ArrowLine( 10, 15)( 25, 30)
\ArrowLine( 25, 30)( 25, 65)
\ArrowLine( 25, 65)( 25,100)
\ArrowLine( 25,100)( 10,115)
\ArrowLine(110, 15)( 95, 30)
\ArrowLine( 95, 30)(110, 40)
\ArrowLine(110, 55)( 95, 65)
\ArrowLine( 95, 65)(110, 75)
\ArrowLine(110, 90)( 95,100)
\ArrowLine( 95,100)(110,115)
\Photon(25, 30)(95, 30){2}{8}
\Photon(25, 65)(95, 65){2}{8}
\Photon(25,100)(95,100){2}{8}
\Vertex(25, 30){2.0}
\Vertex(25, 65){2.0}
\Vertex(25,100){2.0}
\Vertex(95, 30){2.0}
\Vertex(95, 65){2.0}
\Vertex(95,100){2.0}
\put( 9,81){$F_1$}
\put( 9,43){$F_2$}
\put(58,86){$V_1$}
\put(58,50){$V_2$}
\put(58,14){$V_3$}
\put(  0,115){$a$}
\put(  0, 10){$b$}
\put(115,115){$c$}
\put(115, 89){$d$}
\put(115, 71){$e$}
\put(115, 54){$f$}
\put(115, 36){$g$}
\put(115, 10){$h$}
\put(  0,135){\bf (3a)}
\end{picture}
\hspace{1em}
\setlength{\unitlength}{1pt}
\begin{picture}(130,150)(0,0)
\ArrowLine( 10, 50)( 25, 65)
\ArrowLine( 25, 65)( 25,100)
\ArrowLine( 25,100)( 10,115)
\ArrowLine(110, 15)( 95, 30)
\ArrowLine( 95, 30)(110, 40)
\ArrowLine(110, 55)( 95, 65)
\ArrowLine( 95, 65)(110, 75)
\ArrowLine(110, 90)( 95,100)
\ArrowLine( 95,100)(110,115)
\Photon(25,100)(95,100){2}{8}
\Photon(25, 65)(60, 48){2}{5}
\Photon(60, 48)(95, 65){2}{5}
\Photon(60, 48)(95, 30){2}{5}
\Vertex(60, 48){2.0}
\Vertex(25, 65){2.0}
\Vertex(25,100){2.0}
\Vertex(95, 30){2.0}
\Vertex(95, 65){2.0}
\Vertex(95,100){2.0}
\put( 9,81){$F$}
\put(58,86){$V_1$}
\put(41,65){$V_2$}
\put(68,65){$\PW$}
\put(68,24){$V_3$}
\put(  0,115){$a$}
\put(  0, 45){$b$}
\put(115,115){$c$}
\put(115, 89){$d$}
\put(115, 71){$e$}
\put(115, 54){$f$}
\put(115, 36){$g$}
\put(115, 10){$h$}
\put(  0,135){\bf (3b)}
\end{picture}
\hspace{1em}
\setlength{\unitlength}{1pt}
\begin{picture}(130,150)(0,0)
\ArrowLine( 10, 15)( 25, 30)
\ArrowLine( 25, 30)( 25, 65)
\ArrowLine( 25, 65)( 10, 80)
\ArrowLine(110, 15)( 95, 30)
\ArrowLine( 95, 30)(110, 40)
\ArrowLine(110, 55)( 95, 65)
\ArrowLine( 95, 65)(110, 75)
\ArrowLine(110, 90)( 95,100)
\ArrowLine( 95,100)(110,115)
\Photon(25, 30)(95, 30){2}{8}
\Photon(25, 65)(60, 82){2}{5}
\Photon(60, 82)(95,100){2}{5}
\Photon(60, 82)(95, 65){2}{5}
\Vertex(25, 30){2.0}
\Vertex(25, 65){2.0}
\Vertex(60, 82){2.0}
\Vertex(95, 30){2.0}
\Vertex(95, 65){2.0}
\Vertex(95,100){2.0}
\put( 9,43){$F$}
\put(58,14){$V_1$}
\put(41,60){$V_2$}
\put(68,99){$\PW$}
\put(68,60){$V_3$}
\put(  0, 80){$a$}
\put(  0, 10){$b$}
\put(115,115){$e$}
\put(115, 89){$f$}
\put(115, 71){$g$}
\put(115, 54){$h$}
\put(115, 36){$c$}
\put(115, 10){$d$}
\put(  0,135){\bf (3c)}
\end{picture}
} 
\vspace*{.5em}
\centerline{
\setlength{\unitlength}{1pt}
\begin{picture}(130,150)(0,0)
\ArrowLine( 10, 50)( 25, 65)
\ArrowLine( 25, 65)( 25,100)
\ArrowLine( 25,100)( 10,115)
\ArrowLine(110, 15)( 95, 30)
\ArrowLine( 95, 30)(110, 40)
\ArrowLine(110, 55)( 95, 65)
\ArrowLine( 95, 65)(110, 75)
\ArrowLine(110, 90)( 95,100)
\ArrowLine( 95,100)(110,115)
\Photon(25,100)(95,100){2}{8}
\DashLine(25, 65)(60, 48){5}
\Photon(60, 48)(95, 65){2}{5}
\Photon(60, 48)(95, 30){2}{5}
\Vertex(60, 48){2.0}
\Vertex(25, 65){2.0}
\Vertex(25,100){2.0}
\Vertex(95, 30){2.0}
\Vertex(95, 65){2.0}
\Vertex(95,100){2.0}
\put( 2,81){top}
\put(58,86){$\PW$}
\put(41,65){$\phi$}
\put(68,65){$\PW$}
\put(68,24){$V$}
\put(  0,115){$a$}
\put(  0, 45){$b$}
\put(115,115){$c$}
\put(115, 89){$d$}
\put(115, 71){$e$}
\put(115, 54){$f$}
\put(115, 36){$g$}
\put(115, 10){$h$}
\put(  0,135){\bf (3d)}
\end{picture}
\hspace{1em}
\setlength{\unitlength}{1pt}
\begin{picture}(130,150)(0,0)
\ArrowLine( 10, 15)( 25, 30)
\ArrowLine( 25, 30)( 25, 65)
\ArrowLine( 25, 65)( 10, 80)
\ArrowLine(110, 15)( 95, 30)
\ArrowLine( 95, 30)(110, 40)
\ArrowLine(110, 55)( 95, 65)
\ArrowLine( 95, 65)(110, 75)
\ArrowLine(110, 90)( 95,100)
\ArrowLine( 95,100)(110,115)
\Photon(25, 30)(95, 30){2}{8}
\DashLine(25, 65)(60, 82){5}
\Photon(60, 82)(95,100){2}{5}
\Photon(60, 82)(95, 65){2}{5}
\Vertex(25, 30){2.0}
\Vertex(25, 65){2.0}
\Vertex(60, 82){2.0}
\Vertex(95, 30){2.0}
\Vertex(95, 65){2.0}
\Vertex(95,100){2.0}
\put( 2,43){top}
\put(58,14){$\PW$}
\put(41,60){$\phi$}
\put(68,99){$\PW$}
\put(68,60){$V$}
\put(  0, 80){$a$}
\put(  0, 10){$b$}
\put(115,115){$e$}
\put(115, 89){$f$}
\put(115, 71){$g$}
\put(115, 54){$h$}
\put(115, 36){$c$}
\put(115, 10){$d$}
\put(  0,135){\bf (3e)}
\end{picture}
\hspace{1em}
\setlength{\unitlength}{1pt}
\begin{picture}(130,150)(0,0)
\end{picture}
} 
\vspace*{-.5em}
\caption{Generic diagrams with 3 fermion currents}
\label{fig:3Vdiags}
\efi
\bfi
\centerline{
\setlength{\unitlength}{1pt}
\begin{picture}(130,150)(0,0)
\ArrowLine( 10, 55)( 25, 65)
\ArrowLine( 25, 65)( 25,100)
\ArrowLine( 25,100)( 10,115)
\ArrowLine( 10, 15)( 25, 30)
\ArrowLine( 25, 30)( 10, 40)
\ArrowLine(110, 15)( 95, 30)
\ArrowLine( 95, 30)( 95, 65)
\ArrowLine( 95, 65)(110, 75)
\ArrowLine(110, 90)( 95,100)
\ArrowLine( 95,100)(110,115)
\Photon(25,100)(95,100){2}{8}
\Photon(25, 65)(95, 65){2}{8}
\Photon(25, 30)(95, 30){2}{8}
\Vertex(25, 30){2.0}
\Vertex(25, 65){2.0}
\Vertex(25,100){2.0}
\Vertex(95, 30){2.0}
\Vertex(95, 65){2.0}
\Vertex(95,100){2.0}
\put(  9,81){$F_1$}
\put(101,43){$F_2$}
\put( 58,86){$V_1$}
\put( 58,50){$V_2$}
\put( 58,14){$V_3$}
\put(  0,115){$a$}
\put(  0, 55){$b$}
\put(  0, 36){$g$}
\put(  0, 10){$h$}
\put(115,115){$e$}
\put(115, 89){$f$}
\put(115, 71){$c$}
\put(115, 10){$d$}
\put(  0,135){\bf (2a)}
\end{picture}
\hspace{1em}
\setlength{\unitlength}{1pt}
\begin{picture}(130,150)(0,0)
\ArrowLine( 10, 55)( 25, 65)
\ArrowLine( 25, 65)( 25,100)
\ArrowLine( 25,100)( 10,115)
\ArrowLine( 10, 15)( 25, 30)
\ArrowLine( 25, 30)( 10, 40)
\ArrowLine( 95, 30)(110, 15)
\ArrowLine( 95, 65)( 95, 30)
\ArrowLine(110, 75)( 95, 65)
\ArrowLine(110, 90)( 95,100)
\ArrowLine( 95,100)(110,115)
\Photon(25,100)(95,100){2}{8}
\Photon(25, 65)(95, 65){2}{8}
\Photon(25, 30)(95, 30){2}{8}
\Vertex(25, 30){2.0}
\Vertex(25, 65){2.0}
\Vertex(25,100){2.0}
\Vertex(95, 30){2.0}
\Vertex(95, 65){2.0}
\Vertex(95,100){2.0}
\put(  9,81){$F_1$}
\put(101,43){$F_2$}
\put( 58,86){$V_1$}
\put( 58,50){$V_2$}
\put( 58,14){$V_3$}
\put(  0,115){$a$}
\put(  0, 55){$b$}
\put(  0, 36){$g$}
\put(  0, 10){$h$}
\put(115,115){$e$}
\put(115, 89){$f$}
\put(115, 71){$d$}
\put(115, 10){$c$}
\put(  0,135){\bf (2b)}
\end{picture}
\hspace{1em}
\setlength{\unitlength}{1pt}
\begin{picture}(130,150)(0,0)
\ArrowLine( 25, 65)( 10, 55)
\ArrowLine( 25,100)( 25, 65)
\ArrowLine( 10,115)( 25,100)
\ArrowLine( 10, 15)( 25, 30)
\ArrowLine( 25, 30)( 10, 40)
\ArrowLine(110, 15)( 95, 30)
\ArrowLine( 95, 30)( 95, 65)
\ArrowLine( 95, 65)(110, 75)
\ArrowLine(110, 90)( 95,100)
\ArrowLine( 95,100)(110,115)
\Photon(25,100)(95,100){2}{8}
\Photon(25, 65)(95, 65){2}{8}
\Photon(25, 30)(95, 30){2}{8}
\Vertex(25, 30){2.0}
\Vertex(25, 65){2.0}
\Vertex(25,100){2.0}
\Vertex(95, 30){2.0}
\Vertex(95, 65){2.0}
\Vertex(95,100){2.0}
\put(  9,81){$F_1$}
\put(101,43){$F_2$}
\put( 58,86){$V_1$}
\put( 58,50){$V_2$}
\put( 58,14){$V_3$}
\put(  0,115){$b$}
\put(  0, 55){$a$}
\put(  0, 36){$g$}
\put(  0, 10){$h$}
\put(115,115){$e$}
\put(115, 89){$f$}
\put(115, 71){$c$}
\put(115, 10){$d$}
\put(  0,135){\bf (2c)}
\end{picture}
} 
\vspace*{.5em}
\centerline{
\setlength{\unitlength}{1pt}
\begin{picture}(130,150)(0,0)
\ArrowLine( 10, 55)( 25, 65)
\ArrowLine( 25, 65)( 25,100)
\ArrowLine( 25,100)( 10,115)
\ArrowLine( 10, 15)( 25, 30)
\ArrowLine( 25, 30)( 10, 40)
\ArrowLine(110, 15)( 95, 30)
\ArrowLine( 95, 30)( 95, 65)
\ArrowLine( 95, 65)(110, 75)
\ArrowLine(110, 90)( 95,100)
\ArrowLine( 95,100)(110,115)
\Photon(25,100)(95,100){2}{8}
\DashLine(25, 65)(95, 65){5}
\Photon(25, 30)(95, 30){2}{8}
\Vertex(25, 30){2.0}
\Vertex(25, 65){2.0}
\Vertex(25,100){2.0}
\Vertex(95, 30){2.0}
\Vertex(95, 65){2.0}
\Vertex(95,100){2.0}
\put(  2,81){top}
\put(101,43){top}
\put( 58,86){$\PW$}
\put( 58,50){$\phi$}
\put( 58,14){$\PW$}
\put(  0,115){$a$}
\put(  0, 55){$b$}
\put(  0, 36){$g$}
\put(  0, 10){$h$}
\put(115,115){$e$}
\put(115, 89){$f$}
\put(115, 71){$c$}
\put(115, 10){$d$}
\put(  0,135){\bf (2d)}
\end{picture}
\hspace{1em}
\setlength{\unitlength}{1pt}
\begin{picture}(130,150)(0,0)
\end{picture}
\hspace{1em}
\setlength{\unitlength}{1pt}
\begin{picture}(130,150)(0,0)
\end{picture}
} 
\vspace*{-.5em}
\caption{Generic diagrams with 2 fermion currents}
\label{fig:2Vdiags}
\efi
The vector bosons $V_{\dots}$ represent all gauge-boson fields 
$\gamma$, $\PZ$, $\PW^\pm$, $\Pg$ that are allowed by the
quantum numbers of the external fermions. Whenever a gauge
boson has to be electrically charged, it is already denoted by $\PW$;
if a top quark is present, the charge flow is automatically fixed in the 
diagram.
All purely electroweak diagrams are supported for arbitrary six-fermion final
states. {\sc Lusifer} optionally includes also gluon-exchange diagrams,
but in its first version such diagrams are only included for up to
four quarks in the final state.
This, in particular, implies that there are no gluonic diagrams
of the type shown in \reffi{fig:4Vdiags}.
The scalar boson $S$ stands both for the Higgs-boson field
$\PH$ and for the would-be Goldstone partners $\chi$ and $\phi^\pm$
of the gauge bosons $\PZ$ and $\PW^\pm$. 
The amplitudes given below are all evaluated within 
the `t~Hooft--Feynman gauge.
Among the internal fermions $F_{\dots}$ 
the top quark plays a special role, since 
it is the only fermion that receives a mass, $\Mt$.
Note that the first three diagrams in each class appear already
in a theory with only massless fermions, while diagrams
(3d), (3e), and (2d) are proportional to a power of the top-quark mass
due to the appearance of the top-quark Yukawa coupling in the 
amplitude.

In the generic diagrams the external fermions $f_{a,\dots,h}$,
which are simply denoted as $a,\dots,h$, carry incoming 
momenta $p_{a,\dots,h}$ and helicities $\si_{a,\dots,h}$, respectively.
The helicity amplitudes of these diagrams are calculated within the
Weyl--van~der~Waerden (WvdW) formalism following the conventions of
\citere{Dittmaier:1999nn}.
The amplitudes are expressed in terms of WvdW spinor products,
\beq
\langle pq\rangle=\epsilon^{AB}p_A q_B
=2\sqrt{p_0 q_0} \,\Biggl[
{\mathrm{e}}^{-\ri\phi_p}\cos\frac{\theta_p}{2}\sin\frac{\theta_q}{2}
-{\mathrm{e}}^{-\ri\phi_q}\cos\frac{\theta_q}{2}\sin\frac{\theta_p}{2}
\Biggr],
\eeq
where $p_A$, $q_A$ are the associated momentum spinors for the 
massless momenta
\beqar
p^\mu&=&p_0(1,\sin\theta_p\cos\phi_p,\sin\theta_p\sin\phi_p,\cos\theta_p),\nl
q^\mu&=&q_0(1,\sin\theta_q\cos\phi_q,\sin\theta_q\sin\phi_q,\cos\theta_q).
\eeqar
Fermions are assumed to be incoming and, if necessary, 
turned into outgoing ones by crossing, which
is performed by inverting the corresponding fermion momenta and
helicities.
If spinor products appear with negative momenta $-p$, $-q$ as arguments, 
it is understood that only the complex conjugate spinor products get the
corresponding sign change. We illustrate this by simple examples:
\beq
\begin{array}[b]{rllll}
A(p,q)   &= \langle pq \rangle
&=\phantom{-}A(p,-q)
&=\phantom{-}A(-p,q)
&=A(-p,-q), \\
B(p,q)   &= \langle pq \rangle^*
&=-B(p,-q)
&=-B(-p,q)
&=B(-p,-q).
\end{array}
\eeq
The denominator parts
of the propagators for vector bosons $V$ and fermions $F$
are abbreviated by
\beq
P_V(p) = \frac{1}{p^2-\MV^2+\ri\MV\GV(p^2)}, \qquad
P_F(p) = \frac{1}{p^2-\MF^2+\ri\MF\Gamma_\PF(p^2)},
\label{eq:props}
\eeq
where the fermion mass $\MF$ is only non-zero for the top quark.
The introduction of finite decay widths $\GV(p^2)$ and $\Gamma_\PF(p^2)$
in the propagators is described below.
Moreover, we introduce abbreviations for sums of external momenta,
\beq
p_{ab} = p_a+p_b, \qquad 
p_{abc} = p_a+p_b+p_c, \qquad 
p_{abcd} = p_a+p_b+p_c+p_d.
\eeq
For the Feynman rules and coupling factors we follow the conventions
of \citere{Bohm:1986rj}. For the gauge-boson self-interactions
we define the following constants,
\beqar
C_{\PW\PW^\mp V} &=& C_{\PW V\PW^\pm} = \pm g_{V\PW\PW}, 
\qquad V = \gamma,\PZ,
\nn\\
C_{\PW\PW V_1V_2} &=& -g_{V_1\PW\PW} \, g_{V_2\PW\PW}, \qquad 
V_{1,2} = \gamma,\PZ,
\nn\\
C_{\PW\PW\PW\PW} &=& \frac{1}{\sw^2},
\eeqar
with the abbreviations
\beq
g_{\gamma \PW\PW} = 1, \qquad g_{\PZ\PW\PW} = -\frac{\cw}{\sw}.
\eeq
The sine and cosine of the weak mixing angle are defined by 
the masses of the $\PZ$ and $\PW$ bosons as follows:
\beq
\cw^2 = 1-\sw^2 = \frac{\MW^2}{\MZ^2}.
\eeq
Here and in the following the fields denoted in the subscripts are
assumed to be incoming. If the charge flow does not matter, we 
simply write $\PW$ and $\phi$ instead of $\PW^\pm$ and $\phi^\pm$;
the corresponding coupling is, of course, understood as zero
if charge conservation would be violated.
For the couplings of scalar to gauge bosons we introduce
\beqar
C_{\PH\PZ\PZ} &=& \frac{\MW}{\cw^2\sw},
\nn\\
C_{\PH\PW\PW} &=& \frac{\MW}{\sw},
\nn\\
C_{\phi\gamma\PW} &=& C_{\phi\PW\gamma} = -\MW,
\nn\\
C_{\phi\PZ\PW} &=& C_{\phi\PW\PZ} = -\frac{\sw\MW}{\cw},
\eeqar
and for the couplings of the electroweak gauge bosons to fermions
\beq
g^\sigma_{\gamma\bar f_i f_i} = -Q_i, \qquad
g^\sigma_{\PZ\bar f_i f_i} =
-\frac{\sw}{\cw}Q_i+\frac{I^3_{{\mathrm{w}},i}}{\cw\sw}\delta_{\sigma,-},
\qquad
g^\sigma_{\PW\bar f_i f'_i} = \frac{1}{\sqrt{2}\sw}\delta_{\sigma,-},
\eeq
where $Q_i$ and $I^3_{{\mathrm{w}},i}=\pm{1/2}$ denote the
relative charge and the weak isospin of the fermion $f_i$,
respectively.  
The field $f'_i$ corresponds to the weak-isospin partner of $f_i$.
For the gluon coupling to quarks $q$ we introduce
\beq
g^\sigma_{\Pg\bar qq} = \frac{g_{\mathrm{s}}}{e}, 
\eeq
where $g_{\mathrm{s}}=\sqrt{4\pi\alpha_{\mathrm{s}}}$ is the strong 
gauge coupling and $e=\sqrt{4\pi\al}$ is the electromagnetic coupling. 
The factor $e$ in the denominator is introduced, because we will
separate the global factor $e^6$ from the $\Pep\Pem\to 6f$ 
helicity amplitudes.
By convention, the colour operator as well as the colour
indices of external quarks are split off from the generic amplitudes 
given below; the whole colour structure will be reinserted when
squaring the amplitudes.
Finally, we need the Yukawa coupling of the top quark, for which 
we define
\beq
g^\si_{\phi\bar\Pb\Pt} = \frac{\Mt}{\sqrt{2}\sw\MW}\delta_{\sigma,+}, 
\qquad
g^\si_{\phi\bar\Pt\Pb} = \frac{\Mt}{\sqrt{2}\sw\MW}\delta_{\sigma,-}.
\eeq
If any of the constants $C_{\dots}$ or $g^\si_{\dots}$
appears in the following with subscripts
not listed here, it is understood to vanish.

The amplitudes for the generic graphs with 4 fermion currents 
(see \reffi{fig:4Vdiags}) are given by
\beqar
\M^{\si_a,\dots,\si_h}_{\mathrm{(4a)}}(p_a,\dots,p_h) &=& 4e^6 \,
\de_{\si_a,+} \de_{\si_b,-} \de_{\si_c,+} \de_{\si_d,-}
\de_{\si_e,-\si_f} \de_{\si_g,-\si_h} 
\nn\\
&& {} \times
g^{-}_{\PW\bar f_a f_b}\,       g^{-}_{\PW\bar f_c f_d}\, 
g^{\si_f}_{V_1\bar f_e f_f}\, g^{\si_h}_{V_2\bar f_g f_h}\,
C_{\PW\PW V_1V_2} \,
\nn\\
&& {} \times
P_{\PW}(p_{ab}) P_{\PW}(p_{cd}) P_{V_1}(p_{ef}) P_{V_2}(p_{gh}) \,
\nn\\
&& {} \times
A^{\si_a,\si_c,\si_e,\si_g}_{\mathrm{(4a)}}(p_a,\dots,p_h),
\\[.5em]
\M^{\si_a,\dots,\si_h}_{\mathrm{(4b)}}(p_a,\dots,p_h) &=& 4e^6 \,
\de_{\si_a,+} \de_{\si_b,-} \de_{\si_c,-\si_d}
\de_{\si_e,+} \de_{\si_f,-} \de_{\si_g,-\si_h} 
\nn\\
&& {} \times
g^{-}_{\PW\bar f_a f_b}\, g^{\si_d}_{V_1\bar f_c f_d}\, 
g^{-}_{\PW\bar f_e f_f}\, g^{\si_h}_{V_2\bar f_g f_h}\,
C_{\PW V_1V_3} \,C_{\PW V_2V_3} \,
\nn\\
&& {} \times
P_{\PW}(p_{ab}) P_{V_1}(p_{cd}) P_{\PW}(p_{ef}) P_{V_2}(p_{gh}) 
P_{V_3}(p_{abcd}) \,
\nn\\
&& {} \times
A^{\si_a,\si_c,\si_e,\si_g}_{\mathrm{(4b)}}(p_a,\dots,p_h),
\\[.5em]
\M^{\si_a,\dots,\si_h}_{\mathrm{(4c)}}(p_a,\dots,p_h) &=& -4e^6 \,
\de_{\si_a,-\si_b}\de_{\si_c,-\si_d}\de_{\si_e,-\si_f}\de_{\si_g,-\si_h} 
\nn\\
&& {} \times
g^{\si_b}_{V_1\bar f_a f_b}\, g^{\si_d}_{V_2\bar f_c f_d}\, 
g^{\si_f}_{V_3\bar f_e f_f}\, g^{\si_h}_{V_4\bar f_g f_h}\,
C_{SV_1V_2} \, C_{SV_3V_4} \,
\nn\\
&& {} \times
P_{V_1}(p_{ab}) P_{V_2}(p_{cd}) P_{V_3}(p_{ef}) P_{V_4}(p_{gh}) 
P_S(p_{abcd}) \,
\nn\\
&& {} \times
A^{\si_a,\si_c,\si_e,\si_g}_{\mathrm{(4c)}}(p_a,\dots,p_h),
\eeqar
where the auxiliary functions $A^{\dots}_{\dots}$ contain the
WvdW spinor products. 
Explicit results for the auxiliary functions read
\beqar
A^{{+}{+}{+}{+}}_{\mathrm{(4a)}}(p_a,\dots,p_h) &=&
2\cspbd\cspfh\spac\speg-\cspbf\cspdh\spae\spcg
\nn\\ && {}
-\cspbh\cspdf\spag\spce,
\label{eq:A4a}
\\[.5em]
\label{eq:cdot}
A^{{+}{+}{+}{+}}_{\mathrm{(4b)}}(p_a,\dots,p_h) &=&
\cspbd\cspfh\spac\speg\, (p_{ab}-p_{cd})\cdot(p_{ef}-p_{gh})
\nn\\ && {}
+2\cspbd\spac\Big[
  \langle p_f P_{ab} p_e \rangle \langle p_h P_{cd} p_g \rangle
- \langle p_f P_{cd} p_e \rangle \langle p_h P_{ab} p_g \rangle \Big]
\nn\\ && {}
+2\cspfh\speg\Big[
  \langle p_b P_{ef} p_a \rangle \langle p_d P_{gh} p_c \rangle
- \langle p_b P_{gh} p_a \rangle \langle p_d P_{ef} p_c \rangle \Big]
\nn\\ && {}
+2\langle p_b P_{cd} p_a \rangle \Big[
  \cspdh\spcg \langle p_f P_{gh} p_e \rangle
- \cspdf\spce \langle p_h P_{ef} p_g \rangle \Big]
\nn\\ && {}
-2\langle p_d P_{ab} p_c \rangle \Big[
  \cspbh\spag \langle p_f P_{gh} p_e \rangle
- \cspbf\spae \langle p_h P_{ef} p_g \rangle \Big],
\nn\\
\label{eq:A4b}
\\[.5em]
A^{{+}{+}{+}{+}}_{\mathrm{(4c)}}(p_a,\dots,p_h) &=&
\cspbd\cspfh\spac\speg.
\label{eq:A4c}
\eeqar
Here the dot in the first line of Eq.~\refeq{eq:cdot} 
indicates the usual Lorentz product of four-vectors,
and the extended brackets are shorthands for the expressions
\beq
\langle p_a P_{bc} p_d \rangle = \cspba\spbd+\cspca\spcd.
\label{eq:Pabbracket}
\eeq
The remaining helicity configurations of the auxiliary functions 
follow from discrete symmetries. All non-vanishing
$A^{\dots}_{\dots}$ are obtained from Eqs.~\refeq{eq:A4a}--\refeq{eq:A4c}
by the simple substitutions
\beqar
A^{-\si_a,\si_c,\si_e,\si_g}_{\mathrm{(4\dots)}}
        (p_a,p_b,p_c,p_d,p_e,p_f,p_g,p_h) &=&
A^{\si_a,\si_c,\si_e,\si_g}_{\mathrm{(4\dots)}}
        (p_b,p_a,p_c,p_d,p_e,p_f,p_g,p_h),
\nn\\
A^{\si_a,-\si_c,\si_e,\si_g}_{\mathrm{(4\dots)}}
        (p_a,p_b,p_c,p_d,p_e,p_f,p_g,p_h) &=&
A^{\si_a,\si_c,\si_e,\si_g}_{\mathrm{(4\dots)}}
        (p_a,p_b,p_d,p_c,p_e,p_f,p_g,p_h),
\nn\\
A^{\si_a,\si_c,-\si_e,\si_g}_{\mathrm{(4\dots)}}
        (p_a,p_b,p_c,p_d,p_e,p_f,p_g,p_h) &=&
A^{\si_a,\si_c,\si_e,\si_g}_{\mathrm{(4\dots)}}
        (p_a,p_b,p_c,p_d,p_f,p_e,p_g,p_h),
\nn\\
A^{\si_a,\si_c,\si_e,-\si_g}_{\mathrm{(4\dots)}}
        (p_a,p_b,p_c,p_d,p_e,p_f,p_g,p_h) &=&
A^{\si_a,\si_c,\si_e,\si_g}_{\mathrm{(4\dots)}}
        (p_a,p_b,p_c,p_d,p_e,p_f,p_h,p_g).
\hspace{2em}
\eeqar
Another useful relation is provided by the fact that taking the
complex conjugate of $A^{\dots}_{\dots}$ reverses all helicities,
\beq
A^{-\si_a,-\si_c,-\si_e,-\si_g}_{\mathrm{(4\dots)}}(p_a,\dots,p_h) =
\Big(A^{\si_a,\si_c,\si_e,\si_g}_{\mathrm{(4\dots)}}(p_a,\dots,p_h)\Big)^*,
\eeq
which is useful for checking the amplitudes.
For the class of diagrams with 4 fermion currents it is obvious that
we have helicity conservation for the respective (massless)
fermion--antifermion pairs, i.e.\ each diagram only contributes if
\beq
\si_a = -\si_b, \qquad
\si_c = -\si_d, \qquad
\si_e = -\si_f, \qquad
\si_g = -\si_h.
\label{eq:helcons}
\eeq

The amplitudes for the generic graphs with 3 fermion currents 
(see \reffi{fig:3Vdiags}) are given by
\beqar
\label{eq:A3a}
\M^{\si_a,\dots,\si_h}_{\mathrm{(3a)}}(p_a,\dots,p_h) &=& 8e^6 \,
\de_{\si_a,-\si_b} \de_{\si_c,-\si_d} \de_{\si_e,-\si_f} \de_{\si_g,-\si_h} 
\nn\\
&& {} \times
g^{-\si_a}_{V_1\bar f_a F_1}\, 
 g^{\si_b}_{V_3\bar F_2 f_b}\, g^{\si_d}_{V_1\bar f_c f_d}\, 
 g^{\si_f}_{V_2\bar f_e f_f}\, g^{\si_h}_{V_3\bar f_g f_h}\,
\nn\\
&& {} \times
P_{F_1}(p_{acd}) P_{F_2}(p_{bgh}) 
P_{V_1}(p_{cd}) P_{V_2}(p_{ef}) P_{V_3}(p_{gh}) \,
\nn\\
&& {} \times
\sum_{\tau=\pm}  g^{\tau}_{V_2\bar F_1 F_2}\,
A^{\si_a,\si_c,\si_e,\si_g}_{\mathrm{(3a)},\tau}(p_a,\dots,p_h),
\\[.5em]
\M^{\si_a,\dots,\si_h}_{\mathrm{(3b)}}(p_a,\dots,p_h) &=& -4e^6 \,
\de_{\si_a,-\si_b} \de_{\si_c,-\si_d} \de_{\si_e,+}  \de_{\si_f,-} 
\de_{\si_g,-\si_h} 
\nn\\
&& {} \times
g^{-\si_a}_{V_1\bar f_a F}\, g^{\si_b}_{V_2\bar F f_b}\, 
 g^{\si_d}_{V_1\bar f_c f_d}\, g^{-}_{\PW\bar f_e f_f}\, 
 g^{\si_h}_{V_3\bar f_g f_h}\, C_{\PW V_2V_3}\,
\nn\\
&& {} \times
P_{F}(p_{acd}) P_{V_1}(p_{cd}) 
P_{V_2}(p_{efgh}) P_{\PW}(p_{ef}) P_{V_3}(p_{gh}) \,
\nn\\
&& {} \times
A^{\si_a,\si_c,\si_e,\si_g}_{\mathrm{(3b)}}(p_a,\dots,p_h),
\\[.5em]
\M^{\si_a,\dots,\si_h}_{\mathrm{(3c)}}(p_a,\dots,p_h) &=& -4e^6 \,
\de_{\si_a,-\si_b} \de_{\si_c,-\si_d} \de_{\si_e,+}  \de_{\si_f,-} 
\de_{\si_g,-\si_h} 
\nn\\
&& {} \times
g^{-\si_a}_{V_2\bar f_a F}\, g^{\si_b}_{V_1\bar F f_b}\, 
 g^{\si_d}_{V_1\bar f_c f_d}\, g^{-}_{\PW\bar f_e f_f}\, 
 g^{\si_h}_{V_3\bar f_g f_h}\, C_{\PW V_2V_3}\,
\nn\\
&& {} \times
P_{F}(p_{bcd}) P_{V_1}(p_{cd}) 
P_{V_2}(p_{efgh}) P_{\PW}(p_{ef}) P_{V_3}(p_{gh}) \,
\nn\\
&& {} \times
A^{\si_a,\si_c,\si_e,\si_g}_{\mathrm{(3c)}}(p_a,\dots,p_h),
\\[.5em]
\M^{\si_a,\dots,\si_h}_{\mathrm{(3d)}}(p_a,\dots,p_h) &=& 4e^6 \,
\de_{\si_a,+}  \de_{\si_b,-} \de_{\si_c,+} \de_{\si_d,-} 
\de_{\si_e,+}  \de_{\si_f,-} \de_{\si_g,-\si_h} 
\nn\\
&& {} \times
g^{-}_{\PW\bar f_a \Pt}\, g^{-}_{\phi\bar\Pt f_b}\, 
g^{-}_{\PW\bar f_c f_d}\, g^{-}_{\PW\bar f_e f_f}\, 
g^{\si_h}_{V\bar f_g f_h}\, C_{\phi \PW V}\,
\nn\\
&& {} \times
P_{\Pt}(p_{acd}) P_{\PW}(p_{cd}) 
P_{\PW}(p_{efgh}) P_{\PW}(p_{ef}) P_V(p_{gh}) \,
\nn\\
&& {} \times
A^{\si_a,\si_c,\si_e,\si_g}_{\mathrm{(3d)}}(p_a,\dots,p_h),
\\[.5em]
\label{eq:A3e}
\M^{\si_a,\dots,\si_h}_{\mathrm{(3e)}}(p_a,\dots,p_h) &=& 4e^6 \,
\de_{\si_a,+}  \de_{\si_b,-} \de_{\si_c,+} \de_{\si_d,-} 
\de_{\si_e,+}  \de_{\si_f,-} \de_{\si_g,-\si_h} 
\nn\\
&& {} \times
g^{+}_{\phi\bar f_a \Pt}\, g^{-}_{\PW\bar\Pt f_b}\, 
g^{-}_{\PW\bar f_c f_d}\, g^{-}_{\PW\bar f_e f_f}\, 
g^{\si_h}_{V\bar f_g f_h}\, C_{\phi \PW V}\,
\nn\\
&& {} \times
P_{\Pt}(p_{bcd}) P_{\PW}(p_{cd}) 
P_{\PW}(p_{efgh}) P_{\PW}(p_{ef}) P_V(p_{gh}) \,
\nn\\
&& {} \times
A^{\si_a,\si_c,\si_e,\si_g}_{\mathrm{(3e)}}(p_a,\dots,p_h).
\eeqar
At first sight it seems that the occurrence of a top quark in the
spinor chain between $f_a$ and $f_b$ can violate helicity conservation
in this line. However, this is not true since the top-quark propagator
always appears between two chirality projectors, restoring
helicity conservation in the whole fermion line. 
Thus, as in the case of 4 fermion currents all graphs with
3 fermion currents vanish if Eq.~\refeq{eq:helcons} is not fulfilled.
We have already made use of Eq.~\refeq{eq:helcons}
in the expressions for the matrix elements Eqs.~\refeq{eq:A3a}--\refeq{eq:A3e}.
The auxiliary functions $A^{\dots}_{\dots}$ read
\beqar
A^{{+}{+}{+}{+}}_{\mathrm{(3a)},+}(p_a,\dots,p_h) &=&
-m_{\PF_1} m_{\PF_2} \cspbh\cspdf\spac\speg,
\nn\\ 
A^{{+}{+}{+}{+}}_{\mathrm{(3a)},-}(p_a,\dots,p_h) &=&
-\cspbh\spac \langle p_d P_{ac}p_e \rangle \langle p_f P_{bh}p_g \rangle,
\\[.5em]
A^{{+}{+}{+}{+}}_{\mathrm{(3b)}}(p_a,\dots,p_h) &=&
\spac\Big\{ 
\cspfh\speg\Big[ \phantom{{}+{}} 
   \cspad \Bigl( \langle p_b P_{ef}p_a \rangle
               - \langle p_b P_{gh} p_a \rangle \Bigr)
\nn\\ && \phantom{\spac\Big\{\cspfh\speg\Big[} {}
                +\cspcd \Bigl( \langle p_b P_{ef} p_c \rangle
                             - \langle p_b P_{gh} p_c \rangle \Bigr) \Big]
\nn\\ && \phantom{\spac\Big\{} {}
+2\cspbh \langle p_d P_{ac} p_g \rangle
         \langle p_f P_{gh}  p_e \rangle
\nn\\ && \phantom{\spac\Big\{} {}
-2\cspbf \langle p_d P_{ac} p_e \rangle
         \langle p_h P_{ef}  p_g \rangle \Big\},
\\[.5em]
A^{{+}{+}{+}{+}}_{\mathrm{(3c)}}(p_a,\dots,p_h) &=&
\cspbd\Big\{ 
\cspfh\speg\Big[ 
  -\spbc \Bigl( \langle p_b P_{ef}p_a \rangle
              - \langle p_b P_{gh} p_a \rangle \Bigr)
\nn\\ && \phantom{\cspbd\Big\{\cspfh\speg\Big[} {}
                +\spcd \Bigl( \langle p_d P_{ef} p_a \rangle
                            - \langle p_d P_{gh} p_a \rangle  \Bigr)\Big]
\nn\\ && \phantom{\cspbd\Big\{} {}
-2\spag \langle p_h P_{bd} p_c \rangle
         \langle p_f P_{gh}  p_e \rangle
\nn\\ && \phantom{\cspbd\Big\{} {}
+2\spae \langle p_f P_{bd} p_c \rangle
         \langle p_h P_{ef}  p_g \rangle \Big\},
\\[.5em]
A^{{+}{+}{+}{+}}_{\mathrm{(3d)}}(p_a,\dots,p_h) &=&
A^{{+}{+}{+}{+}}_{\mathrm{(3e)}}(p_a,\dots,p_h) =
\Mt \cspbd\cspfh\spac\speg,
\eeqar
where the expressions 
for the remaining polarizations
are again obtained by simple substitutions,
\beqar
A^{\si_a,-\si_c,\si_e,\si_g}_{\mathrm{(3\dots),\tau}}
        (p_a,p_b,p_c,p_d,p_e,p_f,p_g,p_h) &=&
A^{\si_a,\si_c,\si_e,\si_g}_{\mathrm{(3\dots),\tau}}
        (p_a,p_b,p_d,p_c,p_e,p_f,p_g,p_h),
\nn\\
A^{\si_a,\si_c,-\si_e,\si_g}_{\mathrm{(3\dots),\tau}}
        (p_a,p_b,p_c,p_d,p_e,p_f,p_g,p_h) &=&
A^{\si_a,\si_c,\si_e,\si_g}_{\mathrm{(3\dots),\tau}}
        (p_a,p_b,p_c,p_d,p_f,p_e,p_g,p_h),
\nn\\
A^{\si_a,\si_c,\si_e,-\si_g}_{\mathrm{(3\dots),\tau}}
        (p_a,p_b,p_c,p_d,p_e,p_f,p_g,p_h) &=&
A^{\si_a,\si_c,\si_e,\si_g}_{\mathrm{(3\dots),\tau}}
        (p_a,p_b,p_c,p_d,p_e,p_f,p_h,p_g),
\nn\\
A^{-\si_a,-\si_c,-\si_e,-\si_g}_{\mathrm{(3\dots),-\tau}}
(p_a,\dots,p_h) &=&
\Big(A^{\si_a,\si_c,\si_e,\si_g}_{\mathrm{(3\dots),\tau}}
(p_a,\dots,p_h)\Big)^* \Big|_{m_{\Pt}^*\to m_{\Pt}},
\hspace{3em}
\eeqar
where the index $\tau$ has to be ignored for all functions other
than $A^{\dots}_{\mathrm{(3a),\tau}}$ and the substitution
$m_{\Pt}^*\to m_{\Pt}$ ensures that the top-quark mass (if taken complex)
remains unaffected by the complex conjugation.

The amplitudes for the generic graphs with 2 fermion currents 
(see \reffi{fig:2Vdiags}) are given by
\beqar
\M^{\si_a,\dots,\si_h}_{\mathrm{(2a)}}(p_a,\dots,p_h) &=& 8e^6 \,
\de_{\si_a,-\si_b} \de_{\si_c,-\si_d} \de_{\si_e,-\si_f} \de_{\si_g,-\si_h} 
\nn\\
&& {} \times
g^{-\si_a}_{V_1\bar f_a F_1}\, g^{\si_b}_{V_2\bar F_1 f_b}\, 
g^{-\si_c}_{V_2\bar f_c F_2}\, g^{\si_d}_{V_3\bar F_2 f_d}\, 
 g^{\si_f}_{V_1\bar f_e f_f}\, g^{\si_h}_{V_3\bar f_g f_h}\,
\nn\\
&& {} \times
P_{F_1}(p_{aef}) P_{F_2}(p_{dgh}) 
P_{V_1}(p_{ef}) P_{V_2}(p_{abef}) P_{V_3}(p_{gh}) \,
\nn\\
&& {} \times
A^{\si_a,\si_c,\si_e,\si_g}_{\mathrm{(2a)}}(p_a,\dots,p_h),
\label{eq:M2a}
\\[.5em]
\M^{\si_a,\dots,\si_h}_{\mathrm{(2b)}}(p_a,\dots,p_h) &=& 8e^6 \,
\de_{\si_a,-\si_b} \de_{\si_c,-\si_d} \de_{\si_e,-\si_f} \de_{\si_g,-\si_h} 
\nn\\
&& {} \times
g^{-\si_a}_{V_1\bar f_a F_1}\, g^{\si_b}_{V_2\bar F_1 f_b}\, 
g^{-\si_c}_{V_3\bar f_c F_2}\, g^{\si_d}_{V_2\bar F_2 f_d}\, 
 g^{\si_f}_{V_1\bar f_e f_f}\, g^{\si_h}_{V_3\bar f_g f_h}\,
\nn\\
&& {} \times
P_{F_1}(p_{aef}) P_{F_2}(p_{cgh}) 
P_{V_1}(p_{ef}) P_{V_2}(p_{abef}) P_{V_3}(p_{gh}) \,
\nn\\
&& {} \times
A^{\si_a,\si_c,\si_e,\si_g}_{\mathrm{(2b)}}(p_a,\dots,p_h),
\label{eq:M2b}
\\[.5em]
\M^{\si_a,\dots,\si_h}_{\mathrm{(2c)}}(p_a,\dots,p_h) &=& 8e^6 \,
\de_{\si_a,-\si_b} \de_{\si_c,-\si_d} \de_{\si_e,-\si_f} \de_{\si_g,-\si_h} 
\nn\\
&& {} \times
g^{-\si_a}_{V_2\bar f_a F_1}\, g^{\si_b}_{V_1\bar F_1 f_b}\, 
g^{-\si_c}_{V_2\bar f_c F_2}\, g^{\si_d}_{V_3\bar F_2 f_d}\, 
 g^{\si_f}_{V_1\bar f_e f_f}\, g^{\si_h}_{V_3\bar f_g f_h}\,
\nn\\
&& {} \times
P_{F_1}(p_{bef}) P_{F_2}(p_{dgh}) 
P_{V_1}(p_{ef}) P_{V_2}(p_{abef}) P_{V_3}(p_{gh}) \,
\nn\\
&& {} \times
A^{\si_a,\si_c,\si_e,\si_g}_{\mathrm{(2c)}}(p_a,\dots,p_h),
\label{eq:M2c}
\\[.5em]
\M^{\si_a,\dots,\si_h}_{\mathrm{(2d)}}(p_a,\dots,p_h) &=& 4e^6 \,
\de_{\si_a,+}  \de_{\si_b,-} \de_{\si_c,+} \de_{\si_d,-} 
\de_{\si_e,+}  \de_{\si_f,-} \de_{\si_g,+} \de_{\si_h,-} 
\nn\\
&& {} \times
g^{-}_{\PW\bar f_a \Pt}\,    g^{-}_{\phi\bar \Pt f_b}\, 
g^{+}_{\phi\bar f_c \Pt}\, g^{-}_{\PW\bar \Pt f_d}\, 
g^{-}_{\PW\bar f_e f_f}\,    g^{-}_{\PW\bar f_g f_h}\,
\nn\\
&& {} \times
P_{\Pt}(p_{aef}) P_{\Pt}(p_{dgh}) 
P_{\PW}(p_{ef}) P_{\PW}(p_{abef}) P_{\PW}(p_{gh}) \,
\nn\\
&& {} \times
A^{\si_a,\si_c,\si_e,\si_g}_{\mathrm{(2d)}}(p_a,\dots,p_h).
\label{eq:M2d}
\eeqar
Again we exploited the fact that helicity conservation in each spinor
chain containing top quarks is restored by chirality projectors, i.e.\
that Eq.~\refeq{eq:helcons} is necessary for non-vanishing
contributions.
The functions $A^{\dots}_{\dots}$ read
\beqar
A^{{+}{+}{+}{+}}_{\mathrm{(2a)}}(p_a,\dots,p_h) &=&
\cspdh\spae \langle p_b P_{dh} p_g \rangle
            \langle p_f P_{ae} p_c \rangle,
\nn\\ 
A^{{-}{+}{+}{+}}_{\mathrm{(2a)}}(p_a,\dots,p_h) &=&
\cspaf\cspdh\spbc \Big[
  \spae \langle p_a P_{dh} p_g \rangle
 -\spef \langle p_f P_{dh} p_g \rangle \Big],
\hspace{2em}
\\[.5em]
A^{{+}{+}{+}{+}}_{\mathrm{(2b)}}(p_a,\dots,p_h) &=&
-\cspbd\spae\spcg \Big[
  \cspaf \langle p_h P_{cg} p_a \rangle
 +\cspef \langle p_h P_{cg} p_e \rangle \Big],
\nn\\ 
A^{{-}{+}{+}{+}}_{\mathrm{(2b)}}(p_a,\dots,p_h) &=&
-\cspaf\spcg \langle p_d P_{af} p_e \rangle
             \langle p_h P_{cg} p_b \rangle,
\\[.5em]
A^{{+}{+}{+}{+}}_{\mathrm{(2c)}}(p_a,\dots,p_h) &=&
-\cspbf\cspdh\spac \Big[
  \spbe \langle p_b P_{dh} p_g \rangle
 -\spef \langle p_f P_{dh} p_g \rangle \Big],
\nn\\ 
A^{{-}{+}{+}{+}}_{\mathrm{(2c)}}(p_a,\dots,p_h) &=&
-\cspdh\spbe \langle p_a P_{dh} p_g \rangle
             \langle p_f P_{be} p_c \rangle,
\\[.5em]
A^{{+}{+}{+}{+}}_{\mathrm{(2d)}}(p_a,\dots,p_h) &=&
-\Mt^2 \cspbf\cspdh\spae\spcg,
\eeqar
with the following substitutions for the remaining polarizations,
\beqar
A^{\si_a,\si_c,-\si_e,\si_g}_{\mathrm{(2\dots)}}
        (p_a,p_b,p_c,p_d,p_e,p_f,p_g,p_h) &=&
A^{\si_a,\si_c,\si_e,\si_g}_{\mathrm{(2\dots)}}
        (p_a,p_b,p_c,p_d,p_f,p_e,p_g,p_h),
\nn\\
A^{\si_a,\si_c,\si_e,-\si_g}_{\mathrm{(2\dots)}}
        (p_a,p_b,p_c,p_d,p_e,p_f,p_g,p_h) &=&
A^{\si_a,\si_c,\si_e,\si_g}_{\mathrm{(2\dots)}}
        (p_a,p_b,p_c,p_d,p_e,p_f,p_h,p_g),
\nn\\
A^{-\si_a,-\si_c,-\si_e,-\si_g}_{\mathrm{(2\dots)}}
(p_a,\dots,p_h) &=&
\Big(A^{\si_a,\si_c,\si_e,\si_g}_{\mathrm{(2\dots)}}
(p_a,\dots,p_h)\Big)^* \Big|_{m_{\Pt}^*\to m_{\Pt}}.
\hspace{3em}
\eeqar

\subsection{Squared matrix elements from generic amplitudes}

Having defined all amplitudes involving 8 external massless fermions
in the previous section, we now turn to the evaluation of the squared matrix 
elements for a given process $2f\to 6f$.
Since the adopted approach is based on helicity eigenstates of the
external fermions, 
in a first step we calculate the squares of all helicity matrix 
elements and take the incoherent sum over the relevant helicity
configurations at the end. We widely suppress the helicity labels 
in the following.

The fermion lines with an outgoing arrow are enumerated by $i=1,3,5,7$; they 
correspond to incoming antifermions $\bar f_i$ or outgoing fermions $f_i$. 
The fermion lines with an incoming arrow are enumerated by $i=2,4,6,8$; they 
correspond to incoming fermions $f_i$ or outgoing antifermions $\bar f_i$. 
The respective incoming momenta and helicities are denoted by 
$p_i$ and $\si_i$,
i.e.\ these quantities receive a minus sign by crossing if they belong
to final-state particles. For instance, for a scattering reaction
of the incoming $\bar f_7 f_8$ pair, our notation is
\beqar
\bar f_7(p_7,\si_7) + f_8(p_8,\si_8) &\;\to\;& \phantom{{}+{}}
f_1(-p_1,-\si_1) + \bar f_2(-p_2,-\si_2) + f_3(-p_3,-\si_3) 
\nn\\
&& {} 
+ \bar f_4(-p_4,-\si_4) + f_5(-p_5,-\si_5) + \bar f_6(-p_6,-\si_6),
\eeqar
where the momenta and helicities within parentheses correspond to incoming 
fields, i.e.\ they are identified with appropriate permutations of the 
momenta $p_a,\dots,p_h$ and helicities $\sigma_a,\dots,\sigma_h$ 
in the generic amplitudes $\M^{\si_a,\dots,\si_h}_{(\dots)}(p_a,\dots,p_h)$
of the previous section.
Finally, all relevant Feynman graphs
have to be summed and squared taking into account colour correlations.

In detail the construction proceeds in three steps:

\setcounter{paragraph}{0}
\paragraph{Fermion permutations}
First we write down all permutations 
of $(1,3,5,7)$ and $(2,4,6,8)$ for $(a,c,e,g)$ and $(b,d,f,h)$, respectively,
resulting in $(4!)^2=576$ different combinations $(a,\dots,h)$.
The generic amplitudes vanish if at least one of the pairs
$\bar f_a f_b$, $\bar f_c f_d$, $\bar f_e f_f$, and $\bar f_g f_h$
does not appear in a vertex $V\bar f f$
with a gauge
boson $V$. Therefore, we can omit such combinations $(a,\dots,h)$
from the beginning. This excludes, for instance, combinations where
at least one of the pairs consists of a lepton and a quark. However,
there are also channels, such as $\Pep\Pem\to\Pep\Pem\Pep\Pem\Pep\Pem$,
where all 576 combinations contribute. In the following we enumerate the
relevant combinations by $n=1,\dots,N$ ($N\le576$) and denote them by
$(a_n,\dots,h_n)$.
Next we determine the relative sign factors $\eta_n$ between diagrams
that differ in their fermion-number flow by an interchange of 
external lines. Obviously
\beq
\eta_n = \eps_{a_n,\dots,h_n} = \left\{ 
\begin{array}{rl}
+1 & \quad \mbox{for even permutations $(a_n,\dots,h_n)$ of $(1,\dots,8)$,} \\
-1 & \quad \mbox{for odd  permutations $(a_n,\dots,h_n)$ of $(1,\dots,8)$,} \\
 0 & \quad \mbox{otherwise,}
\end{array}
\right.
\eeq
yields the correct signs. 

\paragraph{Amplitude evaluation}
In the next step we evaluate the generic amplitudes 
$\M^{\si_a,\dots,\si_h}_{(\dots)}(p_a,\dots,p_h)$ with the identifications
$(a,\dots,h)=(a_n,\dots,h_n)$, i.e.\ we calculate all graphs shown in
\reffis{fig:4Vdiags}--\ref{fig:2Vdiags} with all possible internal bosons.
More precisely, in order to avoid double counting we require appropriate 
conditions for $(a,\dots,h)$ and the internal bosons to ensure that each 
Feynman diagram is calculated only once. These conditions are listed in 
\refta{tab:genampcond}, where $Q_{ab}=Q_b-Q_a$ is the total incoming charge
of the pair $\bar f_a f_b$.%
\begin{table}
\centerline{
\begin{tabular}{|c|l|l|l|}
\hline
Amplitude & Bosons & Fermions & Charge flow
\\ \hline\hline
4a & $V_{\{1,2\}}=\{\PW,\PW\}$ & $a<c$, $e<g$ & $Q_{ab}=Q_{cd}=-1$
\\ \cline{2-4}
   & $V_{\{1,2\}}=\{\ga/\PZ,\ga/\PZ\}$ & $e<g$ & $Q_{ab}=-1$
\\ \hline
4b & $V_{\{1,2,3\}}=\{\PW,\PW,\ga/\PZ\}$ & $a<g$ & $Q_{ab}=Q_{gh}=-1$
\\ \cline{2-4}
   & $V_{\{1,2,3\}}=\{\ga/\PZ,\ga/\PZ,\PW\}$ & none & $Q_{ab}=-1$
\\ \hline
4c & $V_{\{1,2,3,4\}}=\{\PW,\PW,\PW,\PW\}$, $S=\PH$ & $a<g$ & 
$Q_{ab}=Q_{gh}=-1$
\\ \cline{2-4}
   & $V_{\{1,2,3,4\}}=\{\PW,\PW,\PZ,\PZ\}$, $S=\PH$ & $e<g$ & $Q_{ab}=-1$
\\ \cline{2-4}
   & $V_{\{1,2,3,4\}}=\{\PZ,\PZ,\PZ,\PZ\}$, $S=\PH$  & $a=1$, $e<g$ & none
\\ \cline{2-4}
   & $V_{\{1,2,3,4\}}=\{\PW,\ga/\PZ,\PW,\ga/\PZ\}$, $S=\phi$  & none & 
$Q_{ab}=-1$
\\ \hline
3b, 3c & $V_{\{1,2,3\}}=\{\PW,\PW,\ga/\PZ\}$ & none & none
\\ \cline{2-4}
       & $V_{\{1,2,3\}}=\{\PW,\ga/\PZ,\PW\}$   & none & $Q_{ef}=-1$
\\ \cline{2-4}
       & $V_{\{1,2,3\}}=\{\ga/\PZ/\Pg,\PW,\ga/\PZ\}$ & none & none
\\ \cline{2-4}
       & $V_{\{1,2,3\}}=\{\ga/\PZ/\Pg,\ga/\PZ,\PW\}$   & none & $Q_{ef}=-1$
\\ \hline
2b, 2c & \mbox{none} & $a<c$ & none
\\ \hline
3a, 3d, 3e, & \mbox{none} & none & none
\\ 
2a, 2d & & & 
\\ \hline
\end{tabular} }
\caption{Restrictions on the field insertions and on the charge flow in
the generic amplitudes (The lists for the field insertions are ordered.)}
\label{tab:genampcond}
\end{table}
An entry ``none'' means that all insertions allowed by charge
conservation are possible.
For example, the second line in \refta{tab:genampcond} for diagram ``4b''
means that the vector bosons $V_1$ and $V_2$ each can be either a photon
or Z~boson, $V_3$ is a W~boson with a negative charge $-e$ flowing
through the diagram from the $\bar f_a f_b$ pair to the $\bar f_e f_f$ 
pair, while no restriction on the numbers $a,\dots,h$ is imposed.

At this stage, we can already sum up diagrams that belong to the same
fermion permutation and have the same colour structure. In this way we
obtain the sums $\M^{(\mathrm{ew})}_n$ and $\M^{(1\Pg)}_n$ which
include all purely electroweak and one-gluon exchange diagrams
for $(a_n,\dots,h_n)$, respectively.
Here one should realize that the number of diagrams can be rather large,
ranging from typically $\sim 10^2$--$10^3$ up to 13896 for 
$\Pep\Pem\to\Pep\Pem\Pep\Pem\Pep\Pem$.

\paragraph{Squared matrix element}
If no external quarks are involved the squared matrix element 
$|\M_{\mathrm{lept}}|^2$ is simply obtained by squaring the sum over all
contributions $\eta_n\M^{(\mathrm{ew})}_n$,
\beq
|\M_{\mathrm{lept}}|^2 = 
\left| \sum_{n=1}^N \eta_n \M^{(\mathrm{ew})}_n \right|^2.
\eeq
For two external quarks the situation is as simple as in the purely
leptonic case. The quark line always closes to the same loop in the
squared ``interference'' diagrams, leading to a global factor 3
from the colour trace,
\beq
\sum_{\mathrm{colour}}|\M_{2q}|^2 =
3 \left| \sum_{n=1}^N \eta_n \M^{(\mathrm{ew})}_n \right|^2.
\eeq
For four external quarks non-trivial colour interferences occur.
Denoting the four quarks by $\bar q_1 q_2\bar q_3 q_4$, there are
two types of diagrams: one type in which the pairs $\bar q_1 q_2$ 
and $\bar q_3 q_4$ define the two quark lines, and another one
in which $\bar q_1 q_4$ and $\bar q_3 q_2$ define the lines.
According to this criterion we divide the set of $N$ permutations 
$(a_n,\dots,h_n)$ into two subsets of $N/2$ permutations for each of the 
two diagram types and
label the permutations by the pair of indices $(n,\al)$ with
$n=1,\dots,N/2$, $\al=1,2$ instead of taking $n=1,\dots,N$.
Hence, $\M^{(\mathrm{ew})}_{n,\al}$ and $\M^{(1\Pg)}_{n,\al}$
denote the sums of purely electroweak and one-gluon diagrams
for $(a_{n,\al},\dots,h_{n,\al})$, respectively, where
the index $\al$ determines whether $\bar q_1 q_2$ and $\bar q_3 q_4$
or $\bar q_1 q_4$ and $\bar q_3 q_2$ 
correspond to the two quark lines.
The squared matrix element is obtained from terms of the form
$\M^{(X)}_{n,\al} (\M^{(Y)}_{m,\be})^*$.
If $\al=\be$, diagrammatically such a term represents interference
graphs in which the two quark lines close separately; for purely
electroweak diagrams ($X=Y=\mathrm{ew}$) each closed quark yields 
a colour factor 3, 
leading to a colour factor $3^2=9$ for the whole interference term.
If $\al\ne\be$, the two quark lines close to a single loop, 
leading to a colour factor 3 for purely electroweak diagrams.
If at least one of the terms $\M^{(X)}_{n,\al}$ and
$(\M^{(Y)}_{m,\be})^*$ stands for one-gluon diagrams, the
colour factors result from some simple traces over Gell--Mann 
matrices. We summarize the colour factors in terms of $2\times2$
matrices $C^{(X,Y)}_{4q}$:
\beq
C^{(\mathrm{ew,ew})}_{4q} = \pmatrix{9 & 3 \cr 3 & 9}, \qquad
C^{(1\Pg,1\Pg)}_{4q} = \pmatrix{2 & -\frac{2}{3} \cr -\frac{2}{3} & 2}, \qquad
C^{(\mathrm{ew},1\Pg)}_{4q} = \pmatrix{0 & 4 \cr 4 & 0}.
\eeq
Using these matrices, the squared matrix element reads
\beqar
\sum_{\mathrm{colour}}|\M_{4q}|^2 &=&
\sum_{\al,\be=1}^2 
\left(\sum_{n=1}^{N/2} 
  \eta_{n,\al} \M^{(\mathrm{ew})}_{n,\al}\right) 
C^{(\mathrm{ew,ew})}_{4q,\al\be}
\left(\sum_{m=1}^{N/2} 
  \eta_{m,\be} \M^{(\mathrm{ew})}_{m,\be}\right)^* 
\nn\\
&& {} +
\sum_{\al,\be=1}^2
\left(\sum_{n=1}^{N/2} 
  \eta_{n,\al} \M^{(1\Pg)}_{n,\al}\right) 
C^{(1\Pg,1\Pg)}_{4q,\al\be}
\left(\sum_{m=1}^{N/2}
  \eta_{m,\be} \M^{(1\Pg)}_{m,\be}\right)^* 
\nn\\
&& {} + 2\Re\left\{
\sum_{\al,\be=1}^2 
\left(\sum_{n=1}^{N/2} 
  \eta_{n,\al} \M^{(\mathrm{ew})}_{n,\al}\right) 
C^{(\mathrm{ew},1\Pg)}_{4q,\al\be}
\left(\sum_{m=1}^{N/2} 
  \eta_{m,\be} \M^{(1\Pg)}_{m,\be}\right)^* 
\right\}.
\eeqar

\subsection{Introduction of finite decay widths}
\label{se:width}

For the finite decay widths introduced in the propagator functions
\refeq{eq:props} we provide three possibilities as options:
\beq
\Gamma(p^2) \;=\; \left\{
\barr{ll}
\Gamma, & \mbox{fixed width,} \\
\label{eq:stepwidth}
\Gamma\times\theta(p^2), & \mbox{step width,} \\
\disp\Gamma\times\frac{p^2}{M^2}\,\theta(p^2), \qquad & \mbox{running width,}
\earr \right.
\eeq
where $M$ and $\Gamma$ denote the mass and on-shell decay width of the 
propagating particle, respectively, and $\theta(x)$ is the usual step function.
In the {\it fixed-width scheme}, the constant width $\Gamma$
is introduced in each propagator, $s$-channel- 
or $t$-channel-like, i.e.\
for time-like and space-like momentum transfer, respectively. 
In the {\it step-width scheme} and in the {\it running-width scheme}
only $s$-channel propagators receive a finite width, as actually demanded 
from field-theoretical principles, and the factor $p^2/M^2$ in the 
running width reproduces the correct momentum dependence of the imaginary 
part of a one-loop self-energy for a particle that decays into massless
decay products. 

Note that none of these schemes preserves gauge invariance, as
for instance explained in 
\citeres{Grunewald:2000ju,Beenakker:1996kt,Stuart:1991xk,Argyres:1995ym,Dittmaier:1997mc,Beenakker:1999hi}.
Nevertheless for the processes $\Pep\Pem\to 4f(+\gamma)$ the 
gauge-invariance-violating effects in the fixed-width scheme turned out 
to be sufficiently suppressed \cite{Denner:1999gp,Argyres:1995ym}, 
rendering this simple scheme very useful. Moreover, it was pointed out in 
\citere{Denner:1999gp} that gauge invariance is restored in the
fixed-width approach if complex gauge-boson masses are used in all
Feynman rules,%
\footnote{It is not clear whether this scheme, where masses and couplings are
complex, is consistent also in higher perturbative orders.}
i.e.\ in particular the weak mixing angle is derived from the
complex gauge-boson masses,
\beq
\cw^2 = 1-\sw^2 = \frac{\MW^2-\ri\MW\GW}{\MZ^2-\ri\MZ\GZ}.
\eeq
This {\it complex-mass scheme} is included as a fourth option in
{\sc Lusifer}.

The so-called {\it fermion-loop scheme} \cite{Argyres:1995ym}, which
introduces gauge-boson widths by a gauge-invariant fermion-loop resummation,
is not sufficient for six-fermion production in general
and thus not considered in the following, since it
does not provide an introduction of the decay widths of the top quark
and the Higgs boson, 
because these particles do not (or not entirely) decay into 
fermion--antifermion pairs.
On the other hand, the effective Lagrangian approach of
\citere{Beenakker:1999hi} or appropriate expansions
\cite{Stuart:1991xk} about resonance poles could also be used for
$\Pep\Pem\to 6f$ processes, but this task goes beyond the scope of this paper.

\subsection{Cross-checks}
\label{se:checkamp}

Apart from performing internal checks, we have compared the squared
matrix elements for some phase-space points 
with the results obtained with
{\sc Madgraph} \cite{Stelzer:1994ta}. Since {\sc Madgraph} calculates
amplitudes within the unitary gauge without providing a gauge-invariant
introduction of decay widths, we performed the comparison for
vanishing decay widths consistently. We find very good numerical agreement
for non-exceptional phase-space points
between our results and the 
ones of {\sc Madgraph}, whenever the latter
program delivers an amplitude. For a recent test version of {\sc Madgraph}, 
directly obtained from the authors \cite{stelzer}, there are only 
two $\Pep\Pem\to 6f$ channels that are not covered, 
$\Pep\Pem\to\Pep\Pem\Pep\Pem\Pep\Pem$ and 
$\Pep\Pem\to\Pep\Pem\Pep\Pem\nu_\Pe\bar\nu_\Pe$.
We note that the amplitude check was done with and without (coherent) 
inclusion of gluon background diagrams.

\section{Multi-channel phase-space integration}
\setcounter{paragraph}{0}
\label{se:psint}

Monte Carlo generators are widely used for experimental analyses
since they provide realistic event samples that can be compared with 
experimental observations after detector simulation. 
Another advantage of a Monte Carlo generator is its flexibility; all 
possible observables can be studied easily by choosing suitable
separation cuts. However, for a proper use of the Monte Carlo program a 
basic knowledge of the numerical integration techniques is helpful. 
Therefore, the numerical integration of {\sc Lusifer} is briefly
outlined in the following. The general method is very similar to 
the one used in {\sc RacoonWW}, as described in
\citeres{Denner:1999gp,Roth:kk}, 
but generalized to six-fermion processes.

For six-fermion production, $\Pep \Pem \to 6 f$, an integration over
a 14-dimensional phase space has to be performed.
Denoting the incoming $\Pe^\pm$ momenta $p_\pm$ and the outgoing
fermion momenta $k_i$ ($i=1,\dots,6$), the phase-space integral reads
\beqar
\nn
\label{eq:int}
\int\rd \sigma_{\born}&=&
\int\rd \Phi^{2\to6} \, \frac{|\M(p_+,p_-,k_1,\ldots,k_6)|^2}
{ 8 (2 \pi)^{14} E_{\mathrm{CM}}^2}, \\
\rd \Phi^{2\to6}&=& 
\prod_{i=1}^6\frac{\rd^3 {\bk}_i}{2 k^0_i} \,
\delta^{(4)}\bigg(p_++p_--\sum_{j=1}^6k_j\bigg)
\bigg|_{k^0_i=\sqrt{{\bk}_i^2+m_i^2}},
\eeqar
where the fermion masses $m_i$ are zero in our case ($m_i=0$)
and $E_{\mathrm{CM}}$ is the total CM energy.
The numerical integration \refeq{eq:int} is rather complicated 
owing to the rich peaking structure of the integrand. 
The amplitude involves a huge number of 
propagators that become resonant or are enhanced in various phase-space 
regions. The problem is even more serious if the separation cuts, which are
required to exclude infrared and collinear singularities from the physical 
integration domain, are small.

To obtain reliable numerical results,
events have to be sampled more frequently 
in the integration domains where the integrand is large. This
redistribution of events is called {\it importance sampling}. In 
practice, this means that the mapping of the pseudo-random numbers 
$r_i$, $0\le r_i \le 1$, into the space of final-state momenta has to be 
chosen such that the corresponding Jacobian $1/g({\bfr})$ 
cancels the propagators of the differential cross section at least 
partially:
\beq
\rd \Phi^{2\to6}=\rd^{14} \bfr \, \frac{1}{g(\bfr)}.
\eeq

\subsection{Propagator mappings}

In order to smooth a single propagator with momentum transfer $q$
in the square of a matrix element,
\beq
\label{eq:prop}
|\M|^2\propto \frac{1}{(q^2-M^2)^2+M^2 \Ga^2},
\eeq
the phase space is parametrized in such a way that the virtuality $q^2$
of the propagator is chosen as an integration variable. 
The integral over $q^2$ is
transformed into an integral over the random number $r$:
\beq
\int^{q^2_{\max}}_{q^2_{\min}} \rd q^2 =\int_0^1 \rd r \, 
\frac{1}{g_{\mathrm{prop}}(q^2(r))}. 
\eeq
The dependence of $q^2$ on the pseudo-random number $r$ has to be chosen
such that the Jacobian $1/g_{\mathrm{prop}}$ smoothes
the propagator \refeq{eq:prop}.
In {\sc Lusifer} two types of mappings are used according to the mass $M$
of the propagating particle. Since all occurring massive particles
possess a finite decay width $\Gamma$, a suitable mapping is of
Breit--Wigner type ($M\ne 0$, $\Gamma\ne 0$):
\beqar
\nn
\label{eq:impsampling1}
q^2(r)&=&M \Ga \tan [y_1 +(y_2 - y_1 ) r]+M^2,\\
g_{\mathrm{prop}}(q^2)&=&\frac{M \Ga}{(y_2 - y_1 ) [(q^2-M^2)^2+M^2 \Ga^2]}
\eeqar 
with
\beq
y_{1,2}={\mathrm {arctan}}
\left(\frac{q^2_{\min,\max}-M^2}{M \Ga}\right).
\eeq
For massless particles the following mappings are appropriate
($M=\Gamma=0$):
\beqar
\nn
\label{eq:impsampling2}
q^2(r)&=&\left[r (q^2_{\max})^{1-\nu}+
(1-r) (q^2_{\min})^{1-\nu}\right]^{\frac{1}{1-\nu}},\\
g_{\mathrm{prop}}(q^2)&=&\frac{1-\nu}
{(q^2)^\nu\left[(q^2_{\max})^{1-\nu}-(q^2_{\min})^{1-\nu}\right]} \,;\\[1em]
\nn
\label{eq:impsampling3}
q^2(r)&=& (q^2_{\max})^r \, (q^2_{\min})^{1-r},\\
g_{\mathrm{prop}}(q^2)&=&
\frac{1}{q^2\ln (q^2_{\max}/q^2_{\min})}. 
\eeqar
The mapping \refeq{eq:impsampling2} is only valid for $\nu\ne 1$, 
while Eq.~\refeq{eq:impsampling3} is a substitute of Eq.~\refeq{eq:impsampling2}
for $\nu=1$. The naive expectation $\nu=2$ from the squared matrix element 
is not necessarily the best choice because the propagator poles are 
partially cancelled in the collinear limit. It turns out that a proper 
value is $\nu\lsim 1$. 

\subsection{Multi-channel approach}

Obviously, importance sampling of all propagators appearing in the 
amplitude is not possible by a single phase-space
mapping. Therefore, we apply the {\it multi-channel approach} 
\cite{Berends:1994pv,Berends:gf} where $N$ phase-space parametrizations
with appropriate mappings, called {\it channels}, are used simultaneously.
To this end, the phase-space integral is rewritten into the form
\beq
\label{eq:multichannel}
\int \rd \Phi^{2\to6}=
\int_0^1 \rd r \, \sum_{i=1}^N \theta(r-\beta_{i-1}) \theta(\beta_{i}-r) 
\int_0^1 \rd^{14} \bfr \, \frac{1}{g_{\mathrm{tot}}(\bfr)},
\eeq
where 
the $\beta_i$ define a partition of unity, i.e.\
$\beta_i=\sum_{j=1}^i \alpha_j$, $\beta_0=0$, 
$\beta_N=\sum_{j=1}^N\alpha_j=1$.
In Eq.~\refeq{eq:multichannel} an additional random number $r$ is introduced 
in order to select a {\it channel} $i$ randomly with probability 
$\alpha_i\ge 0$. 
A channel is defined by its mapping 
from the pseudo-random numbers into the space of final-state momenta.
The {\it total density} $g_{\mathrm{tot}}$ is composed of the 
single densities $g_i$ of the different {\it channels} weighted by
the {\it a priori weights} $\alpha_i$:
\beq
g_{\mathrm{tot}}=\sum_{i=1}^N \alpha_i g_i.
\eeq
In this way, it is possible to simultaneously include 
different phase-space mappings 
which are suitable for different parts of the integrand. 
To be specific, for each diagram a channel exists, so that all 
propagators of the squared diagram are smoothed by the corresponding 
local density $g_i$. No special channels are provided for interference 
contributions. 

The a priori weights $\alpha_i$ are adapted in the early phase of the 
Monte Carlo run several times to optimize the convergence behaviour 
of the numerical integration.
This {\it adaptive weight optimization} \cite{Kleiss:qy} 
increases the a priori weights $\alpha_i$ for such channels that correspond to 
important diagrams of the process, i.e.\ to diagrams that give large 
contributions to the total cross section.

The actual calculation is performed in two steps: First, 
a channel is chosen and the momenta of the final-state particles are 
calculated with the corresponding phase-space generator. Secondly, the 
total density is determined. Besides the evaluation of the matrix
elements, the second step is the most time-consuming part 
since the local
densities of all channels have to be calculated. 

\subsection{Generic construction of phase-space generators}

\bfi
\centerline{
\setlength{\unitlength}{1pt}
\begin{picture}(230,170)(0,20)
\Line( 25, 50)(100, 50)
\ArrowLine( 25, 25)( 25, 50)
\Line( 25, 50)( 25, 60)
\ArrowLine(200, 25)(200, 50)
\Line(180, 50)(200, 50)
\put( 50, 55){$t_1$}
\put(180, 55){$t_\tau$}
\Vertex(110, 50){.5}
\Vertex(120, 50){.5}
\Vertex(130, 50){.5}
\Vertex(140, 50){.5}
\Vertex(150, 50){.5}
\Vertex(160, 50){.5}
\Vertex(170, 50){.5}
\Vertex( 25, 50){3}
\Line( 25, 50)( 25, 60)
\Line( 25, 90)( 25,150)
\ArrowLine( 25,150)( 10,180)
\ArrowLine( 25,150)( 40,180)
\ArrowLine( 25,125)( 40,155)
\ArrowLine( 25,100)( 40,130)
\Vertex( 25, 70){.5}
\Vertex( 25, 80){.5}
\Vertex( 25,100){3}
\Vertex( 25,125){3}
\Vertex( 25,150){3}
\put(  6,135){$s_1^{(0)}$}
\put(  6,110){$s_2^{(0)}$}
\Vertex( 80, 50){3}
\Line( 80, 50)( 80, 60)
\Line( 80, 90)( 80,125)
\ArrowLine( 95,150)( 83,180)
\ArrowLine( 95,150)(110,180)
\ArrowLine( 65,150)( 50,180)
\ArrowLine( 65,150)( 77,180)
\Line( 80,125)( 95,150)
\Line( 80,125)( 65,150)
\ArrowLine( 80,100)( 95,120)
\Vertex( 80, 70){.5}
\Vertex( 80, 80){.5}
\Vertex( 80,100){3}
\Vertex( 80,125){3}
\Vertex( 95,150){3}
\Vertex( 65,150){3}
\put( 51,132){$s_1^{(1)}$}
\put( 92,132){$s_2^{(1)}$}
\put( 58,110){$s_3^{(1)}$}
\Vertex(200, 50){3}
\Line(200, 50)(200, 60)
\Line(200, 90)(200,125)
\ArrowLine(215,150)(203,180)
\ArrowLine(215,150)(230,180)
\ArrowLine(185,150)(170,180)
\ArrowLine(185,150)(197,180)
\Line(200,125)(185,150)
\Line(200,125)(215,150)
\ArrowLine(200,100)(215,120)
\Vertex(200, 70){.5}
\Vertex(200, 80){.5}
\Vertex(200,100){3}
\Vertex(200,125){3}
\Vertex(185,150){3}
\Vertex(215,150){3}
\put(171,132){$s_1^{(\tau)}$}
\put(212,132){$s_2^{(\tau)}$}
\put(178,110){$s_3^{(\tau)}$}
\end{picture}
}
\vspace*{-.5em}
\caption{Topological diagram for the generic phase-space generator}
\label{fig:GenPS}
\efi 

Since six-fermion production processes involve of the order of $10^2$--$10^4$
diagrams, it is very important to handle the different phase-space mappings 
in a generic way. 
All different channels can be obtained from one generic phase-space 
generator by choosing the topological 
structure of the corresponding diagram, as illustrated in \reffi{fig:GenPS}, 
the order of the incoming and outgoing particles, and
the masses and widths of the internal particles.
In contrast to the calculation of the matrix elements as discussed in 
\refse{se:amps}, there is a fundamental difference between the incoming 
and outgoing particles, and $s$- and $t$-channel propagators.

For each class of diagrams with the same propagator structure we adopt an 
own channel for the numerical integration. Diagrams with four-particle vertices 
also fit into the generic diagram of \reffi{fig:GenPS} if one of the 
propagators is contracted and the corresponding invariant is sampled
uniformly. 
The same is true for diagrams with exclusively $s$-channel propagators, since 
the $s$-channel propagator resulting from a fusion of the two incoming lines
is constant.

The actual calculation of the event kinematics is decomposed into three steps: 
the calculation of {\it time-like invariants}, 
of {\it $2\to 2$ scattering processes with $t$-channel propagators}, and 
of {\it $1\to 2$ particle decays}. 
The phase-space integration reads
\beqar
\int \rd \Phi^{2\to6}&=& 
\prod_{i=1}^4 \int_{s_{i,\min}}^{s_{i,\max}} \rd s_i \, 
\prod_{j=1}^\tau \int \rd \Phi^{2\to 2}_j \, 
\prod_{k=1}^{5-\tau} \int \rd \Phi^{1\to 2}_k,
\label{eq:PSsplit}
\eeqar
where the phase space of the $2 \to 2$ scattering processes and the 
$1 \to 2$ decays are denoted by $\Phi^{2\to 2}_j$ and $\Phi^{1\to 2}_k$,
respectively. Note that the number $\tau$ of $2 \to 2$ scattering processes,
which is the number of $t$-channel lines in \reffi{fig:GenPS}, 
can range from 0 to 5. The four invariants $s_i$ ($i=1,\dots,4$)
result from the phase-space factorization into scattering processes and
decays. More details on phase-space parametrizations can be found in
\citeres{Roth:kk,By73}. 
If several $t$-channel propagators are present, 
i.e.\ for $\tau\ge2$, 
some of the $s_i$ ($i=1,\dots,4$) do not correspond to 
virtualities of propagators in the diagram; for such variables no 
mappings are introduced, they are generated uniformly.

In detail we proceed as follows:

\paragraph{Calculation of time-like invariants}

First of all, the time-like invariants $s^{(j)}_i$ are determined. 
The invariants 
corresponding to $s$-channel propagators are calculated from 
Eqs.~\refeq{eq:impsampling1}--\refeq{eq:impsampling3}. All other time-like
invariants 
are sampled uniformly, i.e.\ calculated from Eq.~\refeq{eq:impsampling2} 
with $\nu=0$. The determination of the invariants $s^{(j)}_i$ is ordered
in such a way that the $s^{(j)}_i$ nearest to final-state particles are 
calculated first, followed by the next-to-nearest $s$-channel propagators,
and so on.

The lower and upper limits $s_{i,\min/\max}^{(j)}$, in general, are 
functions of the already determined invariants.
To enhance the efficiency of the numerical integration, separation cuts 
have to be taken into account. We include invariant-mass cuts of 
final-state particles in the evaluation of 
$s^{(j)}_{i,\min/\max}$ whenever possible. 
 
\paragraph{$2\to2$ scattering processes with $t$-channel propagators}

The subpart of the diagram involving $t$-channel propagators is
decomposed into several $2\to2$ scattering processes. 
The phase-space integration reads
\beq
\int \rd \Phi^{2\to2}(p_1,p_2;q_1,q_2)
=\frac{1}{8 \sqrt{(p_1 p_2)^2 -p_1^2 p_2^2}}
\int_0^{2\pi} \rd \phi^* \,
\int_{t_{\min}}^{t_{\max}} \rd t,\\
\eeq
where $p_{1,2}$ are the incoming and $q_{1,2}$ are the outgoing momenta, and
$\phi^*$ is the azimuthal angle 
defined by $p_1$ and $q_1$ in the CM frame 
of the subprocess, where ${\bf p}_1+{\bf p}_2=0$. The argument 
$t=(p_1-q_1)^2$ of the $t$-channel propagator is calculated according 
to Eqs.~\refeq{eq:impsampling1}--\refeq{eq:impsampling3}. The azimuthal angle
$\phi^*$ is sampled uniformly. Since the corresponding polar angle 
$\cos\theta^*$ of this scattering process depends on the invariant 
$t$ linearly,
\beq
t=\frac{p_1^2 q_2^2+p_2^2 q_1^2-2 (p_1 p_2)(q_1 q_2)
+2 \sqrt{(p_1 p_2)^2-p_1^2 p_2^2} \sqrt{(q_1 q_2)^2-q_1^2 q_2^2} \cos \theta^*}
{(p_1+p_2)^2},
\eeq
the calculation of the momenta and the Lorentz transformation  
into the laboratory (LAB) frame is simple and, thus, skipped here. 
Explicit formulas can for instance be found in \citere{Roth:kk}.
Subprocesses that are nearest to the incoming particles
of the whole $\Pep \Pem\to 6 f$
scattering process are calculated first, followed
by the next-to-nearest, and so on. 

For the $2\to2$ scattering processes that are attached to the
original incoming particles, we already
take into account possible cuts on angles
between outgoing particles and the beams that effectively reduce the
integration range 
of $t$, i.e.\ we include these cuts in the
determination of $t_{\min/\max}$. 

\paragraph{$1\to2$ particle decays}

It remains to perform the decays of the $s$-channel
particles. The phase-space integration reads
\beq
\int \rd \Phi^{1\to 2}(q_1,q_2)
=\frac{\sqrt{(q_1 q_2)^2-q_1^2 q_2^2}}{4 (q_1+q_2)^2}
\int_0^{2 \pi} \rd \phi^* \, 
\int_{-1}^1 \rd \cos \theta^*,
\eeq
where $\phi^*$ and $\theta^*$ are the azimuthal and polar angles 
in the rest frame of the decaying particle, respectively.
The variables $\phi^*$ and $\cos \theta^*$ are sampled uniformly.
The momenta of the outgoing particles are $q_1$ and $q_2$.
As in the former step, the calculation of the momenta and the Lorentz 
transformation into the LAB frame is straightforward 
(see e.g.\ \citere{Roth:kk}). 

\begin{sloppypar}
\vspace{.8em}
We illustrate the general strategy by considering the specific
topology shown in \reffi{fig:SpecificPS}, where a particular choice for
the incoming and outgoing momenta $p_\pm$ and $k_i$ is made.
\bfi
\centerline{
\setlength{\unitlength}{1pt}
\begin{picture}(210,160)(0,20)
\ArrowLine( 25, 25)( 25, 50)
\Line( 25, 50)(175, 50)
\ArrowLine(175, 25)(175, 50)
\ArrowLine( 25, 50)( 25,100)
\Line( 75, 50)( 75,140)
\ArrowLine( 75,140)( 60,175)
\ArrowLine( 75,140)( 90,175)
\ArrowLine( 75, 95)( 90,130)
\ArrowLine(125, 50)(125,100)
\ArrowLine(175, 50)(175,100)
\Vertex( 25, 50){3}
\Vertex( 75, 50){3}
\Vertex( 75, 95){3}
\Vertex( 75,140){3}
\Vertex(125, 50){3}
\Vertex(175, 50){3}
\put( 45, 55){$t_1$}
\put( 95, 55){$t_2$}
\put(145, 55){$t_3$}
\put( 60,115){$s_1$}
\put( 60, 70){$s_2$}
\put(  5, 35){$p_+$}
\put(185, 35){$p_-$}
\put(  5, 75){$k_4$}
\put(185, 75){$k_1$}
\put(130, 75){$k_3$}
\put( 50,155){$k_5$}
\put( 90,110){$k_2$}
\put( 90,155){$k_6$}
\end{picture}
}
\vspace*{-.5em}
\caption{Topological diagram for a specific phase-space generator}
\label{fig:SpecificPS}
\efi 
For this example the phase-space parametrization \refeq{eq:PSsplit}
reads
\beqar
\int \rd \Phi^{2\to6}\Big|_{\mathrm{Fig.\ref{fig:SpecificPS}}}&=& 
\prod_{i=1}^4 \int_{s_{i,\min}}^{s_{i,\max}} \rd s_i \, 
\int \rd \Phi^{2\to 2}(p_+,p_-;k_4,k_{12356}) \, 
\nn\\ && {}\times
\int \rd \Phi^{2\to 2}(p_-,p_+-k_4;k_1,k_{2356}) \,
\int \rd \Phi^{2\to 2}(p_+-k_4,p_--k_1;k_{256},k_{3}) \,
\nn\\ && {}\times
\int \rd \Phi^{1\to 2}(k_2,k_{56})\,
\int \rd \Phi^{1\to 2}(k_5,k_6),
\hspace*{2em}
\eeqar
where sums of outgoing momenta are abbreviated by
$k_{ij}=k_i+k_j$, $k_{ijk}=k_i+k_j+k_k$, etc.
Note that only the time-like invariants $s_1=k_{56}^2$ and $s_2=k_{256}^2$ 
corrspond to virtualities of propagators in the diagram, while
$s_3=k_{2356}^2$ and $s_4=k_{12356}^2$ correspond to invariant masses 
of fictitious final-state particles within the first two $2\to 2$ scattering 
processes.
\end{sloppypar}

For an efficient and fast numerical integration {\sc Lusifer} evaluates
the different subparts of the phase-space mappings 
for all channels only once for a single event.
The local densities of individual channels are then easily
obtained from the Jacobians of these subparts. This speeds up the
numerical integration considerably.

\subsection{Cross checks}

The Monte Carlo part of {\sc Lusifer} can, in principle, be applied to 
arbitrary processes. In order to check the phase-space 
integration, we implemented the 
tree-level matrix elements of {\sc RacoonWW}, as given
in \citere{Denner:1999gp}, in {\sc Lusifer} and reproduced the results 
given in Table~1 of \citere{Denner:1999gp} for four-fermion
production, $\eeffff$, and for the radiative processes, $\eeffffg$,
in the fixed width scheme.
We found good agreement between the $4f$ and $4f\gamma$ results
integrated with {\sc Lusifer} and {\sc RacoonWW}.

\section{Higher-order initial-state radiation at the leading logarithmic level}
\label{se:isr}

A first step to improve tree-level predictions by higher-order
radiative corrections consists in including universal corrections
such as the leading logarithms of initial-state radiation.
In {\sc Lusifer} this is done via structure functions
\cite{sf,Beenakker:1996kt}. 
The lowest-order differential cross section is convoluted 
in the energy fractions $x_\pm$ of the incoming electron and positron:
\beq
\label{eq:isr}
\int \rd \, \sigma_{\mathrm{Born+ISR}}= 
\int_0^1 \rd x_+ \, \Ga_{\mathrm{ee}}^{\mathrm{LL}}(x_+,Q^2) 
\int_0^1 \rd x_- \, \Ga_{\mathrm{ee}}^{\mathrm{LL}}(x_-,Q^2) 
\int \rd \sigma_{\mathrm{Born}} (x_+ p_+, x_- p_-),
\eeq
where the splitting scale $Q$ is not fixed at the leading logarithmic
level and quantifies part of the missing radiative corrections.
In {\sc Lusifer},
$Q$ is set to the CM energy $E_{\mathrm{CM}}$ by default.%
\footnote{For scattering processes that are dominated by small 
splitting scales the use of purely $s$-dependent structure
functions is not a good approximation for treating ISR effects.
This is, in particular, the case if $\Pe^\pm$ are scattered in the
very forward direction. Possible improvements for such situations have been,
for instance, described in the context of single-W production
in $\Pep\Pem\to 4f$ (see e.g.\ \citeres{Grunewald:2000ju,Kurihara:1999qz}).}
The Born cross section in 
Eq.~\refeq{eq:isr} is calculated in the same way as in the case without ISR 
but with the momenta $x_\pm p_\pm$ for the incoming $\Pe^\pm$.
The precise form of the structure function 
$\Ga_{\mathrm{ee}}^{\mathrm{LL}}(x,Q^2)$ used in {\sc Lusifer}
is the same as in Eqs.~(5.2)--(5.4) of the second paper in
\citere{Denner:2000kn}.

\section{\boldmath{Classification of $6f$ final states in $\Pep\Pem$ 
collisions}}
\label{se:ee6fprocs}

The derivation of squared matrix elements described in
\refse{se:amps} is valid for all processes involving eight external
fermions and restricted for gluonic diagrams to processes involving 
up to four external quarks. 
Thus, these results can be applied
to various six-fermion processes in lepton--lepton, lepton--hadron, and
hadron--hadron scattering. In this paper, we focus on 
$\Pep\Pem\to 6f$ processes in the following, representing an
important class of reactions at future linear colliders such as
TESLA \cite{Accomando:1998wt}.
We divide the set of $6f$ final states into three (overlapping)
subsets corresponding to different subprocesses of interest:

\setcounter{paragraph}{0}
\paragraph{Processes involving top quarks}

The most interesting processes involving top quarks correspond to
top-pair production with the subsequent decay of the top
quarks into six fermions which proceeds via W~bosons,
$\Pep\Pem\to\Pt\bar\Pt\to\Pb\PW^+\bar\Pb\PW^-\to6f$. 
The corresponding diagram is shown in \reffi{fig:ttgraphs}.
\bfi
\centerline{
\setlength{\unitlength}{1pt}
\begin{picture}(155,125)(0,5)
\ArrowLine( 10, 45)( 30, 65)
\ArrowLine( 30, 65)( 10, 85)
\ArrowLine(145, 30)(130, 45)
\ArrowLine(130, 45)(145, 55)
\ArrowLine(145, 75)(130, 85)
\ArrowLine(130, 85)(145,100)
\ArrowLine( 70, 65)(100, 95)
\ArrowLine(100, 35)( 70, 65)
\ArrowLine(100, 95)(145,120)
\ArrowLine(145, 10)(100, 35)
\Photon(30, 65)( 70, 65){2}{6}
\Photon(100, 35)(130, 45){2}{5}
\Photon(100, 95)(130, 85){2}{5}
\Vertex(30,  65){2.0}
\Vertex(130, 45){2.0}
\Vertex(130, 85){2.0}
\Vertex(70, 65){2.0}
\Vertex(100, 35){2.0}
\Vertex(100, 95){2.0}
\put(40,50){$\ga/\PZ$}
\put(80,90){$\Pt$}
\put(80,34){$\Pt$}
\put(105,73){$\PW$}
\put(105,48){$\PW$}
\put(150,120){$\Pb$}
\put(150, 5){$\bar\Pb$}
\put( -5, 85){$\Pep$}
\put( -5, 40){$\Pem$}
\end{picture}
} 
\caption{Diagram for $\Pt\bar\Pt$ production:
$\Pep\Pem\to\Pt\bar\Pt\to\Pb\PW^+\bar\Pb\PW^-\to6f$}
\label{fig:ttgraphs}
\efi
All $6f$ final states that are relevant for $\Pt\bar\Pt$ production are 
of the form $\Pep\Pem\to \Pb\bar\Pb f_1 \bar f'_1 f_2 \bar f'_2$,
where $f_i \bar f'_i$ ($i=1,2$) are two weak isospin doublets.

There is a second, but less interesting class of $6f$ final states that
involve top-quark diagrams but do not contain $\Pt\bar\Pt$ production
as a subprocess. These processes are of the form
$\Pep\Pem\to \nu_\Pe\bar\nu_\Pe \Pb\bar\Pb f \bar f$, where $f$
is any fermion other than the electron.

\paragraph{Vector-boson scattering and $s$-channel Higgs production}

One of the most interesting class of subprocesses in $\Pep\Pem\to 6f$
is the scattering of weak gauge bosons, 
$\Pep\Pem\to 2f+(V_1 V_2 \to V_3 V_4) \to 6f$.
According to the charges of the ``incoming'' vector-boson pair
$V_1 V_2$, there are three different types of vector-boson
scattering channels, as shown in \reffi{fig:VVVVgraphs}:
\bfi
\centerline{
\setlength{\unitlength}{1pt}
\begin{picture}(130,150)(0,0)
\ArrowLine( 10, 15)( 30, 15)
\ArrowLine( 30, 15)(110, 15)
\ArrowLine(110,115)( 30,115)
\ArrowLine( 30,115)( 10,115)
\ArrowLine(110, 30)( 95, 45)
\ArrowLine( 95, 45)(110, 55)
\ArrowLine(110, 75)( 95, 85)
\ArrowLine( 95, 85)(110,100)
\Photon(30, 15)(60, 65){2}{7}
\Photon(30,115)(60, 65){2}{7}
\Photon(95, 45)(60, 65){2}{6}
\Photon(95, 85)(60, 65){2}{6}
\Vertex(30, 15){2.0}
\Vertex(30,115){2.0}
\Vertex(95, 85){2.0}
\Vertex(95, 45){2.0}
\GCirc(60, 65){10}{.5}
\put(15,84){$\ga/\PZ$}
\put(15,41){$\ga/\PZ$}
\put( -5,115){$\Pep$}
\put( -5, 10){$\Pem$}
\put(115,115){$\Pep$}
\put(115, 10){$\Pem$}
\put(-10,135){\bf (a)}
\end{picture}
\hspace*{1em}
\setlength{\unitlength}{1pt}
\begin{picture}(130,150)(0,0)
\ArrowLine( 10, 15)( 30, 15)
\ArrowLine( 30, 15)(110, 15)
\ArrowLine(110,115)( 30,115)
\ArrowLine( 30,115)( 10,115)
\ArrowLine(110, 30)( 95, 45)
\ArrowLine( 95, 45)(110, 55)
\ArrowLine(110, 75)( 95, 85)
\ArrowLine( 95, 85)(110,100)
\Photon(30, 15)(60, 65){2}{7}
\Photon(30,115)(60, 65){2}{7}
\Photon(95, 45)(60, 65){2}{6}
\Photon(95, 85)(60, 65){2}{6}
\Vertex(30, 15){2.0}
\Vertex(30,115){2.0}
\Vertex(95, 85){2.0}
\Vertex(95, 45){2.0}
\GCirc(60, 65){10}{.5}
\put(25,84){$\PW$}
\put(25,41){$\PW$}
\put( -5,115){$\Pep$}
\put( -5, 10){$\Pem$}
\put(115,115){$\bar\nu_\Pe$}
\put(115, 10){$\nu_\Pe$}
\put(-10,135){\bf (b)}
\end{picture}
\hspace*{1em}
\setlength{\unitlength}{1pt}
\begin{picture}(130,150)(0,0)
\ArrowLine( 10, 15)( 30, 15)
\ArrowLine( 30, 15)(110, 15)
\ArrowLine(110,115)( 30,115)
\ArrowLine( 30,115)( 10,115)
\ArrowLine(110, 30)( 95, 45)
\ArrowLine( 95, 45)(110, 55)
\ArrowLine(110, 75)( 95, 85)
\ArrowLine( 95, 85)(110,100)
\Photon(30, 15)(60, 65){2}{7}
\Photon(30,115)(60, 65){2}{7}
\Photon(95, 45)(60, 65){2}{6}
\Photon(95, 85)(60, 65){2}{6}
\Vertex(30, 15){2.0}
\Vertex(30,115){2.0}
\Vertex(95, 85){2.0}
\Vertex(95, 45){2.0}
\GCirc(60, 65){10}{.5}
\put(15,84){$\ga/\PZ$}
\put(25,41){$\PW$}
\put( -5,115){$\Pep$}
\put( -5, 10){$\Pem$}
\put(115,115){$\Pep$}
\put(115, 10){$\nu_\Pe$}
\put(-10,135){\bf (c)}
\end{picture}
}
\vspace*{.5em}
\centerline{
\setlength{\unitlength}{1pt}
\begin{picture}(170,140)(0,0)
\ArrowLine( 10, 15)( 30, 15)
\ArrowLine( 30, 15)(110, 15)
\ArrowLine(110,115)( 30,115)
\ArrowLine( 30,115)( 10,115)
\ArrowLine(160, 30)(145, 45)
\ArrowLine(145, 45)(160, 55)
\ArrowLine(160, 75)(145, 85)
\ArrowLine(145, 85)(160,100)
\Photon(30, 15)(60, 65){2}{7}
\Photon(30,115)(60, 65){2}{7}
\Photon(145, 45)(110, 65){2}{6}
\Photon(145, 85)(110, 65){2}{6}
\DashLine(60, 65)(110, 65){5}
\Vertex(30, 15){2.0}
\Vertex(30,115){2.0}
\Vertex(145, 85){2.0}
\Vertex(145, 45){2.0}
\Vertex(60, 65){2.0}
\Vertex(110, 65){2.0}
\put(10,84){$\PW/\PZ$}
\put(10,41){$\PW/\PZ$}
\put(80,70){$\PH$}
\put( -5,115){$\Pep$}
\put( -5, 10){$\Pem$}
\put(115,115){$\bar\nu_\Pe/\Pep$}
\put(115, 10){$\nu_\Pe/\Pem$}
\put(-33,115){\bf (d)}
\end{picture}
\vspace*{-.5em}
} 
\caption{Diagram structures for vector-boson scattering:
$\Pep\Pem\to 2f+(V_1 V_2 \to V_3 V_4) \to 6f$}
\label{fig:VVVVgraphs}
\efi
(a) neutral--neutral, (b) charged--charged, and (c) mixed.
Among all $6f$ final states these channels are distinguished by
the appearance of (a) $\Pep\Pem$, (b) $\nu_\Pe\bar\nu_\Pe$, and
(c) $\Pep\nu_\Pe/\Pem\bar\nu_\Pe$ in the final state.

ZZ and WW fusion processes, which are included in types (a) and (b),
are also important in Higgs physics. If the Higgs-boson mass $\MH$
is large enough that 
the decay channel $\PH\to\PW\PW$ opens,
resonant $s$-channel Higgs production 
dominates the vector-boson scattering cross section,
$\PW\PW,\PZ\PZ\to\PH\to\PW\PW/\PZ\PZ$, which is then one of the
two main production mechanisms for the Higgs boson in $\Pep\Pem$
annihilation. The relevant Feynman diagram is shown in 
\reffi{fig:VVVVgraphs}(d).

\paragraph{Three-boson production and Higgs-strahlung off Z bosons}

All $6f$ final states have in common that they contribute to
production processes of three electroweak gauge bosons,
$\Pep\Pem\to V_1 V_2 V_3 \to 6f$.
As illustrated in \reffi{fig:VVVgraphs}, there are two different
types of reactions, distinguished by the charges of the produced bosons:
(a) charged--charged--neutral and (b) neutral--neutral--neutral.
\bfi
\centerline{
\setlength{\unitlength}{1pt}
\begin{picture}(130,115)(0,15)
\ArrowLine( 10, 45)( 30, 65)
\ArrowLine( 30, 65)( 10, 85)
\ArrowLine(125, 15)(110, 30)
\ArrowLine(110, 30)(125, 40)
\ArrowLine(125, 55)(110, 65)
\ArrowLine(110, 65)(125, 75)
\ArrowLine(125, 90)(110,100)
\ArrowLine(110,100)(125,115)
\Photon(30, 65)( 70, 65){2}{6}
\Photon(70, 65)(110, 65){2}{6}
\Photon(70, 65)(110, 30){2}{7}
\Photon(70, 65)(110,100){2}{7}
\Vertex(30,  65){2.0}
\Vertex(110, 65){2.0}
\Vertex(110, 30){2.0}
\Vertex(110,100){2.0}
\GCirc(70, 65){10}{.5}
\put(33,50){$\ga/\PZ$}
\put(80,92){$\PW$}
\put(97,72){$\PW$}
\put(98,47){$\ga/\PZ$}
\put( -5, 85){$\Pep$}
\put( -5, 40){$\Pem$}
\put( -5,115){\bf (a)}
\end{picture}
\hspace*{2em}
\setlength{\unitlength}{1pt}
\begin{picture}(130,115)(0,15)
\ArrowLine( 10, 45)( 30, 65)
\ArrowLine( 30, 65)( 10, 85)
\ArrowLine(125, 15)(110, 30)
\ArrowLine(110, 30)(125, 40)
\ArrowLine(125, 55)(110, 65)
\ArrowLine(110, 65)(125, 75)
\ArrowLine(125, 90)(110,100)
\ArrowLine(110,100)(125,115)
\Photon(30, 65)( 70, 65){2}{6}
\Photon(70, 65)(110, 65){2}{6}
\Photon(70, 65)(110, 30){2}{7}
\Photon(70, 65)(110,100){2}{7}
\Vertex(30,  65){2.0}
\Vertex(110, 65){2.0}
\Vertex(110, 30){2.0}
\Vertex(110,100){2.0}
\GCirc(70, 65){10}{.5}
\put(33,50){$\ga/\PZ$}
\put(76,92){$\ga/\PZ$}
\put(92,73){$\ga/\PZ$}
\put(98,47){$\ga/\PZ$}
\put( -5, 85){$\Pep$}
\put( -5, 40){$\Pem$}
\put( -5,115){\bf (b)}
\end{picture}
} 
\vspace*{.5em}
\centerline{
\setlength{\unitlength}{1pt}
\begin{picture}(190,125)(0,0)
\ArrowLine( 10, 45)( 30, 65)
\ArrowLine( 30, 65)( 10, 85)
\ArrowLine(125, 15)(110, 30)
\ArrowLine(110, 30)(125, 40)
\ArrowLine(165, 55)(150, 65)
\ArrowLine(150, 65)(165, 75)
\ArrowLine(165, 90)(150,100)
\ArrowLine(150,100)(165,115)
\Photon(30, 65)( 70, 65){2}{6}
\Photon(110, 80)(150, 65){2}{6}
\Photon( 70, 65)(110, 30){2}{7}
\Photon(110, 80)(150,100){2}{7}
\DashLine(70, 65)(110, 80){5}
\Vertex(30,  65){2.0}
\Vertex(110, 30){2.0}
\Vertex(150, 65){2.0}
\Vertex(150,100){2.0}
\Vertex(70, 65){2.0}
\Vertex(110, 80){2.0}
\put(43,50){$\PZ$}
\put(98,49){$\PZ$}
\put(83,80){$\PH$}
\put( -5, 85){$\Pep$}
\put( -5, 40){$\Pem$}
\put( -5,115){\bf (c)}
\end{picture}
} 
\vspace*{-1em}
\caption{Diagram structures for three-boson production:
$\Pep\Pem\to V_1 V_2 V_3 \to 6f$}
\label{fig:VVVgraphs}
\efi

For a sufficiently large Higgs-boson mass, i.e.\ when $\PH\to\PW\PW$
becomes possible, not only $s$-channel Higgs production leads to
relatively large contributions to 
$6f$ production but also Higgs-strahlung off Z bosons,
$\Pep\Pem\to\PZ\PH\to \PZ\PW\PW/\PZ\PZ\PZ \to 6f$, which is the
second of the two main Higgs production mechanisms at future
$\Pep\Pem$ colliders. The corresponding Feynman diagram is shown
in \reffi{fig:VVVgraphs}(c).

\section{Numerical results}
\label{se:numres}

\subsection{Input parameters}

For the numerical evaluation we use the following set of 
Standard-Model parameters \cite{Groom:in},
\beq
\begin{array}[b]{lcllcllcl}
\GF & = & 1.16639 \times 10^{-5} \GeV^{-2}, \hspace*{1em}&
\alpha(0) &=& 1/137.0359895, \\
\alpha_{\mathrm{s}}(\MZ) &=& 0.1181, \\
\MW & = & 80.419\GeV, &
\GW & = & 2.12\GeV, \\
\MZ & = & 91.1882\GeV, &
\GZ & = & 2.4952\GeV, \\
\Mt & = & 174.3\;\GeV, &
\Gt & = & 1.6\GeV, \\
\Me & = & 0.51099907\;\MeV, 
\end{array}
\label{eq:SMpar}
\eeq
where the top-quark width $\Gt$ is, of course, only a reasonable estimate.
In order to absorb parts of the renormalization effects 
(i.e.\ some universal radiative corrections)
into the electroweak couplings, such as the running of $\alpha(Q^2)$ 
from $Q^2=0$ to a high-energy scale and some universal effects related
to the $\rho$ parameter, we evaluate amplitudes in the so-called $\GF$ scheme,
i.e.\ we derive the electromagnetic coupling 
$\alpha=e^2/(4\pi)$ from the Fermi constant $\GF$ according to
\beq
\alpha_{\GF} = \frac{\sqrt{2}\GF\MW^2\sw^2}{\pi}.
\eeq
In the structure functions we use $\alpha(0)$ as
coupling parameter, which is the correct effective coupling for
real photon emission. 
For the Higgs-boson mass we take the set of sample values listed
in \refta{tab:mhgh}, including also the corresponding Higgs-boson widths,
which are calculated with the program {\sc HDECAY} \cite{Djouadi:1997yw}.
If not stated otherwise, we take the value $\MH=170\GeV$ by default.
\begin{table}[h]
\centerline{
\begin{tabular}{|c||c|c|c|}
\hline
$\MH[\GeV]$ & 
170 & 190 & 230
\\ \hline
$\Gamma_\PH[\GeV]$ & 
0.3835 & 1.038 & 2.813
\\ \hline 
\end{tabular} }
\caption{Higgs-boson masses and decay widths as provided by {\sc HDECAY}}
\label{tab:mhgh}
\end{table}

With the only exception of \refse{se:widthnum}, where the interplay
between finite-width schemes and gauge invariance is discussed, 
we make use 
of the fixed-width approach, 
as described in \refse{se:width}.

Finally, we specify the following separation cuts:
\beq\label{LCcuts}
\begin{array}[b]{llll}
\theta (l,\mathrm{beam})>  5^\circ, \qquad & 
\theta (q,\mathrm{beam})>  5^\circ, \qquad & 
\theta( l, l^\prime)>  5^\circ, \qquad &
\theta( l, q)>  5^\circ, \\
E_l>  10\GeV, & E_q>  10\GeV, & m(q,q')>  10\GeV, 
\end{array}
\label{eq:canonicalcuts}
\eeq
where $\theta(i,j)$ is the angle between the particles $i$ and
$j$ in the LAB system,
and $l$, $q$, and ``beam'' denote charged final-state 
leptons, quarks, and the beam electrons or positrons, respectively.
The invariant mass of a quark pair $qq'$ is denoted by
$m(q,q')$. 

\subsection{Survey of cross sections}

Only very few results presented in the literature are found to be
appropriate for a tuned comparison 
to our results, because in most cases only plots but no 
numbers are given, fermion mass singularities are not always excluded
by phase-space cuts, or beamstrahlung effects are included.
Therefore, we decided to adopt 
a public version of a multi-purpose generator available from the www
and to perform a tuned comparison based on the
setup described in the previous section as far as possible.
Specifically, we have used the generator {\sc Whizard} \cite{Kilian:2001qz}, 
version 1.21, together with the implemented {\sc Madgraph} 
\cite{Stelzer:1994ta} amplitudes. We did not include beamstrahlung effects
in the following results, since they depend on the details of the
inspected collider and would spoil the usefulness of our results as
reference for future studies. Beamstrahlung can, however, be 
included in {\sc Lusifer} in a straightforward way.
If not stated otherwise the cross section numbers in this section are
based on $10^7$ weighted Monte Carlo events.
In the following tables, the numbers within parentheses correspond to
the statistical errors of the Monte Carlo integrations.

In \refta{tab:6lfs} we show the results for all processes
$\Pep\Pem\to 6\,$leptons with up to two neutrinos in the final state
(three neutrinos are not possible) for a CM energy of $500\GeV$.
\begin{table}
\centerline{
\begin{tabular}{|c||r@{}l|r@{}l||r@{}l|r@{}l|}
\hline
& \multicolumn{4}{c||}{\sc Lusifer} & 
\multicolumn{4}{c|}{\sc Whizard \& Madgraph}
\\
\cline{2-9}
$\Pep\Pem\to$ & 
\multicolumn{2}{c|}{$\sigma_{\born}[\fba]$} &
\multicolumn{2}{c||}{$\sigma_{\born+\ISR}[\fba]$} &
\multicolumn{2}{c|}{$\sigma_{\born}[\fba]$} &
\multicolumn{2}{c|}{$\sigma_{\born+\ISR}[\fba]$}
\\ \hline\hline
$\mu^- \mu^+ \mu^- \mu^+ \nu_\mu \bar\nu_\mu$  & 0.&041382(87) & 0.&040883(81)
 & 0.&04130(19) & 0.&04077(27)
\\ \hline
$\mu^- \mu^+ \mu^- \bar\nu_\mu \nu_\tau \tau^+$  & 0.&040320(80) & 0.&04002(12)
 & 0.&04044(19) & 0.&03961(15)
\\ \hline
$\mu^- \mu^+ \nu_\mu \bar\nu_\mu \tau^- \tau^+$  & 0.&042297(84) & 0.&041966(89)
 & 0.&04212(12) & 0.&04165(33)
\\ \hline
$\Pe^- \bar\nu_\Pe \mu^- \mu^+ \nu_\mu \mu^+$  & 0.&04729(10) & 0.&04640(12)
 & 0.&04713(25) & 0.&04549(20)
\\ \hline
$\Pe^- \bar\nu_\Pe \mu^- \mu^+ \nu_\tau \tau^+$  & 0.&04708(14) & 0.&04629(14)
 & 0.&04705(16) & 0.&04702(62)
\\ \hline
$\Pe^- \Pe^+ \nu_\Pe \bar\nu_\Pe \mu^- \mu^+$  & 0.&06309(20) & 0.&06026(18)
 & 0.&06139(26) & 0.&05980(65)
\\ \hline
$\Pe^- \Pe^+ \mu^- \mu^+ \nu_\mu \bar\nu_\mu$  & 0.&18300(36) & 0.&16838(32)
 & 0.&18302(32) & 0.&16976(36)
\\ \hline
$\Pe^- \Pe^+ \mu^- \bar\nu_\mu \nu_\tau \tau^+$  & 0.&18023(21) & 0.&16555(30)
 & 0.&18101(31) & 0.&16661(43)
\\ \hline
$\Pe^- \Pe^+ \Pe^- \bar\nu_\Pe \nu_\mu \mu^+$  & 0.&18603(25) & 0.&17091(41)
 & 0.&18624(39) & 0.&17118(48)
\\ \hline
$\Pe^- \Pe^+ \Pe^- \Pe^+ \nu_\Pe \bar\nu_\Pe $  & 0.&19702(36) & 0.&18065(35)
 & \multicolumn{2}{c|}{--} & \multicolumn{2}{c|}{--}
\\ \hline \hline
$\mu^- \mu^+ \mu^- \mu^+ \mu^- \mu^+$  & 0.&00026849(54) & 0.&00027067(57)
 & \multicolumn{2}{c|}{--} & \multicolumn{2}{c|}{--}
\\ \hline
$\mu^- \mu^+ \mu^- \mu^+ \tau^- \tau^+$  & 0.&0008088(15) & 0.&0008182(16)
 & 0.&0007925(69) & 0.&0007689(79)
\\ \hline
$\Pe^- \Pe^+ \mu^- \mu^+ \mu^- \mu^+$  & 0.&002514(16) & 0.&002437(16)
 & 0.&00211(18) & 0.&001467(78)
\\ \hline
$\Pe^- \Pe^+ \mu^- \mu^+ \tau^- \tau^+$  & 0.&005102(31) & 0.&005005(37)
 & 0.&00445(16) & 0.&00464(23)
\\ \hline
$\Pe^- \Pe^+ \Pe^- \Pe^+ \mu^- \mu^+$  & 0.&004158(31) & 0.&004017(39)
 & \multicolumn{2}{c|}{--} & \multicolumn{2}{c|}{--}
\\ \hline
$\Pe^- \Pe^+ \Pe^- \Pe^+ \Pe^- \Pe^+ $  & 0.&001803(16) & 0.&001758(15)
 & \multicolumn{2}{c|}{--} & \multicolumn{2}{c|}{--}
\\ \hline
$\mu^- \mu^+ \mu^- \mu^+ \nu_\tau \bar\nu_\tau$  & 0.&0010312(16) & 0.&0010262(16)
 & 0.&0010295(36) & 0.&0010257(43) 
\\ \hline
$\Pe^- \Pe^+ \mu^- \mu^+ \nu_\tau \bar\nu_\tau$  & 0.&0033141(69) & 0.&0031918(72)
 & 0.&003296(14) & 0.&003133(23)
\\ \hline
$\Pe^- \Pe^+ \Pe^- \Pe^+ \nu_\mu \bar\nu_\mu$  & 0.&0022162(60) & 0.&0020908(62)
 & \multicolumn{2}{c|}{--} & \multicolumn{2}{c|}{--}
\\ \hline
$\nu_\Pe \bar\nu_\Pe \mu^- \mu^+ \mu^- \mu^+$  & 0.&003730(14) & 0.&0034166(76)
 & 0.&003613(25) & 0.&003374(34)
\\ \hline
$\nu_\Pe \bar\nu_\Pe \mu^- \mu^+ \tau^- \tau^+$  & 0.&007366(17) & 0.&006785(16)
 & 0.&007301(33) & 0.&006503(67)
\\ \hline
\end{tabular} }
\caption{Cross sections for $\Pep\Pem\to 6f$ channels with purely leptonic
  final states for $\sqrt{s}=500\GeV$}
\label{tab:6lfs}
\end{table}
The first set of final states comprises the reactions that receive 
contributions from resonant $\PW\PW\PZ$ production and various other
mechanisms. The remaining reactions 
entirely proceed via neutral-current interactions, except for the last two
processes that receive contributions from $\PW\PW\to\PZ\PZ$ scattering.
Comparing the two sets of cross sections, the ones for $\PW\PW\PZ$
production are larger. ISR affects the cross sections at the level of a 
few per cent, ranging up to $\sim 8\%$, and this correction is 
negative in almost all cases.
The comparison with {\sc Whizard} and {\sc Madgraph} 
reveals agreement within 1--3$\sigma$ in general with a few exceptions, 
where larger differences occur.
These differences are accompanied with larger errors in the {\sc Whizard}
results and are assumed \cite{kilian} to be due to the fact that the photon
propagators in the splittings $\gamma\to f\bar f$ are not analytically
smoothed by mappings in {\sc Whizard}. 
In most cases the statistical error given by {\sc Lusifer} is smaller
than the one of {\sc Whizard}. 
The missing entries correspond to those final states that
are not supported by {\sc Madgraph}.%
\footnote{In contrast to the {\sc Madgraph} version used for checking the
amplitudes, as described in \refse{se:checkamp}, the {\sc Madgraph} version 
within {\sc Whizard} is neither able to add electroweak and gluon-exchange 
diagrams coherently, nor to deal with more than 999 Feynman graphs for a 
given process.}

Table~\ref{tab:4l2qfs} shows a tuned comparison for $6f$ final states
containing two quarks and at most three neutrinos for a CM energy of $500\GeV$.
\begin{table}
\centerline{
\begin{tabular}{|c||r@{}l|r@{}l||r@{}l|r@{}l|}
\hline
& \multicolumn{4}{c||}{\sc Lusifer} & 
\multicolumn{4}{c|}{\sc Whizard \& Madgraph}
\\
\cline{2-9}
$\Pep\Pem\to$ & 
\multicolumn{2}{c|}{$\sigma_{\born}[\fba]$} &
\multicolumn{2}{c||}{$\sigma_{\born+\ISR}[\fba]$} &
\multicolumn{2}{c|}{$\sigma_{\born}[\fba]$} &
\multicolumn{2}{c|}{$\sigma_{\born+\ISR}[\fba]$}
\\ \hline  \hline 
$\mu^- \mu^+ \mu^- \bar\nu_\mu \Pu \bar\Pd$   & 0.&11835(22) & 0.&11714(22)
 & 0.&11797(26) & 0.&11732(71)
\\ \hline 
$\mu^- \mu^+ \tau^- \bar\nu_\tau \Pu \bar\Pd$ & 0.&11861(42) & 0.&11695(22)
 & 0.&11791(21) & 0.&11664(29)
\\ \hline 
$\Pe^- \bar\nu_\Pe \mu^- \mu^+ \Pu \bar\Pd$  & 0.&13832(30) & 0.&13516(34)
 & 0.&13785(36) & 0.&13510(62)
\\ \hline 
$\Pe^- \Pe^+ \mu^- \bar\nu_\mu \Pu \bar\Pd$  & 0.&53352(70) & 0.&48897(69)
 & 0.&53493(61) & 0.&49378(73)
\\ \hline 
$\Pe^- \Pe^+ \Pe^- \bar\nu_\Pe \Pu \bar\Pd$  & 0.&55089(74) & 0.&50464(72)
 & 0.&5514(13) & 0.&5061(10)
\\ \hline 
$\mu^- \bar\nu_\mu \nu_\mu \bar\nu_\mu \Pu \bar\Pd$   & 0.&18399(11) & 0.&18182(11)
 & 0.&18396(13) & 0.&18204(14)
\\ \hline 
$\mu^- \bar\nu_\mu \nu_\tau \bar\nu_\tau \Pu \bar\Pd$ & 0.&18406(10) & 0.&18201(11)
 & 0.&18430(13) & 0.&18197(14)
\\ \hline 
$\Pe^- \bar\nu_\Pe \nu_\mu \bar\nu_\mu \Pu \bar\Pd$ & 0.&20272(14) & 0.&19847(14)
 & 0.&20288(14) & 0.&19843(16)
\\ \hline 
$\nu_\Pe \bar\nu_\Pe \mu^- \bar\nu_\mu \Pu \bar\Pd$ & 1.&6326(12) & 1.&4743(12)
 & 1.&6313(13) & 1.&4746(13)
\\ \hline 
$\Pe^- \bar\nu_\Pe \nu_\Pe \bar\nu_\Pe \Pu \bar\Pd$  & 1.&6500(17) & 1.&4906(15)
 & 1.&6482(15) & 1.&4914(14)
\\ \hline\hline
$\mu^- \mu^+ \nu_\mu \bar\nu_\mu (\Pu\bar\Pu{+}\Pd\bar\Pd)$  & 0.&26632(36) & 0.&26266(33)
 & 0.&26647(19) & 0.&26242(21)
\\ \hline  
$\mu^- \bar\nu_\mu \nu_\tau \tau^+ (\Pu\bar\Pu{+}\Pd\bar\Pd)$ & 0.&25408(18) & 0.&25068(19)
 & 0.&25427(17) & 0.&25067(17)
\\ \hline 
$\Pe^- \bar\nu_\Pe \nu_\mu \mu^+ (\Pu\bar\Pu{+}\Pd\bar\Pd)$  & 0.&28161(24) & 0.&27514(24)
 & 0.&28189(21) & 0.&27471(21)
\\ \hline    
$\Pe^- \Pe^+ \nu_\Pe \bar\nu_\Pe (\Pu\bar\Pu{+}\Pd\bar\Pd)$  & 0.&35788(62) & 0.&34361(73)
 & 0.&35917(62) & 0.&34366(36)
\\ \hline      
$\mu^- \mu^+ \mu^- \mu^+ (\Pu\bar\Pu{+}\Pd\bar\Pd)$      & 0.&0043727(73) & 0.&0043774(77)
 & 0.&004368(18) & 0.&004303(18)
\\ \hline    
$\mu^- \mu^+ \tau^- \tau^+ (\Pu\bar\Pu{+}\Pd\bar\Pd)$    & 0.&008731(14) & 0.&008736(15)
 & 0.&008652(26) & 0.&008585(41)
\\ \hline    
$\Pe^- \Pe^+ \mu^- \mu^+ (\Pu\bar\Pu{+}\Pd\bar\Pd)$ & 0.&015466(59) & 0.&014886(49)
 & 0.&01523(11) & 0.&01396(27)
\\ \hline      
$\Pe^- \Pe^+ \Pe^- \Pe^+ (\Pu\bar\Pu{+}\Pd\bar\Pd)$ & 0.&010730(58) & 0.&010207(45)
 & \multicolumn{2}{c|}{--} & \multicolumn{2}{c|}{--}
\\ \hline  
$\mu^- \mu^+ \nu_\tau \bar\nu_\tau (\Pu\bar\Pu{+}\Pd\bar\Pd)$ & 0.&012057(17) & 0.&012021(17)
 & 0.&012042(12) & 0.&011964(19)
\\ \hline        
$\Pe^- \Pe^+ \nu_\mu \bar\nu_\mu (\Pu\bar\Pu{+}\Pd\bar\Pd)$  & 0.&016967(30) & 0.&016484(32)
 & 0.&017006(23) & 0.&016417(25)
\\ \hline    
$\nu_\Pe \bar\nu_\Pe \mu^- \mu^+ (\Pu\bar\Pu{+}\Pd\bar\Pd)$  & 0.&044940(85) & 0.&041135(86)
 & 0.&044976(64) & 0.&04083(11)
\\ \hline 
\end{tabular} }
\caption{Cross sections for $\Pep\Pem\to 6f$ channels with four leptons
and two quarks in the final state for $\sqrt{s}=500\GeV$}
\label{tab:4l2qfs}
\end{table}
The table is divided into two parts: the first part 
comprises all final states that receive contributions from a hadronically
decaying W~boson; final states corresponding to hadronically decaying
Z~bosons form the second part, where contributions from $\Pu\bar\Pu$ and
$\Pd\bar\Pd$ pairs are added, in order to keep the table more compact.
In comparison with the purely leptonic final states the cross
sections are larger, which is mainly due to the colour factor 3 resulting
from the two quarks in the final states. Otherwise the various channels
show similar features as their leptonic counterparts. $\PW\PW\PZ$
production channels have larger cross sections than processes that
proceed via neutral-current interactions only, and ISR affects cross
sections at the level of a few per cent, mainly in negative direction.
It should be mentioned that in the second set
of final states the {\sc Lusifer} results have again been obtained from
$10^7$ events, while the {\sc Whizard} runs have been performed with
$10^7$ events for $\Pu\bar\Pu$ and $\Pd\bar\Pd$ each, so that the final
numbers effectively result from $2\times 10^7$ events. In view of this
difference
the statistical error of {\sc Lusifer} is somewhat smaller.
The agreement between the two programs is again
within 1--3$\sigma$ with a few exceptions for the same reason as explained 
above.

A tuned comparison of results for final states containing two leptons 
and four quarks is presented in \refta{tab:2l4qfs}, again for a CM
energy of $500\GeV$. 
\begin{table}
\centerline{
\begin{tabular}{|c|c||r@{}l|r@{}l||r@{}l|r@{}l|}
\hline
&& \multicolumn{4}{c||}{\sc Lusifer} & 
\multicolumn{4}{c|}{\sc Whizard \& Madgraph}
\\
\cline{3-10}
$\Pep\Pem\to$ & QCD? &
\multicolumn{2}{c|}{$\sigma_{\born}[\fba]$} &
\multicolumn{2}{c||}{$\sigma_{\born+\ISR}[\fba]$} &
\multicolumn{2}{c|}{$\sigma_{\born}[\fba]$} &
\multicolumn{2}{c|}{$\sigma_{\born+\ISR}[\fba]$}
\\ \hline \hline   
$\mu^- \bar\nu_\mu + 4q$ & yes & 3.&8170(84) & 3.&7917(86)
 & \multicolumn{2}{c|}{--} & \multicolumn{2}{c|}{--}
\\ 
 & no & 2.&9603(28) & 2.&9172(27) & 2.&9616(13) & 2.&9195(13)
\\ \hline
$\Pe^- \bar\nu_\Pe + 4q$ & yes & 4.&605(13) & 4.&529(14) 
 & \multicolumn{2}{c|}{--} & \multicolumn{2}{c|}{--}
\\ 
 & no & 3.&2798(38) & 3.&1930(40) & 3.&2708(16) & 3.&1916(17)
\\ \hline\hline
$\mu^- \mu^+ + 4q$ & yes    & 1.&5150(30) & 1.&5017(31)
 & \multicolumn{2}{c|}{--} & \multicolumn{2}{c|}{--}
\\ 
 & no & 1.&4406(40) & 1.&4227(29) & 1.&4343(18) & 1.&4192(19)
\\ \hline 
$\nu_\mu \bar\nu_\mu + 4q$ & yes & 2.&3405(29) & 2.&3147(28)
 & \multicolumn{2}{c|}{--} & \multicolumn{2}{c|}{--}
\\ 
 & no & 2.&2466(33) & 2.&2138(26) & 2.&2440(12) & 2.&2153(13)
\\ \hline 
$\Pe^- \Pe^+ + 4q$ & yes      & 6.&570(17) & 6.&053(16) 
 & \multicolumn{2}{c|}{--} & \multicolumn{2}{c|}{--}
\\ 
 & no & 6.&1439(95) & 5.&679(10) & 6.&2001(50) & 5.&7115(60) 
\\ \hline 
$\nu_\Pe \bar\nu_\Pe + 4q$ & yes & 19.&260(40) & 17.&441(37)
 & \multicolumn{2}{c|}{--} & \multicolumn{2}{c|}{--}
\\
 & no & 18.&850(32) & 17.&038(29) & 18.&879(11) & 17.&076(11)
\\ \hline 
\end{tabular} }
\caption{Cross sections for $\Pep\Pem\to 6f$ channels with two leptons
and four quarks in the final state for $\sqrt{s}=500\GeV$}
\label{tab:2l4qfs}
\end{table}
The notation ``$4q$'' stands for the sum over all possible four-quark
configurations of $\Pu$, $\Pd$, $\Pc$, $\Ps$ quarks. In the table we
separately give results that include or exclude the contributions of
gluon-exchange diagrams in the amplitude calculation, as indicated in
the ``QCD'' column. Note that the large impact of these gluon-exchange 
diagrams sensitively depends on the separation cuts, in particular
on the minimal invariant mass of quark pairs. The impact of ISR is
similar to the cases discussed previously. The largest
cross section is observed for $\nu_\Pe \bar\nu_\Pe + 4q$ production,
which is the only channel that receives contributions from 
$\PW\PW\to\PW\PW/\PZ\PZ$ scattering and from W~fusion to a Higgs
boson. The comparison to {\sc Whizard} shows the same features as
in the other cases. However, it should be mentioned that all
$4q$ final states have been integrated individually with {\sc Whizard},
but in a single run with {\sc Lusifer}. This explains the smaller
integration errors of the {\sc Whizard} results. Since the
{\sc Madgraph} version included in {\sc Whizard} is not able to
coherently add gluon-exchange and purely electroweak diagrams, 
a comparison of cross sections
based on the full amplitudes has not been carried out.

Finally, in \refta{tab:bfs} we collect our results on cross sections 
for final states that involve amplitudes with intermediate top quarks.
\begin{table}
\centerline{
\begin{tabular}{|c|c||r@{}l|r@{}l||r@{}l|r@{}l|}
\hline
&& \multicolumn{4}{c||}{\sc Lusifer} & 
\multicolumn{4}{c|}{\sc Whizard \& Madgraph}
\\
\cline{3-10}
$\Pep\Pem\to$ & QCD? &
\multicolumn{2}{c|}{$\sigma_{\born}[\fba]$} &
\multicolumn{2}{c||}{$\sigma_{\born+\ISR}[\fba]$} &
\multicolumn{2}{c|}{$\sigma_{\born}[\fba]$} &
\multicolumn{2}{c|}{$\sigma_{\born+\ISR}[\fba]$}
\\ \hline\hline
$\mu^- \bar\nu_\mu \nu_\mu \mu^+ \Pb \bar\Pb$  &--& 5.&8091(49) & 5.&5887(36) 
 & 5.&8102(26) & 5.&5978(30)
\\ \hline
$\mu^- \bar\nu_\mu \nu_\tau \tau^+ \Pb \bar\Pb$  &--& 5.&7998(36) & 5.&5840(40)
 & 5.&7962(26) & 5.&5893(29)
\\ \hline
$\Pem \bar\nu_\Pe \nu_\mu \mu^+ \Pb \bar\Pb$ &--& 5.&8188(45) & 5.&6042(38)
 & 5.&8266(27) & 5.&6071(30)
\\ \hline
$\Pem \bar\nu_\Pe \nu_\Pe \Pep \Pb \bar\Pb$ &--& 5.&8530(68) & 5.&6465(70)
 & 5.&8751(30) & 5.&6508(36)
\\ \hline
$\mu^- \bar\nu_\mu \Pu \bar\Pd \Pb \bar\Pb$ & yes & 17.&171(24) & 16.&561(24)
 & \multicolumn{2}{c|}{--} & \multicolumn{2}{c|}{--}
\\ 
                                            & no  & 17.&095(11) & 16.&4538(98) 
 & 17.&1025(80) & 16.&4627(87)
\\ \hline
$\Pem \bar\nu_\Pe \Pu \bar\Pd \Pb \bar\Pb$  & yes & 17.&276(45) & 16.&577(21)
 & \multicolumn{2}{c|}{--} & \multicolumn{2}{c|}{--}
\\ 
                                            & no & 17.&187(21)  & 16.&511(12)
 & 17.&1480(82) & 16.&5288(92)
\\ \hline \hline
$\nu_\Pe \bar\nu_\Pe \mu^- \mu^+ \Pb \bar\Pb$  && 0.&024550(45) & 0.&022472(45)
 & 0.&024619(43) & 0.&022398(41)
\\ \hline
$\nu_\Pe\bar\nu_\Pe\Pu\bar\Pu\Pb\bar\Pb$  & yes & 0.&12625(32) & 0.&11703(32)
 & \multicolumn{2}{c|}{--} & \multicolumn{2}{c|}{--}
\\ 
                                          & no  & 0.&06984(15) & 0.&06369(14)
 & 0.&069781(70) & 0.&063635(86)
\\ \hline
$\nu_\Pe\bar\nu_\Pe\Pd\bar\Pd\Pb\bar\Pb$  & yes & 0.&13709(41) & 0.&12636(37)
 & \multicolumn{2}{c|}{--} & \multicolumn{2}{c|}{--}
\\ 
                                          & no  & 0.&08648(20) & 0.&07871(18)
 & 0.&086351(83) & 0.&078533(96)
\\ \hline
$\nu_\Pe\bar\nu_\Pe\Pb\bar\Pb\Pb\bar\Pb$  & yes & 0.&06741(18) & 0.&06226(18)
 & \multicolumn{2}{c|}{--} & \multicolumn{2}{c|}{--}
\\ 
                                          & no  & 0.&04352(10) & 0.&03974(12)
 & 0.&043473(49) & 0.&039721(68)
\\ \hline
\end{tabular} }
\caption{Cross sections for $\Pep\Pem\to 6f$ channels involving
intermediate top quarks for $\sqrt{s}=500\GeV$}
\label{tab:bfs}
\end{table}
The first set of channels comprises all final states that are relevant
for top-quark pair production, $\Pep\Pem\to\Pt\bar\Pt\to 6f$. In the 
second set of processes top quarks appear only on non-resonant lines
in diagrams. The difference between the $\Pt\bar\Pt$ production cross sections 
for two and four quarks in the final states roughly reflects the factor 3
between leptonically and hadronically decaying W~bosons that have been
produced in $\Pt\to\Pb\PW^+$. Since these channels are strongly dominated by
the diagram shown in \reffi{fig:ttgraphs}, the impact of gluon-exchange
diagrams is small. Concerning the ISR effects and the
comparison of the two programs, the same features are observed as in the
other cases discussed above.

Finally, we mention that {\sc Lusifer} runs faster than the
combination of the {\sc Whizard} and {\sc Madgraph} packages.
The factor in speed varies with the $6f$ final state from roughly a factor
of 2 up to an order of magnitude, where the superiority of 
{\sc Lusifer} becomes more apparent if a large number of diagrams
is involved.

\subsection{Results on top-quark pair production}

Figure~\ref{fig:eett_cs} illustrates the (well-known) energy dependence 
of the top-quark pair production cross section for final states where
one of the produced W~bosons decays hadronically and the other
leptonically. 
The results for the total cross section are obtained with $5\times 10^6$ 
events per CM energy. Gluon-exchange diagrams are not taken into account.
\bfi
\setlength{\unitlength}{1cm}
\centerline{
\begin{picture}(16.0,7.0)
\put(-1.2,-17.1){\includegraphics{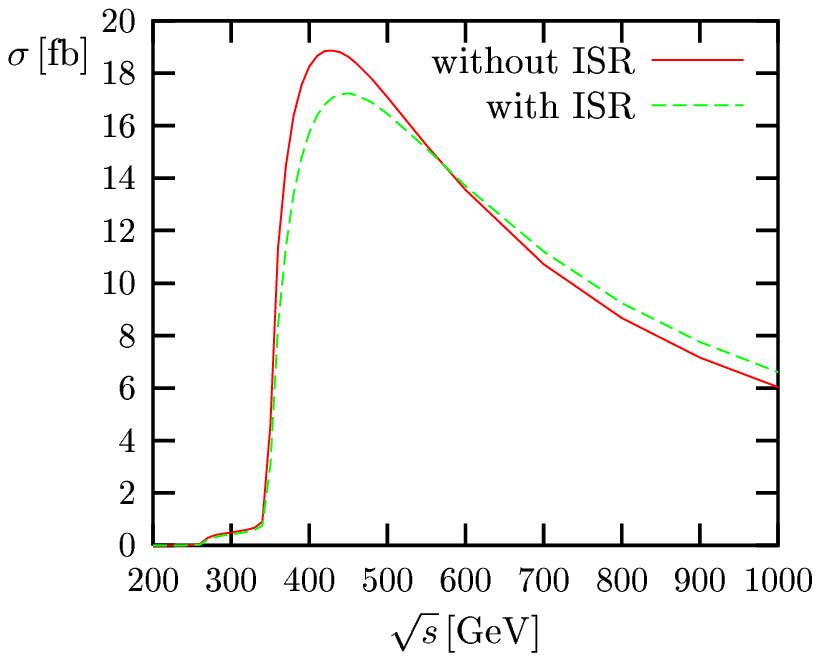}}
\put( 7.0,-17.1){\includegraphics{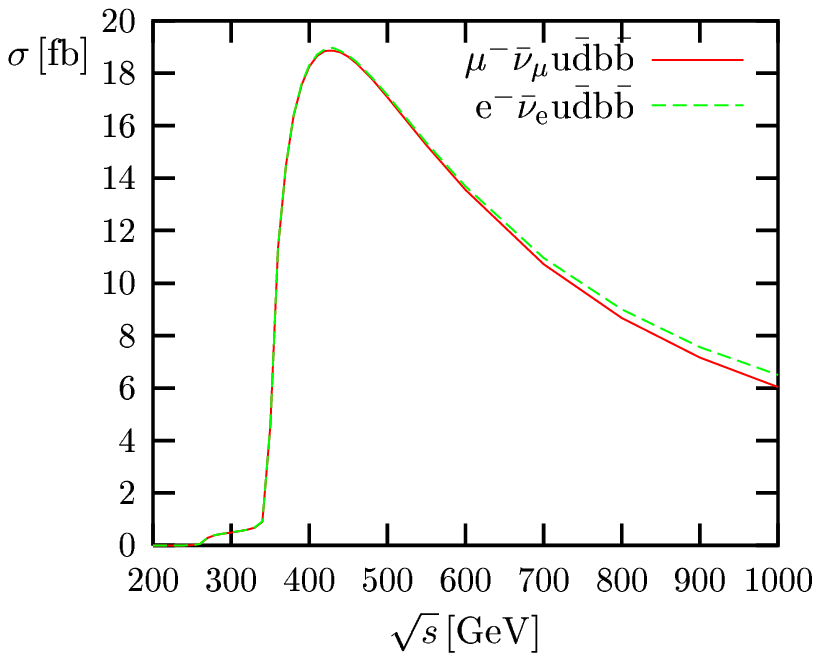}}
\put( 3.7,1.5){$\Pep\Pem\to\mu^-\bar\nu_\mu\Pu\bar\Pd\Pb\bar\Pb$}
\end{picture} } 
\caption{Total cross section of
$\Pep\Pem\to\mu^-\bar\nu_\mu\Pu\bar\Pd\Pb\bar\Pb$ (without
gluon-exchange diagrams)
as function of the CM energy with and without ISR on the l.h.s.\
and in comparison with $\Pep\Pem\to\Pe^-\bar\nu_\Pe\Pu\bar\Pd\Pb\bar\Pb$
(without ISR) on the r.h.s.}
\label{fig:eett_cs}
\efi
The cross section steeply rises at the $\Pt\bar\Pt$ threshold, reaches
its maximum between $400\GeV$ and $500\GeV$, and then decreases with
increasing energy. The l.h.s.\ of the figure shows that ISR reduces the
cross section for energies below its maximum and enhances it above, 
thereby shifting the maximum to a higher energy. This behaviour is
simply due to the radiative energy loss induced by ISR. Near a CM energy
of $250\GeV$ the onset of $\PW\PW\PZ$ production can be observed.
Note that this contribution is entirely furnished by background
diagrams, i.e.\ by diagrams that do not have a resonant top-quark
pair. Another type of background diagrams exists if electrons or
positrons are present in the final state, since an incoming 
$\Pe^\pm$ line can then go through to the final state.
The impact of such diagrams is illustrated on the r.h.s.\ of
\reffi{fig:eett_cs}, where the final states are equal up to the
change of $\mu^-\bar\nu_\mu$ to $\Pe^-\bar\nu_\Pe$. For energies
around $500\GeV$ the difference is of the order of a per cent, but
increasing with energy. It should, however, be noted that this
difference strongly depends on the separation of final-state
$\Pe^\pm$ from the beams. For smaller cut angles, or for particular
regions in distributions, the impact of such background diagrams 
will be much larger.%
\footnote{These features are well-known from $4f$ physics, where 
forward-scattered $\Pe^\pm$, in particular, are used to 
define the so-called single-W production signal 
(see, e.g., \citere{Grunewald:2000ju,Kurihara:1999qz}).}
The problem of working out an optimal strategy to define a clear
$\Pt\bar\Pt$ signal, i.e.\ to systematically suppress background
contributions, obviously goes beyond this study.

In \reffis{fig:topmassdist} and \ref{fig:topcthdist} we consider two
examples of distributions that are interesting for $\Pt\bar\Pt$
production, again focusing on the channel 
$\Pep\Pem\to\mu^-\bar\nu_\mu\Pu\bar\Pd\Pb\bar\Pb$.
\bfi
\setlength{\unitlength}{1cm}
\centerline{
\begin{picture}(16.0,8.0)
\put(-4.2,-15.1){\includegraphics{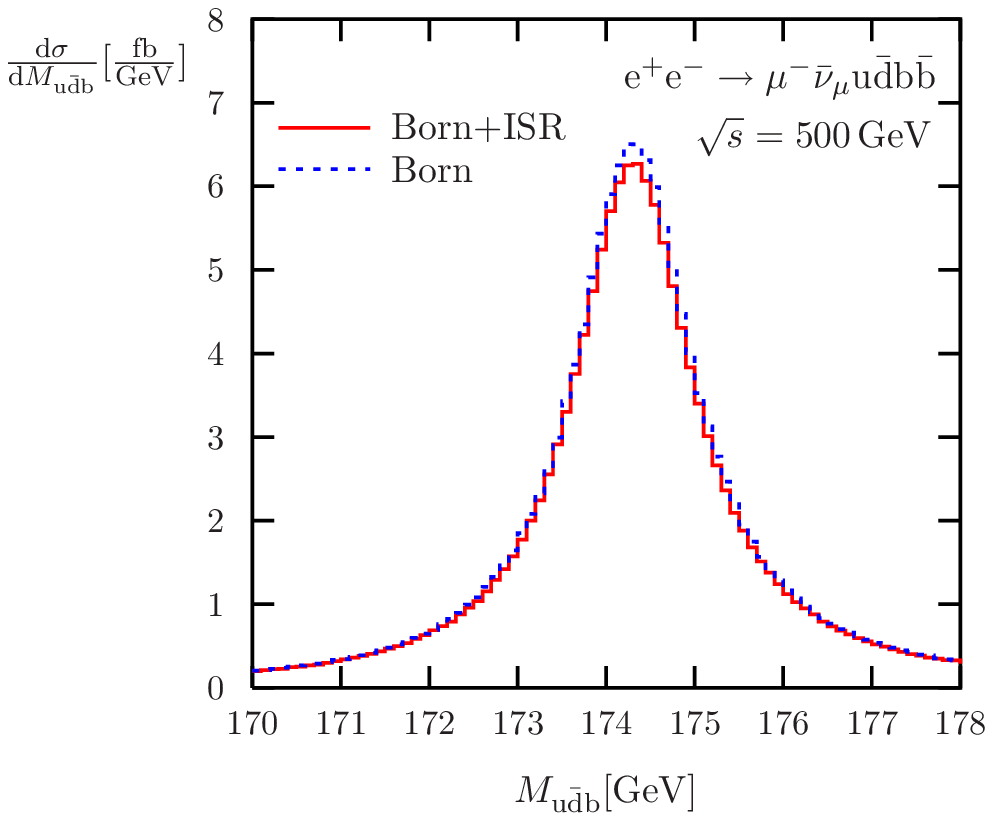}}
\put( 1.8,-15.1){\includegraphics{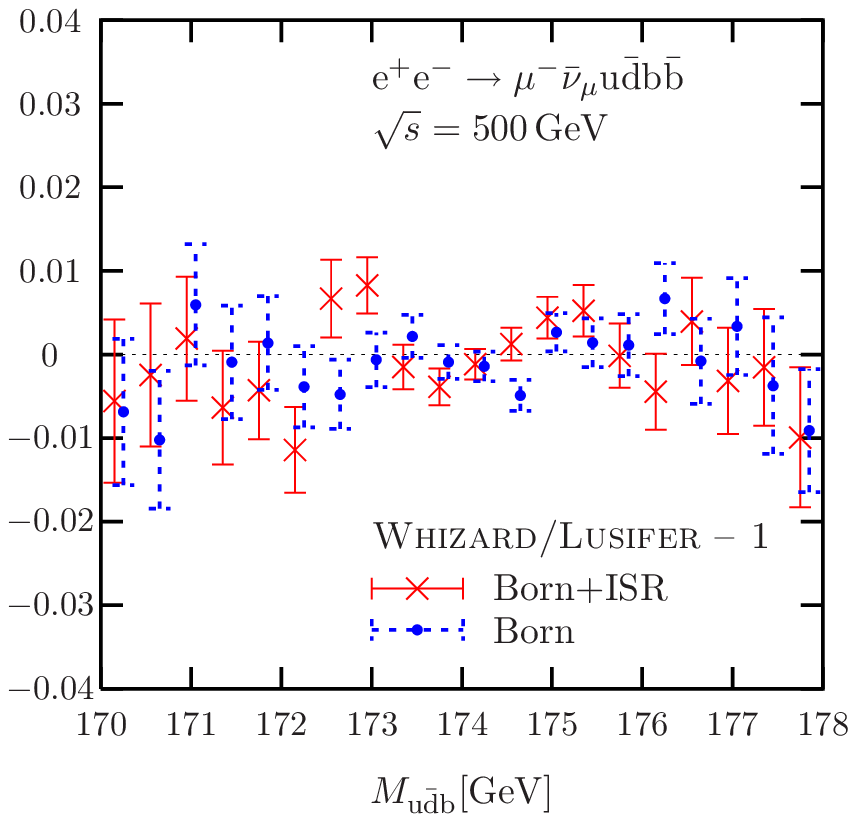}}
\end{picture} } 
\caption{Invariant-mass distribution of the $\Pu\bar\Pd\Pb$ quark triple
in $\Pep\Pem\to\mu^-\bar\nu_\mu\Pu\bar\Pd\Pb\bar\Pb$ 
(without gluon-exchange diagrams):
absolute prediction with and without ISR (l.h.s.) and comparison
between {\sc Lusifer} and {\sc Whizard} (r.h.s.)}
\label{fig:topmassdist}
\vspace*{1em}
\centerline{
\begin{picture}(16.0,8.0)
\put(-4.2,-15.1){\includegraphics{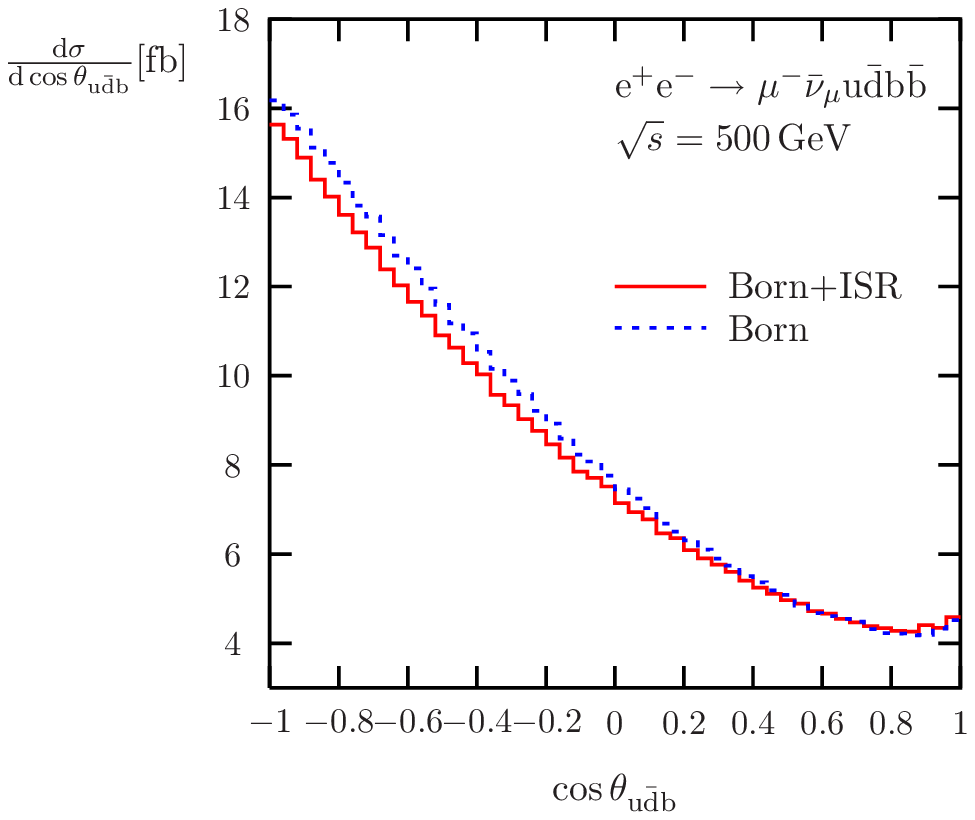}}
\put( 1.8,-15.1){\includegraphics{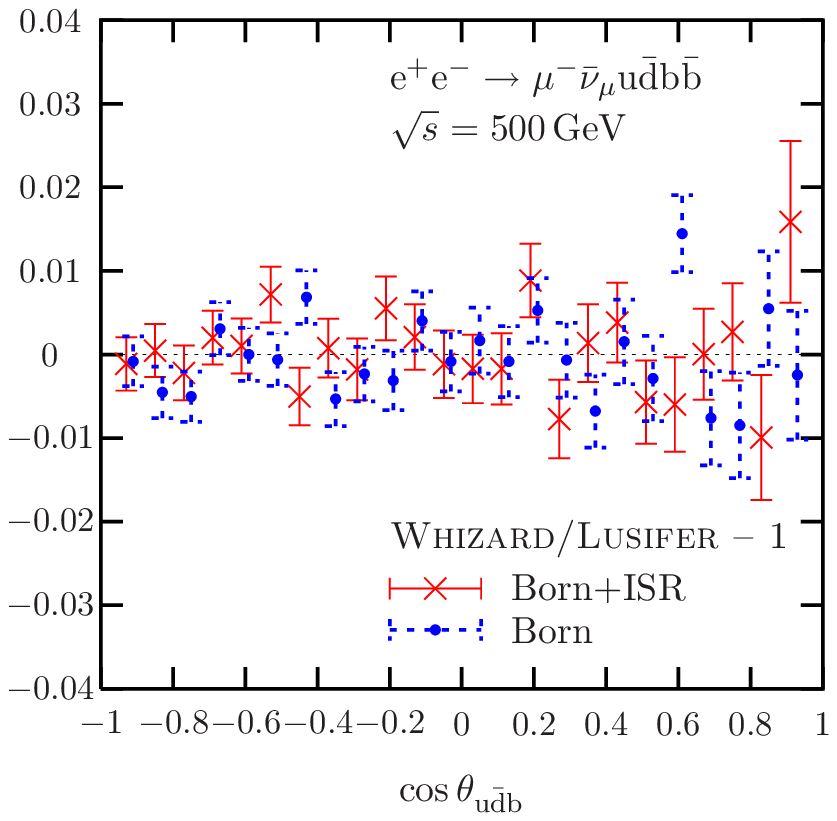}}
\end{picture} } 
\caption{Angular distribution of the $\Pu\bar\Pd\Pb$ quark triple
in $\Pep\Pem\to\mu^-\bar\nu_\mu\Pu\bar\Pd\Pb\bar\Pb$ 
(without gluon-exchange diagrams):
absolute prediction with and without ISR (l.h.s.) and comparison
between {\sc Lusifer} and {\sc Whizard} (r.h.s.)}
\label{fig:topcthdist}
\efi
Figure~\ref{fig:topmassdist} shows the invariant-mass distribution of the 
$\Pu\bar\Pd\Pb$ quark triple that results from the top-quark decay.
As expected, ISR does not distort the resonance shape but merely
rescales the Breit--Wigner-like distribution. More interestingly,
the r.h.s.\ of the figure demonstrates that {\sc Lusifer} and
{\sc Whizard} yield predictions that are fully compatible within
statistical accuracy, both with and without the inclusion of ISR
corrections.
Figure~\ref{fig:topcthdist} shows the production angular distribution
of the $\Pu\bar\Pd\Pb$ quark triple, which is (for resonant top quarks)
equal to the top-quark production angle. ISR tends to flatten
the distribution, which is again due to the impact of effectively reduced
scattering energies where the distribution is less angular dependent.
The r.h.s.\ of the figure reveals agreement between the two
programs within statistical errors.

\subsection{Results on Higgs-boson production}

In this section we discuss some distributions that are relevant for
Higgs-boson production in the intermediate $\MH$ range.
Figures~\ref{fig:higgs-udscmassdist} and \ref{fig:higgs-udsccthdist}
show the four-quark invariant-mass distribution and the related production
angular distribution of the individual channel
$\Pep\Pem\to\nu_\mu\bar\nu_\mu\Pu\bar\Pd\Ps\bar\Pc$ for a CM energy of
$500\GeV$.
\bfi
\setlength{\unitlength}{1cm}
\centerline{
\begin{picture}(16.0,8.0)
\put(-5.5,-15.1){\includegraphics{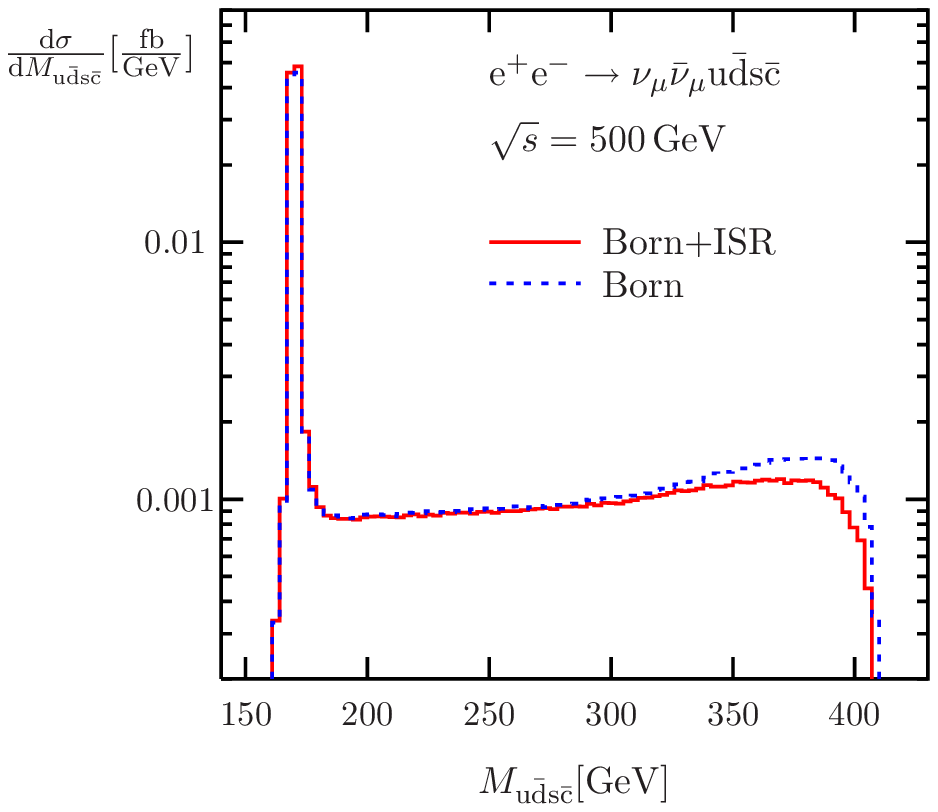}}
\put( 1.9,-15.1){\includegraphics{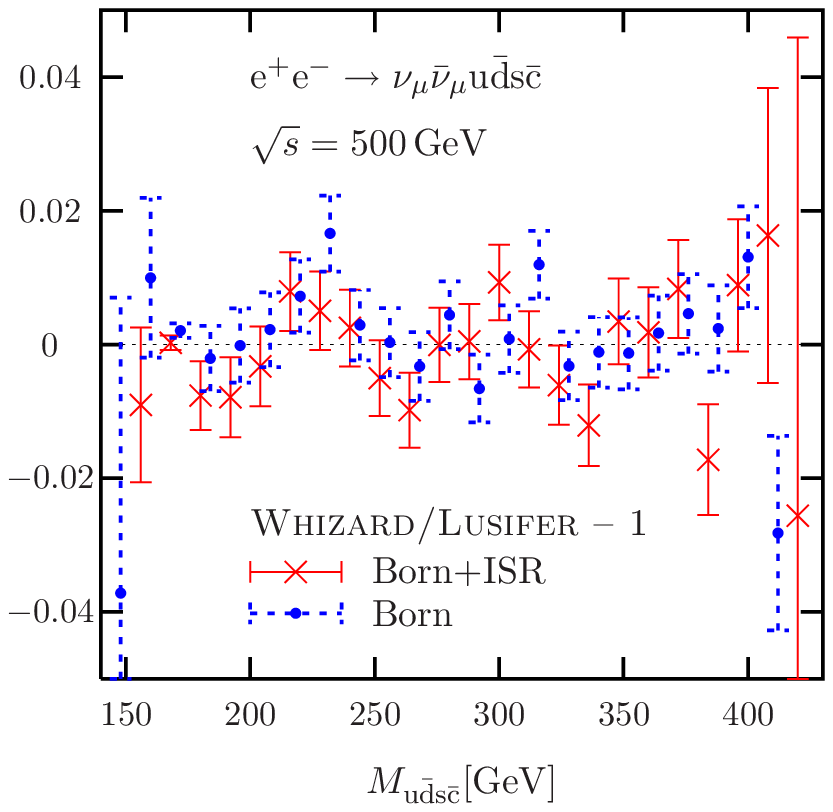}}
\end{picture} } 
\caption{Invariant-mass distribution of the $\Pu\bar\Pd\Ps\bar\Pc$ quark system
in $\Pep\Pem\to\nu_\mu\bar\nu_\mu\Pu\bar\Pd\Ps\bar\Pc$ 
(without gluon-exchange diagrams):
absolute prediction with and without ISR (l.h.s.) and comparison
between {\sc Lusifer} and {\sc Whizard} (r.h.s.)}
\label{fig:higgs-udscmassdist}
\vspace*{1em}
\centerline{
\begin{picture}(16.0,8.0)
\put(-5.1,-15.1){\includegraphics{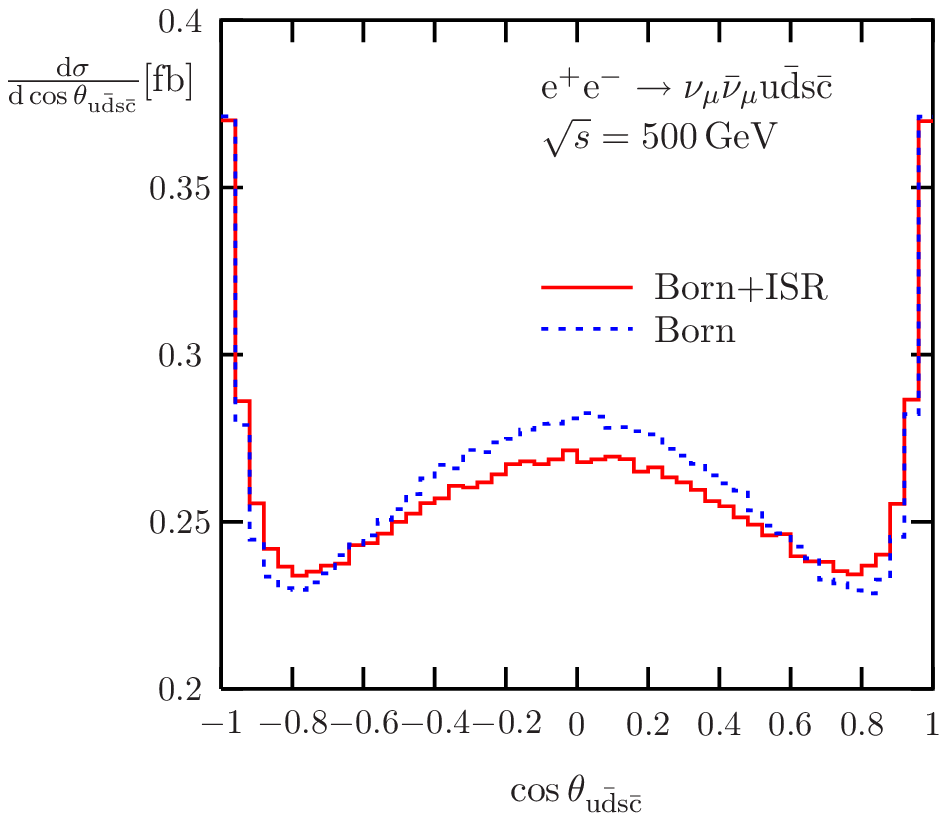}}
\put( 1.9,-15.1){\includegraphics{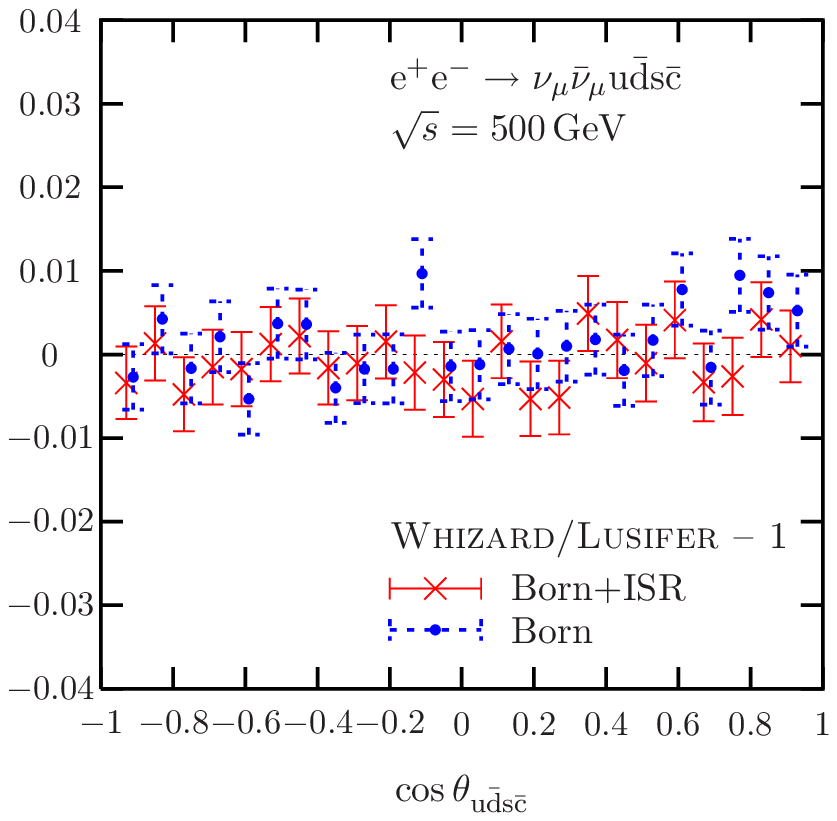}}
\end{picture} } 
\caption{Angular distribution of the $\Pu\bar\Pd\Ps\bar\Pc$ quark system
in $\Pep\Pem\to\nu_\mu\bar\nu_\mu\Pu\bar\Pd\Ps\bar\Pc$ 
(without gluon-exchange diagrams):
absolute prediction with and without ISR (l.h.s.) and comparison
between {\sc Lusifer} and {\sc Whizard} (r.h.s.)}
\label{fig:higgs-udsccthdist}
\efi
Gluon-exchange diagrams 
are not included in these evaluations.
The reaction is dominated by two mechanisms: $\PZ\PH$ production
with the subsequent decays $\PH\to\PW\PW\to 4q$ and $\PZ\to\nu_\mu\bar\nu_\mu$,
and ``continuous'' $\PW\PW\PZ$ production.
In the invariant-mass distribution the narrow Higgs resonance shows up
at $M_{\Pu\bar\Pd\Ps\bar\Pc}=\MH=170\GeV$ over a continuous background
from $\PW\PW\PZ$ production in the range 
$2\MW\lsim M_{\Pu\bar\Pd\Ps\bar\Pc}\lsim\sqrt{s}-\MZ$.
The l.h.s.\ of \reffi{fig:higgs-udscmassdist} shows that ISR does not
influence the resonance structure strongly; the largest ISR effect is
observed at the upper edge of the spectrum, where the effective CM energy loss
by ISR reduces the rate. The r.h.s.\ of the figure illustrates the
agreement between the {\sc Lusifer} and {\sc Whizard} results within
statistical accuracy. Figure~\ref{fig:higgs-udsccthdist} shows that
ISR significantly distorts the four-quark angular distribution at
intermediate angles, where $\PZ\PH$ production dominates; in the very
forward and backward regions the spectrum is mainly due to contributions 
from the subprocess $\Pep\Pem\to(\gamma^*/\PZ^*)\PZ\to\PW\PW\PZ$, 
where the total momentum of  the $\Pu\bar\Pd\Ps\bar\Pc$ quark system 
correspond to the off-shell particles $\gamma^*/\PZ^*$.
The r.h.s.\ of the figure again demonstrates the good agreement between
the two different Monte Carlo programs.

In \reffi{fig:higgs-nmnm4q} we show the analogous distributions for
three different Higgs-boson masses, but now summed over all four-quark 
configurations of the first two generations. 
Figure~\ref{fig:higgs-nene4q} shows the same distributions after
replacing the $\nu_\mu\bar\nu_\mu$ pair in the final state by
$\nu_\Pe\bar\nu_\Pe$.
\bfi
\setlength{\unitlength}{1cm}
\centerline{
\begin{picture}(16.0,8.3)
\put(-6.0,-15.1){\includegraphics{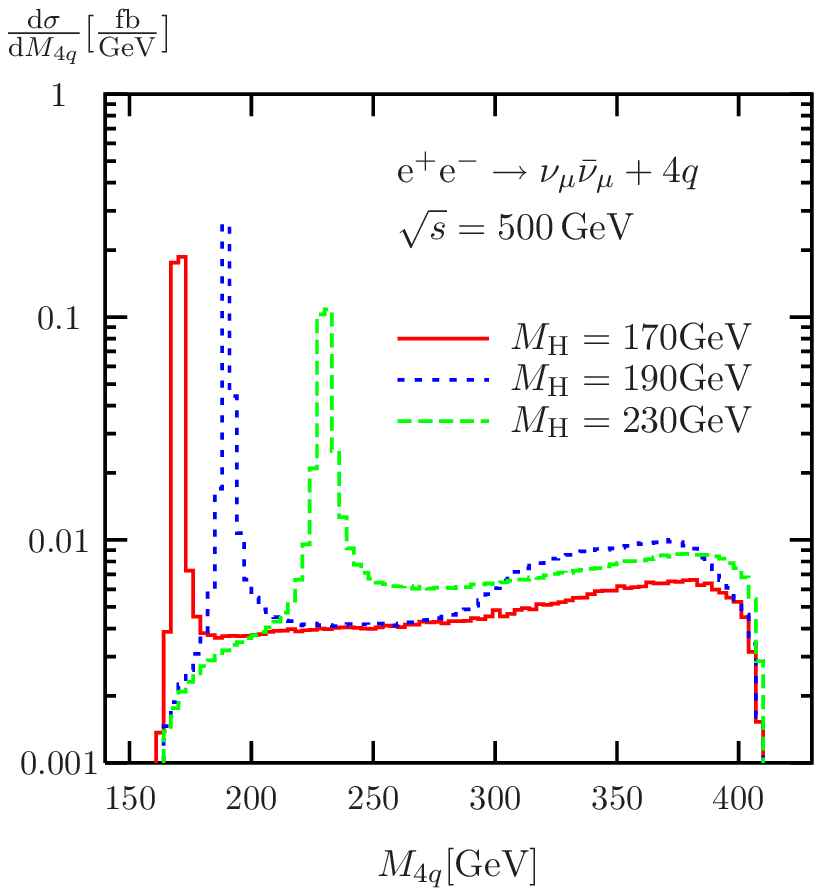}}
\put( 2.7,-15.1){\includegraphics{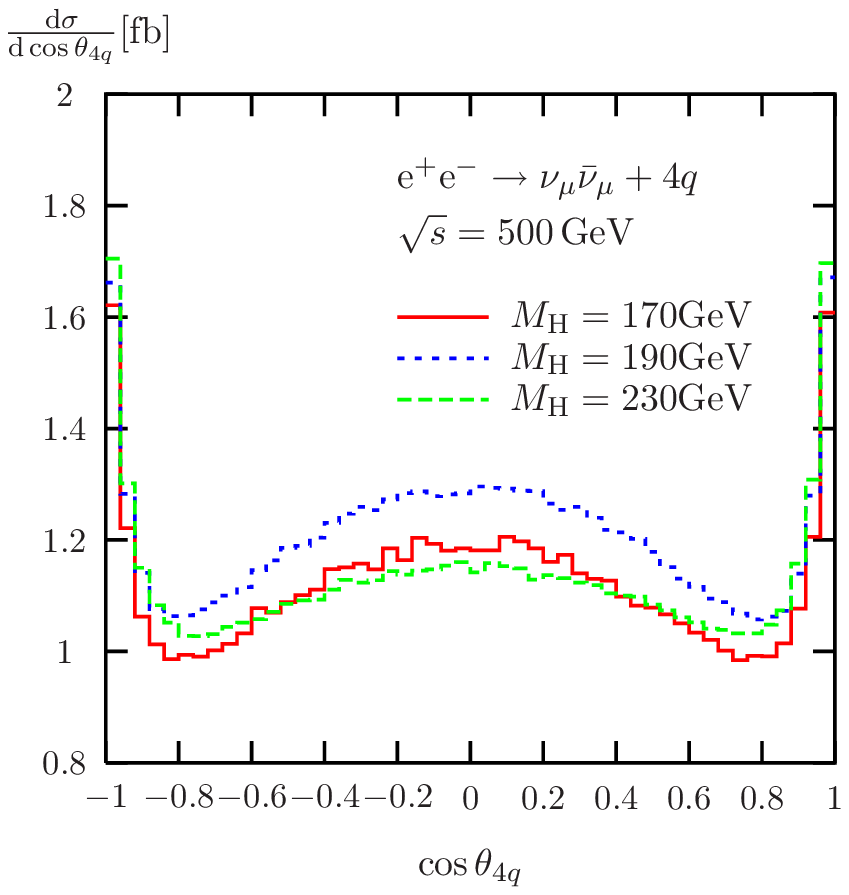}}
\end{picture} } 
\caption{Invariant-mass and angular distributions of the four-quark system
in $\Pep\Pem\to\nu_\mu\bar\nu_\mu+4q$ 
(without ISR and gluon-exchange diagrams) for various Higgs masses and
$\sqrt{s}=500\GeV$}
\label{fig:higgs-nmnm4q}
\vspace*{1em}
\centerline{
\begin{picture}(16.0,8.3)
\put(-6.0,-15.1){\includegraphics{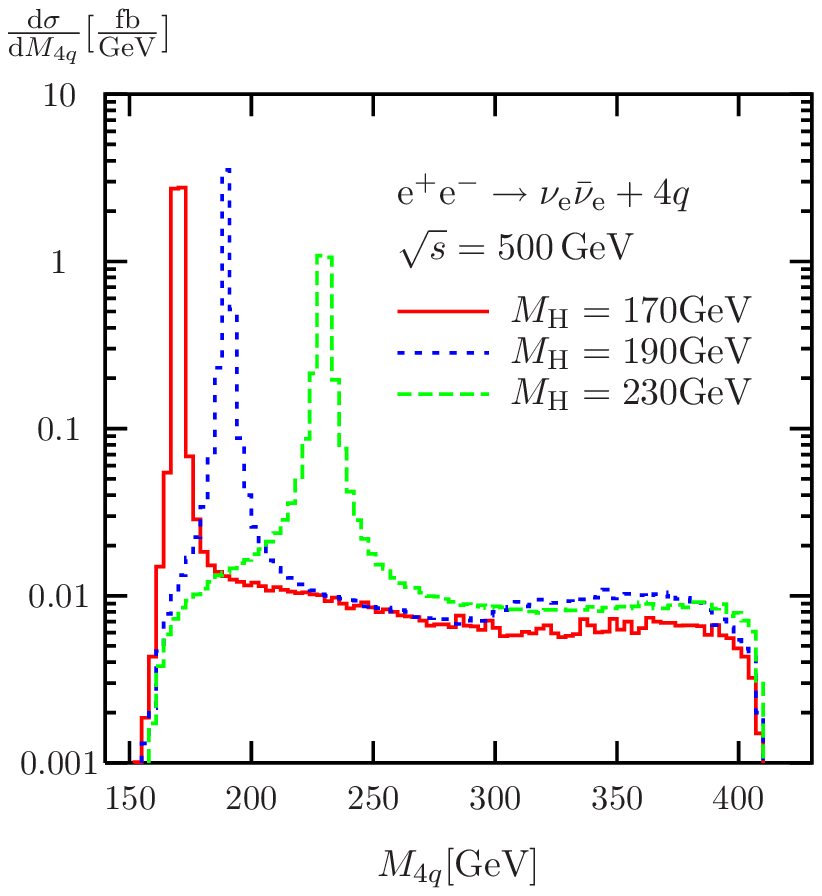}}
\put( 2.7,-15.1){\includegraphics{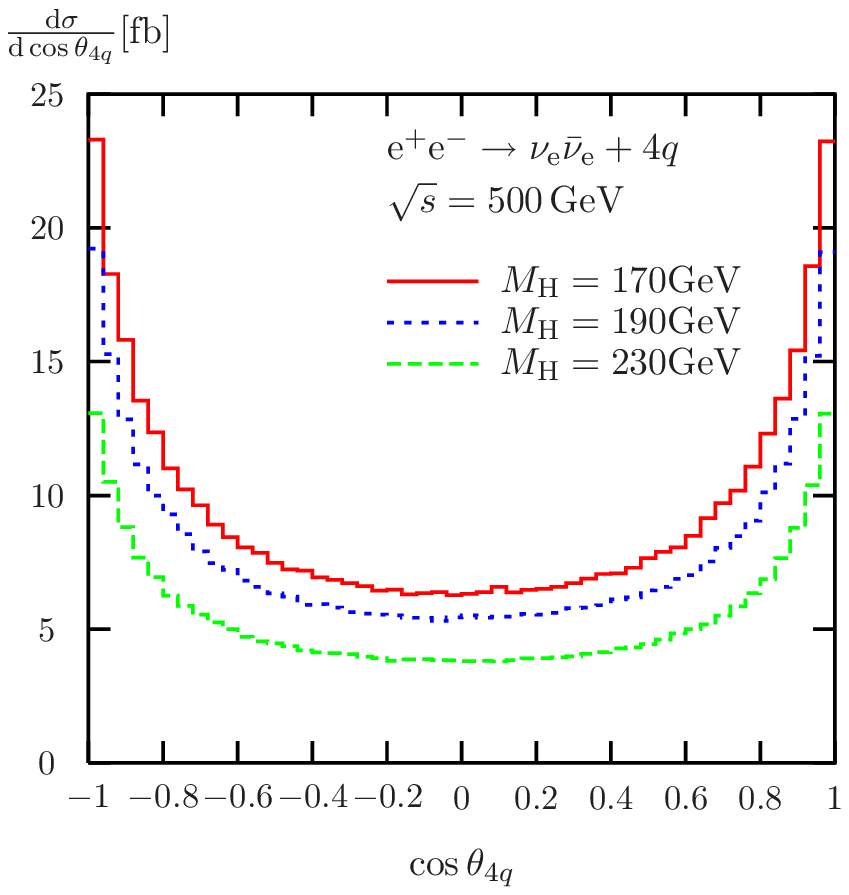}}
\end{picture} } 
\caption{Invariant-mass and angular distributions of the four-quark system
in $\Pep\Pem\to\nu_\Pe\bar\nu_\Pe+4q$ 
(without ISR and gluon-exchange diagrams) for various Higgs masses and
$\sqrt{s}=500\GeV$}
\label{fig:higgs-nene4q}
\efi
ISR and gluon-exchange diagrams are not included in these evaluations.
The crucial difference between the $\nu_\mu\bar\nu_\mu$ and 
$\nu_\Pe\bar\nu_\Pe$ channels lies in the Higgs production mechanisms:
while the former receives only contributions from $\PZ\PH$
production, the latter additionally involves W~fusion,
$\PW\PW\to\PH\to\PW\PW$, which dominates the cross section.
Therefore, the cross section of $\nu_\Pe\bar\nu_\Pe+4q$ is an order
of magnitude larger than the one of $\nu_\mu\bar\nu_\mu+4q$.
The invariant-mass distributions of the two channels look similar,
and for $\MH=170\GeV$ resemble the shape already observed for the
single channel $\Pep\Pem\to\nu_\mu\bar\nu_\mu\Pu\bar\Pd\Ps\bar\Pc$ 
in \reffi{fig:higgs-udscmassdist}. Note that for $\MH=190\GeV$ and
$230\GeV$ the high-energy tails of the distributions show some
Higgs mass dependence. This is due to the subprocess of $\PZ\PH$
production where the Higgs decays via 
$\PH\to\PZ\PZ\to (\nu_\mu\bar\nu_\mu/\nu_\Pe\bar\nu_\Pe)+2q$,
which is not yet possible for the smaller Higgs mass $\MH=170\GeV$.
The corresponding boundary in $M_{4q}$,
which is clearly seen in the plots for $\MH=190\GeV$, 
is determined by the two extreme situations where the decay $\PH\to\PZ\PZ$
proceeds along the $\PZ\PH$ production axis. 
For $\MH=230\GeV$ this boundary is hidden by the Higgs peak and the
upper kinematical limit in the $M_{4q}$ spectrum.
In contrast to the
invariant-mass distributions, the shape of the four-quark angular
distributions of the $\nu_\mu\bar\nu_\mu$ and $\nu_\Pe\bar\nu_\Pe$
channels look very different. For $\nu_\mu\bar\nu_\mu$, i.e.\
for $\PZ\PH$ production, intermediate
production angles dominate, and this dominance is more pronounced
for smaller Higgs-boson masses, where more phase space exists.
For $\nu_\Pe\bar\nu_\Pe$, i.e.\ W-boson fusion, forward and backward 
production of Higgs bosons is preferred, and the $\MH$ dependence is 
mainly visible in the overall scale of the distribution, but not 
in the shape itself.

\subsection{Finite-width decay widths and gauge-invariance violation}
\label{se:widthnum}

We conclude our discussion of numerical results 
by considering the behaviour of various
cross sections in the high-energy limit, using the different schemes
for introducing finite decay widths as described in \refse{se:width}.
Table~\ref{tab:wwww_width} shows the results for the three reactions
$\Pep\Pem\to\mu^-\bar\nu_\mu\Pu\bar\Pd\Pb\bar\Pb$,
$\Pep\Pem\to\mu^-\bar\nu_\mu+4q$, and
$\Pep\Pem\to\nu_\Pe\bar\nu_\Pe\mu^-\bar\nu_\mu\Pu\bar\Pd$, which are
typical representatives for top-quark pair production,
$\PW\PW\PZ$ production, and $\PW\PW\to\PW\PW$ scattering.
\begin{table}
\centerline{
\tabcolsep 3pt
\begin{tabular}{|c|c||c|c|c|c|c|}
\hline
\multicolumn{7}{|c|}{
$\sigma(\Pep\Pem\to\mu^-\bar\nu_\mu\Pu\bar\Pd\Pb\bar\Pb)\,[\fba]$}
\\ \hline \hline
\multicolumn{2}{|c||}{$\sqrt{s}[\GeV]$} & 500 & 800 & 1000 & 2000 & 10000
\\ \hline \hline
{\sc Lusifer} 
& fixed width /& 17.095(11) & 8.6795(83) & 6.0263(76) & 1.8631(31) & 0.08783(28)
\\[-.2em]
& step width  & & & & & 
\\ \cline{2-7} 
& running width & 17.106(10) & 8.6988(85) & 6.0700(73) & 2.3858(31) & 212.61(28)
\\ \cline{2-7} 
& complex mass & 17.085(10) & 8.6773(84) & 6.0249(76) & 1.8627(31) & 0.08800(32)
\\ \hline  \hline 
{\sc W.\&{}M.} & step width & 17.1025(80) & 8.6823(44) & 6.0183(31) & 
1.8657(12) & 0.08837(20)
\\ \hline 
\multicolumn{7}{c}{}
\\
\hline
\multicolumn{7}{|c|}{
$\sigma(\Pep\Pem\to\mu^-\bar\nu_\mu+4q)\,[\fba]$}
\\ \hline \hline
\multicolumn{2}{|c||}{$\sqrt{s}[\GeV]$} & 500 & 800 & 1000 & 2000 & 10000
\\ \hline \hline
{\sc Lusifer} 
& fixed width /& 2.9603(28) & 2.5949(26) & 2.3573(25) & 1.4055(20) & 0.22593(63)
\\[-.2em]
& step width  & & & & & 
\\ \cline{2-7}
& running width & 2.9845(25) & 2.7354(25) & 2.6829(27) & 5.2921(66) & 1623.0(30)
\\ \cline{2-7} 
& complex mass & 2.9600(25) & 2.5948(26) & 2.3559(25) & 1.4048(20) & 0.22532(63)
\\ \hline  \hline 
{\sc W.\&{}M.} & step width & 2.9616(13) & 2.5932(13) & 2.3611(13) & 1.4082(11) & 0.2224(12)
\\ \hline 
\multicolumn{7}{c}{}
\\
\hline
\multicolumn{7}{|c|}{
$\sigma(\Pep\Pem\to\nu_\Pe\bar\nu_\Pe\mu^-\bar\nu_\mu\Pu\bar\Pd)\,[\fba]$}
\\ \hline \hline
\multicolumn{2}{|c||}{$\sqrt{s}[\GeV]$} & 500 & 800 & 1000 & 2000 & 10000
\\ \hline \hline
{\sc Lusifer} 
& fixed width & 1.6326(12) & 4.1046(35) &  5.6795(61) & 11.736(16) & 26.380(55)
\\ \cline{2-7} 
& step width & 1.6333(12) & 4.1044(37) & 5.6720(56) & 11.734(15) & 26.380(55)
\\ \cline{2-7} 
& running width & 1.6398(12) & 4.1324(39) & 5.7206(54) & 12.881(14) & 12965(12)
\\ \cline{2-7} 
& complex mass & 1.6330(12) & 4.1037(34) & 5.6705(54) & 11.730(14) & 26.387(57)
\\ \hline  \hline 
{\sc W.\&{}M.} & step width & 1.6313(13) & 4.1053(35)& 5.6695(49) & 11.741(12) &
26.565(83)
\\ \hline 
\end{tabular} }
\caption{Born cross sections (without ISR and gluon-exchange diagrams) for 
$\Pep\Pem\to\mu^-\bar\nu_\mu\Pu\bar\Pd\Pb\bar\Pb$,
$\Pep\Pem\to\mu^-\bar\nu_\mu+4q$ and
$\Pep\Pem\to\nu_\Pe\bar\nu_\Pe\mu^-\bar\nu_\mu\Pu\bar\Pd$
for various CM energies and schemes for introducing decay widths}
\label{tab:wwww_width}
\end{table}
All three examples confirm the expectation from $4f(+\gamma)$ studies
that the fixed-width scheme, in spite of 
violating gauge invariance, practically yields the same results as
the complex-mass scheme that maintains gauge invariance. 
For the first two processes the step width and the fixed width lead
to the same results, since no propagators of
unstable particles with space-like momenta 
($t$-channel propagators) contribute, 
i.e.\ the widths in the propagators are never
switched off by the step function in Eq.~\refeq{eq:stepwidth}. 
For the last example, the difference between fixed and step widths is
also marginal.
In this context, the comparison with the {\sc Whizard} and {\sc Madgraph} 
results is particularly interesting, since {\sc Madgraph} employs the step 
width within the unitary gauge, in contrast to {\sc Lusifer}, where
the `t~Hooft--Feynman gauge is used. Thus, there is a difference between
the {\sc Madgraph} and {\sc Lusifer} results for the step width, since
gauge invariance is broken in this approach. However, this difference
is not yet numerically significant in the shown numbers.
Finally, all examples of \refta{tab:wwww_width} show
that the running-width scheme breaks gauge invariance so badly that
deviations from the complex-mass scheme are already visible below $1\TeV$.
Above $1\TeV$ these deviations grow rapidly, and the high-energy limit
of the prediction is totally wrong.

From these results we can draw similar conclusions as known from
$4f(+\gamma)$ production. If finite decay widths are introduced 
on cost of gauge invariance, the result is only reliable
if it has been compared to a gauge-invariant calculation, as it is for
instance provided by the complex-mass scheme. Moreover, our 
numerical studies show that the fixed-width scheme is in fact a good 
candidate for reliable results also in six-fermion 
production, although it does not respect 
gauge invariance. Whether this observation generalizes to
all $6f$ final states (or even further) is, however, not clear.

\section{Summary and outlook}
\label{se:sum}

The investigation of six-fermion production
is one of the most important tasks
at a future high-energy $\Pep\Pem$ collider
owing to a variety of interesting subprocesses leading to such
final states.
These, in particular, comprise top-quark pair production, 
massive vector-boson scattering, triple gauge-boson production,
and Higgs-boson production for intermediate Higgs masses.

In this paper, the Monte Carlo event generator {\sc Lusifer} 
has been introduced, which in its first version deals with all 
processes $\Pep\Pem\to 6\,$fermions at tree level in the Standard Model.
In the predictions all Feynman diagrams are included,
the number of which is typically of the order of $10^2$--$10^4$.
Fermions other than top quarks, which are not allowed as external fermions, 
are taken to be massless, and
polarization is fully supported. The helicity amplitudes are
generically calculated with spinor methods and are
presented explicitly.
The phase-space integration is based on the
multi-channel Monte Carlo integration technique. More precisely,
channels and appropriate mappings are provided for each 
individual diagram in a generic way. Owing to the potentially
large number of Feynman diagrams
per final state, an efficient generic approach has been crucial,
in order to gain an acceptable speed and stability of the program.
Initial-state radiation is included at the leading logarithmic
level employing the structure-function approach.
 
The performance of {\sc Lusifer} has been demonstrated in detail.
In particular, a comprehensive survey of cross section results is 
presented, including all final states with up to three neutrinos and
up to four quarks. Moreover, these cross sections are confronted
with results obtained with the multi-purpose packages {\sc Whizard}
and {\sc Madgraph} in a tuned comparison, as far as 
these programs were applicable. 
Apart from a few cases, where
the limitations of {\sc Whizard} and {\sc Madgraph} becomes visible,
we find good numerical agreement. 

We have supplemented the numerical results on cross sections by
presenting some distributions that are phenomenologically 
interesting for top-quark pair and Higgs-boson production. 
A comparison to {\sc Whizard} and {\sc Madgraph} results shows
also in this case good agreement within statistical errors.

Finally, we have numerically investigated 
possible effects from gauge-invariance violation due to the introduction 
of the finite decay widths of unstable particles in the amplitudes. 
Similarly to the known results in four-fermion
production, it turns out that the use of running decay widths
in general leads to a totally wrong high-energy behaviour of the cross
section, since gauge cancellations are disturbed. 
Although an
approach based on fixed gauge-boson widths does not maintain
gauge invariance either, the gauge-breaking effects are found to be
sufficiently suppressed in the considered examples. This conclusion
is based on a comparison with the gauge-invariant result obtained in 
the so-called 
complex-mass scheme, where gauge invariance is restored
by introducing appropriate complex couplings that are derived from 
complex mass parameters.

Apart from emphasizing the issue of speed and stability, a third
motivation in the construction of {\sc Lusifer} lies in the
simplicity and flexibility of both the underlying concept and the
actual computer code. This simplicity considerably facilitates
the task of going a step further in theoretical sophistication, 
i.e.\ beyond a tree-level Monte Carlo program improved by universal
corrections. The high accuracy of future $\Pep\Pem$ colliders 
requires the inclusion of non-universal radiative corrections to 
subprocesses such as top-quark pair and Higgs-boson production, 
or vector-boson scattering. To improve {\sc Lusifer} accordingly
will be the subject of future work.

\section*{Acknowledgements}

We thank Wolfgang Kilian for his aid in the {\sc Whizard} installation
and Tim Stelzer and David Rainwater for their help in the
application of {\sc Madgraph} to six-fermion production processes.
Moreover, Ansgar Denner is gratefully acknowledged for discussions and
for carefully reading the manuscript.
This work was partially supported by the European Commission 5th 
framework contract HPRN-CT-2000-00149.

\end{document}